\documentclass[acmsmall,screen]{acmart}
\acmYear{2024}
\acmMonth{1}
\acmDOI{} 
\startPage{1}

\setcopyright{none}

\usepackage[utf8]{inputenc}
\usepackage{algpseudocode}
\usepackage{listings}
\usepackage{tikz}
\usepackage{xcolor}
\usepackage{float}
\usepackage{multirow}
\usepackage{booktabs}
\usepackage{xspace}
\usepackage{caption}
\usepackage{subcaption}
\usepackage{balance}

\usepackage{amsmath}
\theoremstyle{definition}

\usepackage{pgfplots}
\pgfplotsset{compat=1.17}

\makeatletter
\newenvironment{btHighlight}[1][]
{\begingroup\tikzset{bt@Highlight@par/.style={#1}}\begin{lrbox}{\@tempboxa}}
	{\end{lrbox}\bt@HL@box[bt@Highlight@par]{\@tempboxa}\endgroup}

\newcommand\btHL[1][]{%
	\begin{btHighlight}[#1]\bgroup\aftergroup\bt@HL@endenv%
	}
	\def\bt@HL@endenv{%
	\end{btHighlight}%
	\egroup
}
\newcommand{\bt@HL@box}[2][]{%
	\tikz[#1]{%
		\pgfpathrectangle{\pgfpoint{1pt}{0pt}}{\pgfpoint{\wd #2}{\ht #2}}%
		\pgfusepath{use as bounding box}%
		\node[anchor=base west, fill=orange!25,outer sep=.5pt,inner xsep=0.5pt, inner ysep=0.15pt, rounded corners=1pt, minimum height=\ht\strutbox-.1pt,#1]{\raisebox{.01pt}{\strut}\strut\usebox{#2}};
	}%
}
\makeatother

\newcommand\redout{\bgroup\markoverwith
	{\textcolor{red}{\rule[.45ex]{1.5pt}{1.pt}}}\ULon}

\lstdefinestyle{mystyle}{
	frame=single,
	framexleftmargin=0pt,
	commentstyle=\color{ForestGreen},
	keywordstyle=\color{blue}\bfseries,
	numberstyle=\tiny\color{gray},
	stringstyle=\color{purple},
	basicstyle=\tiny\ttfamily\bfseries,
	breakatwhitespace=false,         
	breaklines=false,                 
	captionpos=b,                    
	keepspaces=true,     
	numbers=none,                    
	numbersep=4pt,                  
	showspaces=false,                
	showstringspaces=false,
	showtabs=false,                  
	tabsize=2,
	language=Java,
	escapechar=|,
        moredelim=**[is][{\btHL[fill=red!40]}]{@}{@},
}
\definecolor{light-gray}{gray}{0.80}
\definecolor{ForestGreen}{RGB}{63,147,88}

\lstdefinestyle{lstinlinestyle}{
	columns=fixed, 
        style=mystyle, 
        basicstyle=\footnotesize\ttfamily\bfseries,
}

\newcommand*\circled[1]{\tikz[baseline=(char.base)]{
            \node[shape=circle,draw,inner sep=0.1pt] (char) {#1};}}

\usepackage{soul}
\usetikzlibrary{calc}

\makeatletter
\newif\if@anonymize

\@anonymizefalse  

\if@anonymize
  \newcommand{\highlight@DoHighlight}{
    \fill [outer sep = -15pt, inner sep = 0pt, color=black]
          ($(begin highlight)+(0,8pt)$) rectangle ($(end highlight)+(0,-3pt)$) ;
  }

  \newcommand{\highlight@BeginHighlight}{
    \coordinate (begin highlight) at (0,0) ;
  }

  \newcommand{\highlight@EndHighlight}{
    \coordinate (end highlight) at (0,0) ;
  }

  \newdimen\highlight@previous
  \newdimen\highlight@current
  \newlength{\item@width}

  \DeclareRobustCommand*\anonymize{%
    \SOUL@setup
    \def\SOUL@preamble{%
      \begin{tikzpicture}[overlay, remember picture]
        \highlight@BeginHighlight
        \highlight@EndHighlight
      \end{tikzpicture}%
    }%
    \def\SOUL@postamble{%
      \begin{tikzpicture}[overlay, remember picture]
        \highlight@EndHighlight
        \highlight@DoHighlight
      \end{tikzpicture}%
    }%
    \def\SOUL@everyhyphen{%
      \discretionary{%
        \SOUL@setkern\SOUL@hyphkern
        \SOUL@sethyphenchar
        \tikz[overlay, remember picture] \highlight@EndHighlight ;%
      }{%
      }{%
        \SOUL@setkern\SOUL@charkern
      }%
    }%
    \def\SOUL@everyexhyphen##1{%
      \SOUL@setkern\SOUL@hyphkern
      \settowidth{\item@width}{##1}%
      \makebox[\item@width]{}%
      \discretionary{%
        \tikz[overlay, remember picture] \highlight@EndHighlight ;%
      }{%
      }{%
        \SOUL@setkern\SOUL@charkern
      }%
    }%
    \def\SOUL@everysyllable{%
      \begin{tikzpicture}[overlay, remember picture]
        \path let \p0 = (begin highlight), \p1 = (0,0) in \pgfextra
          \global\highlight@previous=\y0
          \global\highlight@current =\y1
        \endpgfextra (0,0) ;
        \ifdim\highlight@current < \highlight@previous
          \highlight@DoHighlight
          \highlight@BeginHighlight
        \fi
      \end{tikzpicture}%
      \settowidth{\item@width}{\the\SOUL@syllable}%
      \makebox[\item@width]{}%
      \tikz[overlay, remember picture] \highlight@EndHighlight ;%
    }%
    \SOUL@
  }
\else
  \newcommand{\anonymize}[1]{#1}
\fi
\makeatother

\usepackage{xcolor,amsmath}
\usepackage[linesnumbered,ruled,vlined]{algorithm2e}
\DontPrintSemicolon

\SetKwComment{Comment}{\color{green!50!black}// }{}

\newcommand{\assign}{\leftarrow}

\newcommand{\FuncCall}[2]{\texttt{\bfseries #1(#2)}}
\SetKwProg{Function}{function}{}{}

\newcommand{\eg}{\hbox{\emph{e.g.,}}\xspace}
\newcommand{\ie}{\hbox{\emph{i.e.,}}\xspace}

\title{Finding Cross-rule Optimization Bugs in Datalog Engines}

\author{Chi Zhang}
\affiliation{
  \department{State Key Laboratory for Novel Software Technology, Department of Computer Science and Technology}
  \institution{Nanjing University}
  \city{Nanjing}
  \country{China}
  \thanks{This work was completed during Chi Zhang's visit to National University of Singapore.}
}
\email{zhangchi_seg@smail.nju.edu.cn}

\author{Linzhang Wang}
\authornote{Corresponding authors.}
\affiliation{
  \department{State Key Laboratory for Novel Software Technology, Department of Computer Science and Technology}
  \institution{Nanjing University}
  \city{Nanjing}
  \country{China}
}
\email{lzwang@nju.edu.cn}

\author{Manuel Rigger}
\authornotemark[1]
\affiliation{
  \institution{National University of Singapore}
  \country{Singapore}
}
\email{rigger@nus.edu.sg}

\newcommand*{\tool}{\textsc{Deopt}\xspace}
\newcommand*{\approach}{IRE\xspace}

\newcommand*{\totalbugs}{30\xspace}
\newcommand*{\totallogicbug}{13\xspace}
\newcommand*{\confirmedlogicbug}{6\xspace}
\newcommand*{\needtobeconfirm}{7\xspace}
\newcommand*{\fixedlogicbug}{2\xspace}

\newcommand*{\totalfixedbug}{6\xspace}
\newcommand*{\totalconfirmedbug}{19\xspace}
\newcommand*{\totalneedtobeconfirmed}{11\xspace}

\begin{document}

\begin{abstract}
Datalog is a popular and widely-used declarative logic programming language.
Datalog engines apply many cross-rule optimizations; bugs in them can cause incorrect results.
To detect such optimization bugs, we propose an automated testing approach called \emph{Incremental Rule Evaluation (\approach)}, which synergistically tackles the test oracle and test case generation problem. The core idea behind the test oracle is to compare the results of an optimized program and a program without cross-rule optimization; any difference indicates a bug in the Datalog engine.
Our core insight is that, for an optimized, incrementally-generated Datalog program, we can evaluate all rules individually by constructing a reference program to disable the optimizations that are performed among multiple rules. Incrementally generating test cases not only allows us to apply the test oracle for every new rule generated---we also can ensure that every newly added rule generates a non-empty result with a given probability and eschew recomputing already-known facts.
We implemented \approach as a tool named \tool, and evaluated \tool on four mature Datalog engines, namely Souffl\'{e}, CozoDB, µZ, and DDlog, and discovered a total of \totalbugs bugs. Of these, \totallogicbug were logic bugs, while the remaining were crash and error bugs. 
\tool can detect all bugs found by queryFuzz, a state-of-the-art approach. Out of the bugs identified by \tool, queryFuzz might be unable to detect 5. Our incremental test case generation approach is efficient; for example, for test cases containing 60 rules, our incremental approach can produce 1.17$\times$ (for DDlog) to 31.02$\times$ (for Souffl\'{e}) as many valid test cases with non-empty results as the naive random method. 
We believe that the simplicity and the generality of the approach will lead to its wide adoption in practice.
\end{abstract}

\maketitle

\section{INTRODUCTION}

Datalog is a popular declarative logic programming language supported by deductive database systems~\cite{ramakrishnan1995survey,ceri1989you}, which can conceptually be seen as relational database systems that have also logic programming capabilities.
Initially, most Datalog dialects considered only relations, which are the basic components of rules, as well as facts, which can be used to represent data. 
Many Datalog dialects now also include language features such as constraints~\cite{877512, 10.1007/3-540-63255-7_2}, aggregates~\cite{10.1007/3-540-53507-1_90, 761663}, and subsumptions~\cite{kiessling1994database, 10.1145/3379446}, which have expanded the functionality of Datalog significantly.
In recent years, Datalog has been applied to important domains, including distributed systems~\cite{conway2012logic}, networking~\cite{loo2009declarative}, natural language processing~\cite{mooney1996inductive}, program understanding~\cite{hajiyev2006codequest}, program analysis~\cite{bravenboer2009strictly, lam2005context, whaley2004cloning, bembenek2020formulog, 10.1145/2892208.2892226}, big-data analytics~\cite{shkapsky2016big}, and new programming languages that embed or are based on Datalog~\cite{DBLP:conf/pldi/MadsenYL16, DBLP:books/mc/18/Borraz-SanchezKPA18}. 

Mature Datalog engines such as Souffl\'{e}~\cite{jordan2016souffle}, DDlog~\cite{ryzhyk2019differential}, XSB~\cite{10.1145/191839.191927}, and LogicBox~\cite{aref2015design} are highly optimized in terms of performance~\cite{10.1145/3453483.3454070, 10.1007/978-3-031-16767-6_5} and memory usage~\cite{ryzhyk2019differential}.  Many evaluation strategies and optimizations they implement are applied to efficiently evaluate rules. Commonly, they use bottom-up or top-down evaluation methods~\cite{ullman1989bottom}. In terms of optimizations, magic sets~\cite{10.1145/6012.15399, alviano2012magic} are applied as a logical rewriting technique to transform a Datalog program into an equivalent more efficient one. Join optimizers~\cite{10.1007/978-3-031-16767-6_5} can find an efficient order in which to join the relations in a rule. Automatic index selection~\cite{subotic2018automatic} and index sharing~\cite{ryzhyk2019differential} significantly reduce memory usage. Furthermore, various data structures~\cite{jordan2022specializing, jordan2019specialized, jordan2019brie, nappa2019fast} have been proposed to speed up the execution of Datalog programs.

Due to the complexity of modern Datalog engines, they can be affected by bugs. 
Logic bugs are a particularly notorious category of bugs, as they cause Datalog engines to silently produce incorrect results for a given program. 
As stated by a developer in response to a bug that we have found in their Datalog engine: ``The worst thing a database can do is to give wrong outputs.''\footnote{\url{https://github.com/cozodb/cozo/issues/\anonymize{101}}}
Determining the expected result for a complex query can be difficult for users and even developers, making it challenging to identify logic bugs.
In fact, the main contributor of Souffl\'{e} emphasized in a recent keynote that designing novel ``automated testing tools'' for Datalog engines is one of the main open challenges in this domain~\cite{scholz2022keynote}.
This is also suggested by empirical evidence; we checked 222 issues in the GitHub repository of Souffl\'{e} that were tagged with either ``bug-triage'' or ``bug-identified'' labels prior to February 2023, and discovered that 39 of them were logic bugs.
More significantly, out of the 39 issues, 21 were caused by optimization bugs, which are logic bugs caused by incorrect optimizations. 
To detect logic bugs, queryFuzz~\cite{10.1145/3468264.3468573} was proposed as an automated testing approach for Datalog engines, which has found 8 logic bugs in Souffl\'{e}.
Our manual analysis suggests that all 8 bugs were, in fact, optimization bugs.
Due to the prevalence of optimization bugs, we believe that a general technique for automatically finding logic bugs caused by optimizations in Datalog engines specifically is needed.

Non-optimizing Reference Engine Construction (NoREC)~\cite{10.1145/3368089.3409710} was proposed to find optimization bugs in relational database systems. Its key insight is that a query can be rewritten so that the database engine can no longer effectively optimize it; a discrepancy in the results of the two queries indicates a bug in the database system. The insights that enabled the approach are specific to SQL, 
and it is unclear how the approach could be used to test Datalog engines. Another intuitive idea to detect optimization bugs in Datalog engines would be to apply differential testing~\cite{mckeeman1998differential}, by running the same Datalog program in different Datalog engines, which should all produce the same results.
However, different Datalog engines share only core features, and most of the powerful functionalities of Datalog engines are implemented by unique features; for example, the subsumption and equivalence relation in Souffl\'{e} are not supported by other Datalog engines. In addition, most Datalog engines do not offer optimization options, making it difficult to perform differential testing by disabling optimizations on the same engine.

In this work, we propose \emph{Incremental Rule Evaluation (\approach)}, a novel black-box approach for testing Datalog engines based on the idea of disabling the optimizations that can be applied in the presence of multiple rules (\ie cross-rule optimizations), 
allowing us to compare an unoptimized execution of a Datalog program with an optimized one. 
Our technique is incremental, which allows us to build complex programs with non-empty results while efficiently computing the results of the unoptimized program. 
We have two core insights that enable our approach. First, Datalog programs consist of both facts and rules. While facts represent data and thus can hardly be optimized, many optimizations have been proposed that result in rules being processed more efficiently. Second, given a Datalog program, we can derive all facts from the rules using an existing Datalog engine. Overall, these two insights allow us to provide a conceptual reference engine by incrementally adding rules to a Datalog program---ensuring that they produce non-empty results---and maintaining a version of the program that removes the rules after deriving all possible facts. The modified version cannot be effectively optimized by the Datalog engine, since only one rule is visible at a time, thus enabling the detection of cross-rule optimization bugs by comparing the results of the real optimizing Datalog engine, and the conceptual non-optimizing engine.
Even for Datalog engines that provide optimization flags that allow selectively disabling optimizations, our technique is useful, as the incremental construction of the unoptimized program is more efficient than simply disabling the optimizations of the optimized programs, as all intermediate results---which are needed to generate non-empty results---would need to be recomputed for every version of the optimized program, rather than being re-used as for IRE.

We implemented \approach as a tool called \tool---derived from \textbf{D}atalog \textbf{E}ngine \textbf{Op}timization \textbf{T}ester---and tested four mature and well-tested Datalog engines with it. 
We have found \totalbugs bugs in these engines, out of which \totallogicbug bugs were previously unknown logic bugs.
Of these, \totalconfirmedbug were confirmed and \totalfixedbug of them were fixed. A further \totalneedtobeconfirmed are awaiting confirmation. 
For the logic bugs, \confirmedlogicbug were confirmed, out of which \fixedlogicbug were fixed, and \needtobeconfirm are awaiting confirmation.
Our incremental test case generation approach is more efficient than a random approach when generating large test cases; for the test cases containing 60 rules, our incremental approach can produce 1.17$\times$ (on DDlog) to 31.02$\times$ (on Souffl\'{e}) as many valid test cases with non-empty results. The performance of our method improves significantly as the test case size increases.
Finally, our approach is effective also in comparison to a state-of-the-art approach implemented by queryFuzz;
\approach can detect 12 out of 13 bugs found by queryFuzz; an extension of \tool enables us to detect the remaining bug. Out of the bugs identified by \tool, our analysis suggests that queryFuzz might be unable to detect 5. 

In summary, we make the following main contributions:
\begin{itemize}
    \item We present a novel testing idea for finding cross-rule optimization bugs in Datalog engines termed \emph{Incremental Rule Evaluation (\approach)}.
    \item We realized \approach through an approach that handles complex features such as recursion and negation in a general way.
    \item We implemented \approach as an open-source tool named \tool and conducted an extensive evaluation in terms of effectiveness, efficiency, and a comparison with queryFuzz.
\end{itemize}

\section{BACKGROUND}
\label{sect:background}

This section provides a minimal introduction to Datalog, which is necessary to understand the paper's contribution.
We refer interested readers to relevant textbooks for further details~\cite{ketsman2022modern, green2013datalog}.

\begin{figure}
    \centering
    \parbox{0.7\textwidth}{
        \lstset{style=mystyle, numbers=left, framexleftmargin=7pt}
\begin{lstlisting}[basicstyle=\scriptsize\ttfamily\bfseries]
PointsTo(x, y) :- AddressOf(x, y). |\label{codeline:addressof}|
PointsTo(x, y) :- Assign(x, z), PointsTo(z, y). |\label{codeline:recursion}|
PointsTo(x, w) :- Load(x, z), PointsTo(z, y), PointsTo(y, w). 
PointsTo(x, y) :- Store(z, w), PointsTo(z, x), PointsTo(w, y).
\end{lstlisting}
    }
    \caption{A Datalog program used in points-to analysis.}
    \label{fig:pointsto}
\end{figure}

\paragraph{Datalog}
A Datalog \textit{program} $P$ consists of a set of \textit{rules}, where each rule follows the format:
$$A :- B_1, B_2, ... , B_m.$$
$A$ serves as the \textit{head} of the rule, while $B_1, B_2, ... , B_m$ constitute its \textit{body}.
Both $A$ and $B_m$ are \textit{relations}, where each relation consists of a set of ordered tuples. The number of \textit{variables} within each tuple is referred to as the \textit{arity} of the relation.
A relation is formally denoted as $R(t_1, t_2, ..., t_n)$, where $R$ is the name of this relation, $n$ represents the arity of $R$, and $t_1, ..., t_n$ form the tuple of $R$; note that the actual syntax depends on the system as detailed below.

The Datalog program shown in Figure~\ref{fig:pointsto} demonstrates how to conduct a points-to analysis ---a popular use-case for Datalog~\cite{bravenboer2009strictly}---for a C-like language.
In this Datalog program, the relation \lstinline[style=lstinlinestyle]{AddressOf(a, b)} represents the relationship \lstinline[style=lstinlinestyle]{a = &b}; \lstinline[style=lstinlinestyle]{Assign(a, b)} represents \lstinline[style=lstinlinestyle]{a = b}; \lstinline[style=lstinlinestyle]{Load(a, b)} represents  \lstinline[style=lstinlinestyle]{a = *b}; and \lstinline[style=lstinlinestyle]{Store(a, b)} represents  \lstinline[style=lstinlinestyle]{*a = b}.
The program has four rules, and all of them take the relation \lstinline[style=lstinlinestyle]{PointsTo} as head, which has two variables \lstinline[style=lstinlinestyle]{a} and \lstinline[style=lstinlinestyle]{b}, representing that \lstinline[style=lstinlinestyle]{a} may point to \lstinline[style=lstinlinestyle]{b}.
For a given \textit{fact} \lstinline[style=lstinlinestyle]{AddressOf("var1", "var2")}, which serves as the program input, we can derive the new fact \lstinline[style=lstinlinestyle]{PointsTo("var1", "var2")} based on the first rule.

\paragraph{Recursion}
Recursion in Datalog allows for more intricate applications.
The code presented in line~\ref{codeline:recursion} of Figure~\ref{fig:pointsto} exemplifies the application of recursion in Datalog.
To evaluate a program with recursion, the Datalog engine will repeatedly evaluate program rules until no further facts can be derived, signifying that a \textit{least-fixpoint} has been reached~\cite{abiteboul1995foundations}.

\paragraph{Negation}
Negation enhances Datalog's functionality.
Suppose we want to find the variable pairs that do not have a points-to relationship, we can search for pairs not present in the points-to collection with the following rule: 
\lstinline[style=lstinlinestyle]{NoPointsTo(x, y) :- Var(x), Var(y), not PointsTo(x, y).} 
The relation \lstinline[style=lstinlinestyle]{Var} contains all the variables in the program that are subject to the points-to analysis.
Negation in Datalog has a safety restriction: every variable in a rule's body must appear in at least one positive relation.
Core Datalog does not support negation because it cannot express non-monotonic queries.
To support negation in Datalog, two extensions in evaluation are commonly used: semi-positive~\cite{abiteboul1995foundations} and stratified Datalog~\cite{10.1145/50202.50218}.
Semi-positive Datalog allows negation to only appear on the input relation in rule bodies.
Stratified Datalog permits negation on any relation except within recursion.
Our approach utilizes stratified Datalog to handle cases in the program that contain negation.

\sloppy{}
\paragraph{Variability across Datalog engines}
Various Datalog dialects exist that differ significantly from each other.
Firstly, the syntax for the core language differs widely, including the use of punctuation and case sensitivity requirements.
For example, CozoDB represents the tuples in a relation by enclosing them in \lstinline[style=lstinlinestyle]{[]} rather than the standard notation introduced above, which uses parentheses (\ie \lstinline[style=lstinlinestyle]{()}).
Assuming CozoDB's syntax, the rule shown in line~\ref{codeline:addressof} of Figure~\ref{fig:pointsto} is \lstinline[style=lstinlinestyle]{PointsTo[x, y] : - AddressOf[x ,y]}.
Souffl\'{e}, µZ, and DDlog require a period at the end of a rule, whereas CozoDB does not.
DDlog is case-sensitive for identifiers; for example, the name of a relation must start with an uppercase letter, whereas other Datalog engines do not have this requirement.
Secondly, the syntax and semantics in which various Datalog engines handle inputs and outputs differ. 
Inputs can be sourced using multiple ways, such as files, databases, or direct declarations within the program.
Datalog engines offer varying levels of support for these input methods.
Souffl\'{e} provides multiple methods to load inputs and save outputs, such as CSV or text files, SQLite3 databases, and more.
CozoDB uses text files to load normal values, and leverages the BlobDB feature of RocksDB---a key-value database system---for loading large values.
µZ and DDlog use text files to handle inputs.
Souffl\'{e}, µZ, and DDlog also allow using a rule without a body as the declaration of a fact (\eg 
\lstinline[style=lstinlinestyle]{A(1, 2)}), which can be used as the input of a Datalog program.
In CozoDB, the manner of input declaration stands out. One such declaration form is \lstinline[style=lstinlinestyle]{A[a, b] <-[[1, 2]]}.
CozoDB designates the result of a specific rule as its output. In contrast, 
Souffl\'{e}, µZ, and DDlog use keywords to specify relations as output. 
Thirdly, different Datalog engines support varying language features.
For example, equivalence relations and subsumption are only supported by Souffl\'{e}.
Lastly, the declaration of relations is necessary for Souffl\'{e}, µZ, and DDlog, whereas it is not required for CozoDB.
In Souffl\'{e}, the declaration of relations allows us to dictate the optimizations applied to these relations---a feature other engines lack.
Thus, it is challenging to apply differential testing as a general technique to identify logical bugs within engines by either (1) controlling which optimizations are applied using flags or declarations and (2) comparing the results of multiple Datalog engines.
Our oracle relies on the Datalog engine under test to compute the results of two equivalent programs; thus it can be applied to any dialect, irrespective of whether the Datalog engine provides means to the user to control optimizations.
\section{ILLUSTRATIVE EXAMPLE}
\label{sect:example}

\begin{figure}
    \centering
    \begin{subfigure}{0.6\textwidth}
        \lstset{style=mystyle, numbers=left, framexleftmargin=7pt}
\begin{lstlisting}[morekeywords={decl, output, symbol, number, magic, no_magic, inline},basicstyle=\scriptsize\ttfamily\bfseries]
node[A] := A = 1 |\label{codeline:node}|
edge[A, A] := A = 1 |\label{codeline:edge1}|
edge[A, B] := A = 0, B = 1, node[B] |\label{codeline:edge2}|
result[C] := edge[A, _], edge[C, A] |\label{codeline:output}|
?[C] := result[C]
\end{lstlisting}
        \vspace{-7pt}
        \caption{A test case generated with random method in four iterations.}
        \vspace{5pt}
        \label{fig:mainexampletestcase}
    \end{subfigure}
    
    \begin{subfigure}{0.45\textwidth}
        \lstset{style=mystyle, numbers=left, framexleftmargin=7pt}
\begin{lstlisting}[morekeywords={decl, output, symbol, number, magic},basicstyle=\scriptsize\ttfamily\bfseries]
node[A] := A = 1
?[A] := node[A]
\end{lstlisting}
        \vspace{-7pt}
        \caption{The reference program in the first iteration.}
        \vspace{5pt}
        \label{fig:mainexamplereference1}
    \end{subfigure}
    \qquad\;
    \begin{subfigure}{0.45\textwidth}
        \lstset{style=mystyle, numbers=left, framexleftmargin=7pt}
\begin{lstlisting}[morekeywords={decl, output, symbol, number, magic, no_magic},basicstyle=\scriptsize\ttfamily\bfseries]
edge[A, A] := A = 1
?[A, B] := edge[A, B]
\end{lstlisting}
        \vspace{-7pt}
        \caption{The reference program in the second iteration.}
        \vspace{5pt}
        \label{fig:mainexamplereference2}
    \end{subfigure}

    \begin{subfigure}{0.45\textwidth}
        \lstset{style=mystyle, numbers=left, framexleftmargin=7pt}
\begin{lstlisting}[morekeywords={decl, output, symbol, number, magic, no_magic},basicstyle=\scriptsize\ttfamily\bfseries]
node[A] <- [[1]]
edge[A, B] := A = 0, B = 1, node[B]
?[A, B] := edge[A, B]
\end{lstlisting}
        \vspace{-7pt}
        \caption{The reference program in the third iteration.}
        \vspace{-7pt}
        \label{fig:mainexamplereference3}
    \end{subfigure}
    \qquad\;
    \begin{subfigure}{0.45\textwidth}
        \lstset{style=mystyle, numbers=left, framexleftmargin=7pt}
\begin{lstlisting}[morekeywords={decl, output, symbol, number, magic, no_magic},basicstyle=\scriptsize\ttfamily\bfseries]
edge[A, B] <- [[1, 1], [0, 1]] |\label{codeline:inputs}|
result[C] := edge[A, _], edge[C, A] |\label{codeline:mainrule}|
?[C] := result[C] |\label{codeline:outputrule}|
\end{lstlisting}
        \vspace{-7pt}
        \caption{The reference program in the fourth iteration.}
        \vspace{-7pt}
        \label{fig:mainexamplereference4}
    \end{subfigure}
    \caption{Illustrative example, triggering a bug in CozoDB.}
    \label{fig:mainexample}
\end{figure}


\paragraph{Example program} To illustrate the intuition behind \approach, we present a bug-inducing test case\footnote{\url{https://github.com/cozodb/cozo/issues/\anonymize{101}}} from CozoDB~\cite{CozoDB} that our approach allowed us to find. 
For presentation purposes, we show the bug using an artificial, but illustrative scenario.
Assume that we want to identify all the nodes in a directed graph that have successors two steps away. One step is from a node to its neighbor. For example, given the directed edge \lstinline[style=lstinlinestyle]{edge[0, 1]} and \lstinline[style=lstinlinestyle]{edge[1, 1]}, the result should be nodes \lstinline[style=lstinlinestyle]{0} and \lstinline[style=lstinlinestyle]{1}. 
We use the Datalog program shown in Figure~\ref{fig:mainexampletestcase} to implement this task. 
The first rule (\ie line~\ref{codeline:node}) derives \lstinline[style=lstinlinestyle]{node[1]}.
In practice, users would likely specify nodes using facts (\eg \lstinline[style=lstinlinestyle]{node[A] <- [[1]]}).
However, in this specific case, the bug can only be reproduced when using rules.
Following the same pattern, the second rule (\ie line~\ref{codeline:edge1}) derives a directed edge \lstinline[style=lstinlinestyle]{[1, 1]}. 
The third rule (\ie line~\ref{codeline:edge2}) derives an edge \lstinline[style=lstinlinestyle]{[0, 1]} when \lstinline[style=lstinlinestyle]{node[1]} exists. 
The relation \lstinline[style=lstinlinestyle]{edge} is defined by two rules (\ie line~\ref{codeline:edge1} and \ref{codeline:edge2}), which can be viewed as disjunction rules. 
Finally, the fourth rule (\ie line~\ref{codeline:output}) accomplishes our task by finding a node \lstinline[style=lstinlinestyle]{C} that satisfies the condition of having an edge from \lstinline[style=lstinlinestyle]{C} to \lstinline[style=lstinlinestyle]{A}, while also having an edge from \lstinline[style=lstinlinestyle]{A} to any other nodes. 
Then, node \lstinline[style=lstinlinestyle]{C} is the desired node that has successors two steps away, and encompasses the facts of the relation \lstinline[style=lstinlinestyle]{result}.

\paragraph{Optimization bug} For this program, CozoDB incorrectly produced \lstinline[style=lstinlinestyle]{1}, instead of the expected output of \lstinline[style=lstinlinestyle]{0} and \lstinline[style=lstinlinestyle]{1}, due to an optimization bug. 
A CozoDB developer acknowledged
that this bug was intricate and difficult to explain. Its root cause was an implementation error in the semi-naïve algorithm~\cite{balbin1987generalization}, which is a bottom-up evaluation strategy capable of handling recursive queries.
Differential testing would have failed to detect this bug, due to CozoDB's unique Datalog dialect, which has a different syntax than other Datalog engines, as detailed in Section~\ref{sect:background}. 

\paragraph{Test oracle} The core idea for our test oracle is to individually execute each rule in the original program, disabling all optimizations that consider multiple rules by making only a single rule visible to Datalog engines at any point in time.
Using this bug as an example, we can demonstrate how to generate a test oracle that might detect optimization bugs. 
We refer to the original test case shown in Figure~\ref{fig:mainexampletestcase} as the \textit{optimized program}, because it is receptive to cross-rule optimizations by the Datalog engine. 
In order to generate a test oracle for this program,  we aim to construct a \textit{reference program},
which computes the same result as the optimized program, without being receptive to cross-rule optimizations.
Figure~\ref{fig:mainexamplereference4} derives the facts for the same output relation as the optimized program. 
However, we cannot construct it directly as we do not have access to the necessary information on the facts of \lstinline[style=lstinlinestyle]{edge}, which are required as inputs for the reference program. 
Thus, we build the programs step-by-step as shown in Figure~\ref{fig:mainexamplereference1}, \ref{fig:mainexamplereference2}, and \ref{fig:mainexamplereference3}, which contain only one rule each, preventing any potential cross-rule optimizations. The inputs for these programs are obtained from the outputs of the previous steps, which, in turn, are derived from the Datalog engine under test. 
Specifically, Figure~\ref{fig:mainexamplereference1} derives a fact \lstinline[style=lstinlinestyle]{node[1]}; Figure~\ref{fig:mainexamplereference2} derives a fact \lstinline[style=lstinlinestyle]{edge[1, 1]}; 
Figure~\ref{fig:mainexamplereference3} takes the fact derived from the rule in Figure~\ref{fig:mainexamplereference1} as an input to derive the new fact \lstinline[style=lstinlinestyle]{edge[0, 1]}.
Then, by evaluating these three rules, we can obtain the facts \lstinline[style=lstinlinestyle]{edge[0, 1]} and \lstinline[style=lstinlinestyle]{edge[1, 1]}.
Finally, we evaluate the reference program shown in Figure~\ref{fig:mainexamplereference4} with the Datalog engine under test to derive the test oracle (\ie \lstinline[style=lstinlinestyle]{0} and \lstinline[style=lstinlinestyle]{1}).
We can determine that there is a bug, 
because the output of the optimized program is different from the test oracle provided by the reference program.

\paragraph{Test case generation} 
Incrementally generating test cases is also beneficial for efficiently tackling the test case generation problem. 
A key challenge is to explore interesting behaviors of the Datalog engine under test by producing valid programs whose rules mostly produce non-empty results---non-empty results cause subsequent rules that depend on a previously-generated rule to exercise the logic of the Datalog engine under test.
For example, we would like the rules for \lstinline[style=lstinlinestyle]{edge} to produce non-empty results, so that the output rule might subsequently produce non-empty results. 
Another challenge is that by disabling the engine's optimizations, we might expect a slowdown in the execution of the program.
Incrementally generating rules in the test case allows for tackling both challenges. 
Let us assume that we generate the rule of \lstinline[style=lstinlinestyle]{edge} as shown in Figure~\ref{fig:mainexamplereference3}. Before evaluating the optimized program, we first use the Datalog engine to execute the reference program including the newly-generated rule.
At this stage, we could already discard the rule if it would result in a semantic error; for example, CozoDB produces an error if we introduce a modulo by zero, which is challenging to prevent, since, for a complex expression used as the argument of the modulo operation, determining its value would require a complex analysis.
In addition, if the rule would produce an empty result, we could decide to generate another rule---we only retain rules resulting in empty results with a low, manually-specified probability.
Since the newly generated rule produces non-empty results, we append it to the optimized program and validate that both the optimized and reference programs compute the same results.
Let us assume that we generate the rule in line~\ref{codeline:output} of the program shown in Figure~\ref{fig:mainexample} next.
Retaining all the facts computed from the reference program shown in Figure~\ref{fig:mainexamplereference1}, \ref{fig:mainexamplereference2}, and \ref{fig:mainexamplereference3} allows us to generate the results of the current reference program shown in Figure~\ref{fig:mainexamplereference4} by evaluating only the newly-added rule, achieving high efficiency---as shown in Figure~\ref{fig:ruleratio}, executing the reference program is, in fact, up to 6$\times$ faster than executing the optimized program when we generate test cases with 100 rules, as we will discuss in Section~\ref{sect:efficiency}.
While the complete \emph{optimized} program must be evaluated in every step, doing so is necessary to find optimization bugs spanning across multiple rules as soon as a newly-added rule triggers an optimization bug.

\paragraph{Remaining challenges} In this simple example, applying our idea of deriving the test oracle was straightforward, as no recursive queries or negations were present. Thus, we could use the output of the reference program as the test oracle directly. 
However, when recursion is present, it is necessary to re-evaluate the rules that depend on the new rule, because the facts derived from this rule might result in a new fixpoint for the rules depending on the new rule. 
Subsequently, we detail our approach and how it applies to Datalog programs containing recursion and negation.

\section{APPROACH}\label{sect:approach}
\begin{figure*}
    \centering
    \includegraphics[width=.98\linewidth]{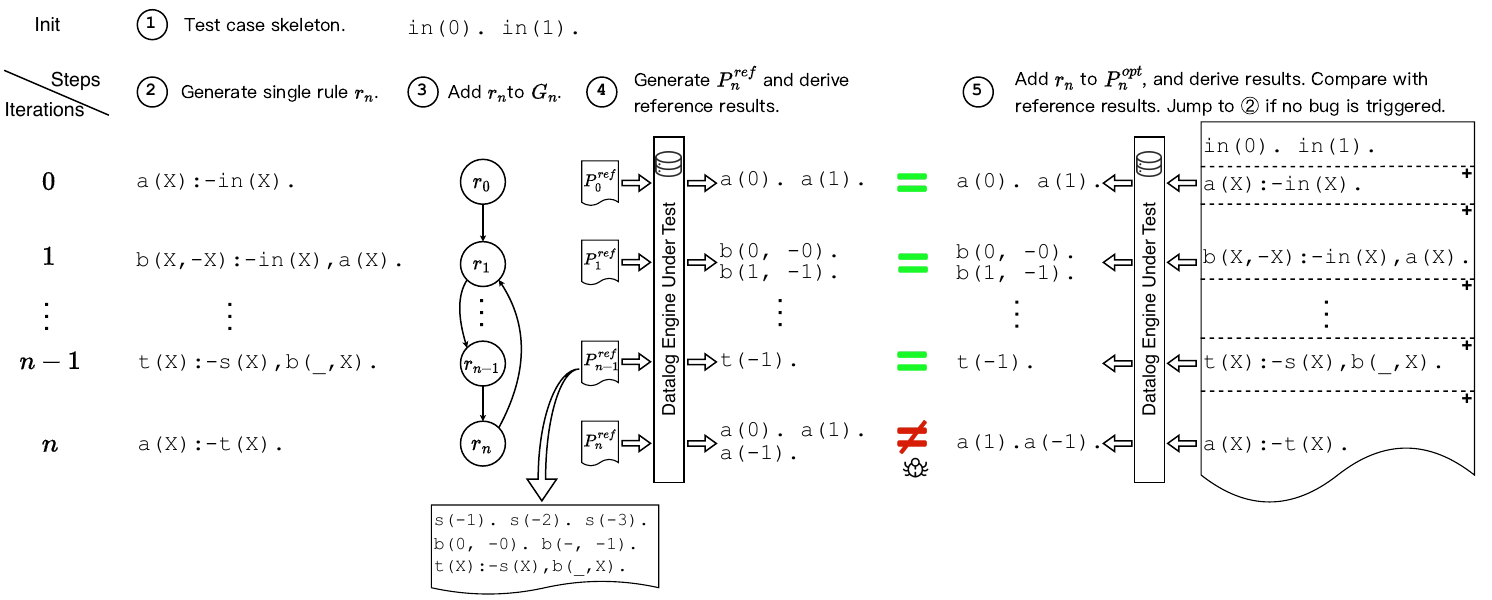}
    \caption{Overview of \approach. The initial step is not included in the iterations, as it is required only once. In iteration $n$, $r_n$ denotes the newly generated rule, $G_n$ denotes the precedence graph, $P^{ref}_{n}$ and $P^{opt}_{n}$ denote the reference and optimized programs respectively. The symbol \textbf{+} indicates that $r_n$ is appended to the optimized program from the previous iteration.}
    \label{fig:approach}
\end{figure*}
We propose \emph{Incremental Rule Evaluation (\approach)}, a novel black-box technique to find optimization bugs in Datalog engines, addressing the two main challenges in automated testing in this context, namely test case generation, and the test oracle problem.
The key insight of our approach is that we can incrementally generate test cases to synergistically generate meaningful test cases as well as evaluate new rules individually to disable many optimizations, in particular, cross-rule optimizations. This allows us to both generate effective test cases and tackle the test oracle problem. 
For test case generation, we incrementally generate a test case by appending rules to a Datalog program; we continue expanding the program ensuring that it is valid (\ie can be executed without errors), and produces non-empty results. 
To provide a test oracle, we execute each rule in the test case individually, preventing optimizations that occur across multiple rules and forcing the Datalog engine to optimize while being able to consider only a single rule. 

Figure~\ref{fig:approach} shows an overview of our approach supporting also complex features involving recursion. 
The example uses features and syntax supported by many Datalog engines such as Souffl\'{e}, µZ, and DDlog. The Datalog dialect used in CozoDB differs in terms of the input facts definition (\eg the definition of the inputs \lstinline[style=lstinlinestyle]{in(0). in(1).} would correspond to \lstinline[style=lstinlinestyle]{in[A]<-[[0], [1]]} in CozoDB) and the connector between the rule head and the rule body (\ie \lstinline[style=lstinlinestyle]{:=} is used in CozoDB).
Our approach starts by generating an initial test case skeleton as the inputs of the test case (step~\circled{1}), and then incrementally adds rules to this test case. In each iteration, we randomly generate a rule (step~\circled{2}) and 
add a corresponding node to the precedence graph 
(step~\circled{3}). The \textit{precedence graph} (\ie $G_n$) specifies the dependence between rules and is used to determine the order in which the rules will be executed. 
Next, we construct a reference program, which contains only this new rule, and evaluate it (step~\circled{4}). If execution results in a semantic error (\ie a syntactically valid rule that might result in an error during execution), we discard the rule. Similarly, if the newly-added rule results in an empty result, we discard it with an empirically specified, high probability. In both cases, we continue execution with step~\circled{2} by generating a new candidate rule. 
If the rule is successfully retained, we append it to the optimized program. The optimized program is then evaluated and any discrepancy between its results and the reference results derived from the reference programs indicates a bug (step~\circled{5}).

\subsection{Test Case Generation}
\label{sect:testcase}
In Figure~\ref{fig:approach}, steps~\circled{1},~\circled{2}, and~\circled{4} are relevant for test case generation.
We maintain two types of programs: an \textit{optimized program} $P^{opt}_{n}$, where a program at iteration $n$ contains the initial test case skeleton and the rules generated in the current iteration as well as those generated in the previous iterations, and will be optimized by the Datalog engine; 
a \textit{reference program} $P^{ref}_{n}$ that each contains only a single rule $r_n$ and the necessary inputs, whose result facts are used as test oracles. The reference programs are not receptive to cross-rule optimizations.
In Figure~\ref{fig:approach}, as an example, consider the reference program $P^{ref}_{n-1}$ below step~\circled{4} and its corresponding optimized program $P^{opt}_{n-1}$ on the right of step~\circled{5}. While the reference program has only a single rule \lstinline[style=lstinlinestyle]{t(X):-s(X),b(_,X).} and its necessary inputs, the optimized program contains the initial input facts \lstinline[style=lstinlinestyle]{in(0). in(1).} as well as the rules up to \lstinline[style=lstinlinestyle]{t(X):-s(X),b(_,X).}.

\paragraph{Test case skeleton}
In step~\circled{1}, we generate an initial test case skeleton including an initial set of facts as well as potentially necessary declarations. Such declarations are necessary for Datalog engines like Souffl\'{e} and DDlog.
These facts serve as the input facts of the optimized programs, and the corresponding relations will be used as optional relations to generate new rules. In Figure~\ref{fig:approach}, \lstinline[style=lstinlinestyle]{in(0). in(1).} are generated as the initial set of facts, and are considered the sole inputs for all optimized programs. The relation \lstinline[style=lstinlinestyle]{in} can be utilized to generate rules in subsequent iterations.

\paragraph{Rule generation}
In step~\circled{2},  in each iteration $n$, we create a new rule $r_n$ based on the existing relations. 
In this step, we first select relations to form the body of $r_n$, and add other language constructs. 
We need to consider only the relations that are currently available and select one or multiple of them to form the body of the new rule. In theory, our approach supports all Datalog language features with deterministic behavior, such as recursive queries, negation, aggregation, subsumption, and functors. We consider adding any feature to a rule based on a pre-defined probability~\cite{SQLsmith, slutz1998massive, 10.1145/3368089.3409710}, allowing us to focus on specific features that we want to test. 
For the head of this new rule, we have the option to either generate a new relation and choose variable types for it from those that appear in the body of the rule, which can introduce a new relation into the test case. Alternatively, 
we can select an existing relation from the test case,
which may introduce recursive queries.  
Generating the rule body before the rule head ensures that the variables in the head are bound~\cite{DBS-017}, with each variable in the rule head appearing at least once in a positive relation within the rule body.


\paragraph{Retaining interesting rules}
We have two key considerations that help us decide when to retain a randomly-generated rule (step~\circled{4}).
The first consideration is that the rule must not result in an error during execution.
By using rule-based generators adapted to a specific Datalog engine, it is straightforward to ensure syntactic correctness.
However, as also highlighted by prior work on testing database engines~\cite{rigger2020testing, zhong2020squirrel, jiangdynsql, liang2022detecting}, it can be difficult to avoid semantic errors. 
To tackle this, we keep a record of expected semantic errors, which allows us to identify and discard invalid rules.
The second consideration is that we would like to generally avoid rules computing empty results, so that subsequent rules might compute a non-empty result and stress the Datalog engine under test. However, we found that approximately ten percent of the bugs detected by queryFuzz and our method were triggered by rules that produced empty results. Therefore, we append rules with a fixed, low probability $P_{empty}$ to allow for empty results. 
Incrementally generating rules is beneficial for both considerations, as it enables us to incrementally build complex, valid test cases; the high-level idea is similar to DynSQL~\cite{jiangdynsql}, which was proposed for fuzzing database engines, which, however, considered only test case validity.

Last, we combine $r_n$ and $P^{opt}_{n-1}$ to form $P^{opt}_{n}$. To form a valid program, we add declarations for the new relations, and set the relation in the head of the rule $r_n$ as the output relation of $P^{opt}_{n}$. 


\subsection{Test Oracle}
\label{sect:oracle}

In steps~\circled{4} and \circled{5}, we derive the test oracle by comparing the derived facts from the reference programs and the optimized program.
To this end, we also require the precedence graph built in step~\circled{3}.

\begin{figure}
    \centering
    \begin{subfigure}{0.50\textwidth}
    \lstset{style=mystyle, numbers=left, framexleftmargin=7pt}
\begin{lstlisting}[morekeywords={decl, output, symbol, number, magic},basicstyle=\scriptsize\ttfamily\bfseries,  mathescape]
a(1). a(2). a(3). b(1). // input facts
c(A) :- a(A), !b(A).    // $r_0$, res: c(2). c(3).
d(A) :- c(A).           // $r_1$, res: d(2). d(3).
c(A) :- d(A).           // $r_2$, res: c(2). c(3).
b(A) :- A=2.            // $r_3$
\end{lstlisting}
    \end{subfigure}
    \hspace{1pt}
    \begin{subfigure}{0.1880\linewidth}
    \raisebox{2.6mm}{\fbox{\includegraphics[width=\textwidth, trim=0 0 0 0]{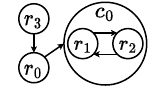}}}
    \end{subfigure}
    \vspace{-7pt}
    \caption{
    A test case in iteration $3$, where care is required to evaluate the rules in the correct order.
    The results displayed in the comments already reach the fixpoint before iteration $3$. The graph at right is the precedence graph of this test case. }
    \vspace*{-7pt}
    \label{fig:challenge}
\end{figure}

\paragraph{Challenge} 
Evaluating the new rule alone is insufficient for generating a reference result for some programs. In iteration $n$, if we generate a new rule with an existing relation as head, it will cause other rules which depend on the new rule to generate additional facts or remove existing facts. 
We need to re-evaluate rules which depend on the new rule and take care that they are executed in a specific order.
Consider the test case illustrated in Figure~\ref{fig:challenge}. In iteration $3$, the new rule $r_3$ derives a new fact (\ie \lstinline[style=lstinlinestyle]{b(2)}) for \lstinline[style=lstinlinestyle]{b}, causing the rule $r_0$ to be re-evaluated to remove \lstinline[style=lstinlinestyle]{c(2)} from the previous results, producing the new result \lstinline[style=lstinlinestyle]{c(3)} for $r_0$. 
Subsequently, the update of \lstinline[style=lstinlinestyle]{c} causes $r_1$ to be re-evaluated to reach the new fixpoint, as the relation \lstinline[style=lstinlinestyle]{c} appears in the rule $r_1$. 
Selecting the correct input facts for $r_1$ requires careful consideration. 
While $r_2$ also derives facts for relation \lstinline[style=lstinlinestyle]{c}, using its results as inputs for $r_1$ would result in an error, as \lstinline[style=lstinlinestyle]{c(2)} of $r_2$ originates from the \lstinline[style=lstinlinestyle]{d(2)} of $r_1$, which was derived from the \lstinline[style=lstinlinestyle]{c(2)} of $r_0$, and should now be discarded.

\begin{algorithm}[tb]
\footnotesize

\Function{TestOracleGen($prec\_graph, stable\_facts, output\_rel$)}{
    $strata \assign$ \FuncCall{GraphStratify}{$prec\_graph$}\;\label{algline:calltostrati}
    \For{$stratum\ \textbf{in}\ strata$}{\label{algline:begineva}
        \For{$node\ \textbf{in}\ stratum$}{
            \If{\FuncCall{isCondensedNode}{$node$}}{
                \FuncCall{HandleRecursion}{$node, stable\_facts$}
            }
            \Else{
                $stable\_facts[node] \assign$ \FuncCall{GenProgAndExec}{$node, stable\_facts$} \label{algline:normalnode}\;
            }
        }
    }
    $test\_oracle \assign$ \FuncCall{GetFacts}{$stable\_facts, output\_rel$} \;\label{algline:getfacts}
    \Return{$test\_oracle$} \;
}
\Function{HandleRecursion($cond\_node, stable\_facts$)}{
    \If{\FuncCall{isNegInCond}{$cond\_node$}}{\label{algline:iscycneg}
        $nodes \assign$ \FuncCall{GetNodesInCond}{$cond\_node$}\;
        $stable\_facts[nodes] \assign$ \FuncCall{GenProgAndExec}{$nodes, stable\_facts$} \;
    }
    \Else{
        $isFixpoint \assign \texttt{False}$\;
        \While{\textbf{not} $isFixpoint$}{\label{algline:fpcomp}
            $isFixpoint \assign \texttt{True}$\;
            \For{$node\ \textbf{in}\ cond\_node$}{
                $res \assign$ \FuncCall{GenProgAndExec}{$node, stable\_facts$} \;
                \If{$res$ != $stable\_facts[node]$}{
                    $isFixpoint \assign \texttt{False}$ \;
                    $stable\_facts[node] \assign res$ \;
                }
            }
        }
    }
}
\caption{Algorithm for test oracle generation}
\label{alg:oraclealg}
\end{algorithm}

\paragraph{Program stratification} 
Stratified Datalog$\urcorner$~\cite{10.1145/50202.50218} provides a solution for evaluating Datalog programs that contain negation in front of intensional-database (\ie IDB) relations, by determining the execution order based on a \textit{precedence graph}, which is a directed graph. Each node in the graph represents a rule in the test case and the destination node of an edge is directly dependent on the source node, specifying the dependence between all the rules. 
For example, in Figure~\ref{fig:approach}, rule $r_{n-1}$ depends on rule $r_1$ because the relation \lstinline[style=lstinlinestyle]{b} is the head of $r_1$ and appears in the rule body of $r_{n-1}$. So, in the precedence graph, there is an edge from the node of $r_1$ to the node of $r_{n-1}$.
Stratified Datalog$\urcorner$ stratifies the precedence graph into different strata, 
where rules in lower strata are executed before those in the higher strata.
Furthermore, execution only progresses to the next stratum when all rules in the current stratum have reached a fixpoint. 
The stratification is guided by two principles: (1) If a relation $R$ is referenced in a positive form in the rule of another relation $R'$, the rules defining $R$ must be placed in a lower or equal stratum compared to the rule of $R'$; (2) If $R$ is referenced in a negated form in the rule of $R'$, the rules defining $R$ must be placed in a lower stratum than the rule of $R'$. 
Recursion without negation can be stratified into a single stratum, whereas recursion involving negation cannot be stratified, as determined by the two principles outlined.
Furthermore, recursion involving negation has a different fixpoint, named alternating fixpoint~\cite{van1989alternating}, which can not be expressed by stratified Datalog$\urcorner$. 

We propose a strategy to address the challenges posed by recursion involving negation in test oracle generation. 
As our approach relies on an existing Datalog engine, we can disregard the execution order of the unstratifiable portions of the Datalog program by placing them in a single reference program; thus, the Datalog engine under test generates the results for these rules, enabling a general, simple solution. 
While this method does not allow us to disable optimization between rules in the recursive query with negation, it might still detect bugs in executing them through their interaction with other rules. 


\paragraph{Reference program execution}
Algorithm~\ref{alg:oraclealg} details how we derive our test oracle. We use the test case shown in Figure~\ref{fig:challenge} as a guiding example to support the explanation of the algorithm.
The function \texttt{TestOracleGen} outlines the procedure for generating test oracles and requires three arguments.
The first argument, $prec\_graph$, is a subgraph of the complete precedence graph, including all nodes that are reachable from the node of the new rule in the precedence graph, as the new rule impacts only the rules that depend on it.
The $prec\_graph$ of the guiding example should be the right graph shown in Figure~\ref{fig:challenge}, as all rules are impacted by the new rule.
The second argument, $stable\_facts$, encompasses both the initial facts generated in step~\circled{1} and those derived by the rules which are not present in $prec\_graph$.
Facts derived from the rules within $prec\_graph$ are now outdated and consequently, they are not applicable for the construction of inputs of reference programs for the rules in $prec\_graph$; thus, $stable\_facts$ does not include the facts derived from the rules within $prec\_graph$.
The $stable\_facts$ of the guiding example includes \lstinline[style=lstinlinestyle]{a(1)}, \lstinline[style=lstinlinestyle]{a(2)}, \lstinline[style=lstinlinestyle]{a(3)} and \lstinline[style=lstinlinestyle]{b(1)}.
The third argument, $output\_rel$, determines which relation is the output relation, which is \lstinline[style=lstinlinestyle]{b} in the guiding example.

A key step is implemented in \texttt{GraphStratify}, which is called in line~\ref{algline:calltostrati}. Stratification is a well-understood concept that has been well-studied in literature~\cite{10.1145/298514.298558}, which is why we only briefly outline it. 
The function stratifies $prec\_graph$, meaning that it determines 
the order of execution and the inputs for the rules contained within it, corresponding to step~\circled{3} in Figure~\ref{fig:approach}. 
During stratification, we first build the condensation graph of $prec\_graph$. 
The \textit{condensation graph} is a directed acyclic graph. It simplifies the structure by amalgamating each strongly connected component of the original graph into a single new node, denoted as $c_{i}$. Subsequently, the edges that were initially connected to the nodes in these strongly connected components are redirected to $c_{i}$.
For the guiding example, we merge the two nodes of $r_1$ and $r_2$ into a single node $c_0$. 
For stratification, we enforce the requirement that the two nodes of each edge must be in different strata. 
We also take note of which edges in the precedence graph express negated dependence, as this information is necessary for the proper handling of recursion involving negation in our approach. 
After stratification, the rules in the guiding example are placed into three strata: $r_3$ is positioned in the first stratum, $r_0$ in the second stratum, and $c_0$---a condensed form of $r_1$ and $r_2$---is placed in the third stratum.

We evaluate the rules following the strata produced by graph stratification, starting from the lowest stratum and progressing to the higher ones (\ie the \lstinline[style=lstinlinestyle]{for} loop in line~\ref{algline:begineva}), corresponding to step~\circled{4} in Figure~\ref{fig:approach}. 
Non-condensed nodes are handled with \texttt{GenProgAndExec} in line~\ref{algline:normalnode}.
\texttt{GenProgAndExec} is used to construct an executable reference program for one or more specific rules, and to execute it subsequently. 
It first generates the required declarations of relations and types, and then specifies the input and output relations. 
Afterward, it collects the input facts for this reference program from $stable\_facts$. 
The facts pertaining to the relations present in the rule body will be considered as inputs.
After the execution of this reference program, the newly derived facts will be stored in $stable\_facts$.
For example, in constructing the reference program for rule $r_0$ in the guiding example, the inputs comprise \lstinline[style=lstinlinestyle]{a(1)}, \lstinline[style=lstinlinestyle]{a(2)}, \lstinline[style=lstinlinestyle]{a(3)}, and \lstinline[style=lstinlinestyle]{b(1)} from initial facts generated in step~\circled{1}, along with \lstinline[style=lstinlinestyle]{b(2)} derived from $r_3$. The result \lstinline[style=lstinlinestyle]{c(3)} will be saved into $stable\_facts$.

\sloppy{}
We handle recursive queries in the function \texttt{HandleRecursion}. The recursive queries are represented by condensed nodes.
We first determine whether the condensed node contains a negated edge as shown in line~\ref{algline:iscycneg}.
If so, this portion is unstratifiable, and we identify all the rules in this recursion, and place these rules into a single reference program (\eg \lstinline[style=lstinlinestyle]{d(a):-c(a). c(a):-c(a),!d(a).}), set all relations that appear in the head of these rules as output relations, and query the answer using the Datalog engine under test.
If there is no negation, we cyclically evaluate every rule in this recursion beginning with a specific node until no new facts are derived (\ie the \lstinline[style=lstinlinestyle]{while} loop in line~\ref{algline:fpcomp}). 
In the guiding example, node $c_0$ exemplifies this case;
we need to cyclically evaluate the two rules $r_1$ and $r_2$ in $c_0$ until both of them reach the fixpoint. 
Before evaluating $r_1$ and $r_2$, $stable\_facts$ contains \lstinline[style=lstinlinestyle]{a(1)}, \lstinline[style=lstinlinestyle]{a(2)}, \lstinline[style=lstinlinestyle]{a(3)}, \lstinline[style=lstinlinestyle]{b(1)}, \lstinline[style=lstinlinestyle]{b(2)}, and  \lstinline[style=lstinlinestyle]{c(3)}.
During the first iteration of evaluating the condensed node $c_0$, \lstinline[style=lstinlinestyle]{d(3)} is derived for rule $r_1$ and \lstinline[style=lstinlinestyle]{c(3)} for rule $r_2$.
Subsequently, in the second iteration, the derived facts for $r_1$ and $r_2$ remain \lstinline[style=lstinlinestyle]{d(3)} and \lstinline[style=lstinlinestyle]{c(3)}, identical to those in the first iteration. Thus, we determine that the condensed node $c_0$ has reached the fixpoint.

\paragraph{Test oracle}
Finally, we gather the facts of the output relation of the optimized program from all reference programs as test oracle (\ie the function call to \texttt{GetFacts} in line~\ref{algline:getfacts}). 
For example, in Figure~\ref{fig:approach}, relation \lstinline[style=lstinlinestyle]{a} is defined by two rules: $r_0$ and $r_n$, so the test oracle in iteration $n$ should be the union of the results derived by these two rules.
By evaluating the optimized program (step~\circled{5}), we can determine whether $P^{opt}_{n}$ triggers a logic bug if its result mismatches the facts by the test oracle. 

\section{EVALUATION}\label{sect:evaluation}
To evaluate \approach, we realized it as a tool named \tool, with its name derived from \textbf{D}atalog \textbf{E}ngine \textbf{Op}timization \textbf{T}ester.
The goal of our evaluation was to investigate whether our approach can effectively and efficiently find bugs in Datalog engines. We selected four mature and well-tested open-source Datalog engines for the evaluation and aimed to answer the following questions:




\begin{table}[H]
\begin{tabular}{l p{.89\textwidth}}
\textbf{Q1:} & How effective is \tool in finding new bugs? \\
\textbf{Q2:} & How do \tool's components contribute to its overall effectiveness and efficiency? \\
\textbf{Q3:} & How effective is \tool's test case generation compared to a random method? \\
\textbf{Q4:} & How does \tool compare to the state-of-the-art approach? \\
\end{tabular}
\end{table}

\paragraph{Implementation}
We implemented \tool based on queryFuzz and modified it in four main aspects. 
First, we added support for another Datalog engine (\ie CozoDB).
Second, for this and the existing engines (\ie Souffl\'{e}, µZ, and DDlog), we modified the random rule generators to generate rules incrementally.
Third, we added additional language features (\eg subsumption, equivalence relation, and query plan).
A potential concern might be that evaluating \tool and queryFuzz might not be fair due to these additional language features. However, queryFuzz's rules do not apply to these language features, and our manual analysis in Section~\ref{sec:compareqf} shows that conceptually, queryFuzz cannot detect the bugs we found using these language features.
Fourth, we implemented our test oracle using Python in only approximately 200 LOC.
Our approach is independent of the Datalog engines under test; we expect it to be compatible with all common Datalog engines. 
Among these four Datalog engines considered, Souffl\'{e} is the only one that supports multiple optimization keywords and execution options. To run the reference program for Souffl\'{e}, we must ignore the \lstinline[style=lstinlinestyle]{inline} keyword as Souffle disallows inline optimization on input and output relations. 
We also added an extension that provides options to ignore other optimization-related keywords and execution options for the reference program.


\paragraph{Tested Datalog engines}
We selected four mature open-source Datalog engines as our target: Souffl\'{e}~\cite{jordan2016souffle}, CozoDB~\cite{CozoDB}, µZ~\cite{hoder2011muz}, and DDlog~\cite{ryzhyk2019differential}. Three of them were the test targets used to evaluate queryFuzz: Souffl\'{e}, µZ, and DDlog.
Souffl\'{e} has been used in numerous works to support analyses, including static analysis~\cite{10.1145/3088515.3088522}, security analysis~\cite{grech2019gigahorse, tsankov2018securify, brent2018vandal}, and binary disassembling~\cite{backes2019reachability}. µZ is part of the well-known SMT solver Z3~\cite{moura2008z3}. DDlog supports incremental computation and has more than one thousand stars on GitHub. 
CozoDB is designed to support large language models and artificial intelligence applications. It has garnered over 2.6 thousand stars on GitHub.

\paragraph{Experimental setup}
We conducted all experiments on a laptop with a 12-core Intel i7-12700 CPU at 2.10 GHz and 16 GB of memory running Ubuntu 22.04. We downloaded Souffl\'{e}, CozoDB, µZ, and DDlog from GitHub. For Souffl\'{e}, we tested it on \texttt{29c5921} and later commit versions. For CozoDB, we tested it on \texttt{0.7.0} and later release versions. For µZ, we tested its code for version \texttt{9a65677}. For DDlog, we tested it on \texttt{dbd6580} and later commit versions. 

\subsection{Effectiveness}
\label{sect:effectiveness}
\paragraph{Methodology}
We evaluated the effectiveness of \tool by testing the four aforementioned, mature Datalog engines. We intermittently ran \tool over a period of four months, while also developing the approach and tool. We reported bugs after determining whether we believed a bug to be unique. For logic bugs, determining whether a bug is a unique bug is an unsolved problem. To avoid reporting duplicate issues, we used a best-effort approach by avoiding reporting bugs in a given feature until the developers fixed previously-reported bugs in this feature, where the developers were responsive. For other bugs, including crashes, internal errors, and hangs, we determined whether a bug is unique based on the stack trace, error message, and/or the features involved. 
As bug-inducing test cases were sometimes large, we reduced them to simpler bug-inducing test cases with C-Reduce~\cite{10.1145/2345156.2254104} and manually.

\begin{table}[htbp]\small
    \centering
    \caption{\tool found \totalbugs unique bugs in three mature engines.}
    \vspace{-5pt}
    \label{tab:bugstatis}
    \begin{tabular}{l r r r r} \toprule
         & \multicolumn{4}{c}{Bug type} \\\cline{2-5}
         Datalog engine & Logic bug & Internal error & Crash & Hang \\\midrule
         Souffl\'{e} & 9 & 9 & 3 & 2 \\
         CozoDB & 2 & 2 & 1 & 0 \\
         µZ & 2 & 0 & 0 & 0 \\
         \midrule
         Total & 13 & 11 & 4 & 2 \\\bottomrule
    \end{tabular}
\end{table}

\begin{table}[htbp]\small
    \centering
    \caption{The status of bugs found by \tool.}
    \vspace*{-7pt}
    \label{tab:logicbugs}
    \begin{tabular}{l r r r} \toprule
         Datalog engine & Fixed & Verified & Open\\\midrule
         Souffl\'{e} & 2 & 10 & 11 \\
         CozoDB & 3 & 2 & 0 \\
         µZ & 1 & 1 & 0\\
         \midrule
         Total & 6 & 13 & 11\\\bottomrule
    \end{tabular}
\end{table}

\paragraph{Results}
Table~\ref{tab:bugstatis} shows the number of bugs found by \tool grouped by the bug type and Datalog engine. 
\tool found a total of \totalbugs previously unknown bugs in three mature Datalog engines. 
Out of all bugs, \totalfixedbug were already fixed and 13 were confirmed as unique bugs by developers.
Among these, \totallogicbug were logic bugs, \fixedlogicbug of which have been fixed, and another \needtobeconfirm are awaiting confirmation.

The number of fixed and confirmed bugs is often considered an important metric of the found bugs' importance. 
For CozoDB, the bugs we reported were promptly confirmed and fixed.
The ratio of fixed and verified bugs in Souffl\'{e} is low.
We received informal feedback that our bug reports are appreciated, but due to lack of resources, the bugs could not be investigated and fixed.
For µZ, one bug was fixed, but we received feedback that the Datalog implementation is no longer maintained.\footnote{\url{https://github.com/Z3Prover/z3/issues/\anonymize{6447}}}

\paragraph{Logic bugs}
Overall, we have found \totallogicbug logic bugs. For the bugs in Souffl\'{e}, we discovered that 8 out of 9 bugs were related to optimizations, because the test cases contained only a limited set of language features after reduction, 
for example, subsumption and equivalence relation, they are language features that enable optimizations implemented in Souffl\'{e}.
One other bug resulted in inconsistent output for specific floating-point values when running in different execution modes. 
7 out of 9 bugs were found by \approach, whereas 
the other 2 logic bugs only occurred under specific execution options. 
One bug resulted from a different result in compiler mode, which need to be executed with \lstinline[style=lstinlinestyle]{-c} option. Another one caused an incorrect result when using provenance, which is a debug method in Souffl\'{e} that needs to be executed with the \lstinline[style=lstinlinestyle]{-t} option. These two bugs were found by randomly ignoring the execution options for the reference program.
For the bugs in CozoDB, we received feedback from the developers indicating that these two logic bugs were caused by errors in the implementation of the semi-naïve algorithm and the magic sets transformation, respectively.
\begin{figure}
    \centering
    \parbox{0.45\textwidth}{
        \lstset{style=mystyle, numbers=left, framexleftmargin=7pt}
\begin{lstlisting}[morekeywords={},basicstyle=\scriptsize\ttfamily\bfseries]
h(29, 29). h(80, 80).
a(A) :- b(A), 43 != A.
c(C) :- d(C), 76 != C.
e(A) :- f(A), 77 < A.
g(E) :- h(D, E), 8 != E, 71 < D.
// expected result: g(80).
// buggy result: g(29). g(80).
\end{lstlisting}
    }
    \vspace{-7pt}
    \caption{A bug-inducing test case for µZ. We have omitted declarations in this test case for simplicity. This bug could be triggered regardless of whether \lstinline[style=lstinlinestyle]{b}, \lstinline[style=lstinlinestyle]{d}, and \lstinline[style=lstinlinestyle]{f} have facts.}
    \label{fig:muzbug}
\end{figure}
For bugs in µZ, we failed to determine the precise cause of the bugs, due to limited developer feedback. Figure~\ref{fig:muzbug} shows one bug-inducing test case for µZ. 
We expected that the output should only be \lstinline[style=lstinlinestyle]{g(80)}, however, both \lstinline[style=lstinlinestyle]{g(29)} and \lstinline[style=lstinlinestyle]{g(80)} were computed as a result.
Interestingly, the bug cannot be triggered when removing any of the individual rules, despite all four rules having no interdependence. 
The process of generating test oracles for all bugs is included in the supplementary materials in the artifact. 


\paragraph{Other bugs}
Our approach was effective also in finding other types of bugs; \tool found a total of 11 internal errors, 4 crashes, and 2 hangs. 
Note that other random-testing approaches without an explicit test oracle could also find these bugs.

\paragraph{Bugs in different engines}
We found most bugs in Souffl\'{e} (\ie 23 out of 30) and 5 bugs in CozoDB. 
For µZ, we have found more bugs without reporting them, because we received feedback that µZ is not actively used in Z3. We decided not to continue testing µZ, to respect the developers' wishes. 
DDlog is known to be stable; queryFuzz found only a single bug in it. While testing the latest version of DDlog, we also repeatedly triggered this bug, since it has not been fixed yet. 

\paragraph{Reception}
Our bug reports have helped developers find many hard-to-find bugs, and they also appreciated our reports in the public issue trackers. For the logic bugs in Souffl\'{e} on magic transformation with subsumption and equivalence relations, one of the developers commented that our finding was \textit{"well spotted!"}.\footnote{\url{https://github.com/souffle-lang/souffle/issues/\anonymize{2322}}} For the assertion failure in Souffl\'{e}, one of the developers thought our finding on subsumption in the provenance system was a \textit{"Great find."}\footnote{\url{https://github.com/souffle-lang/souffle/issues/\anonymize{2321}}}
In addition, we directly communicated with Bernhard Scholz, the main developer of Souffl\'{e}, who told us that he valued our work and encouraged us to continue our testing efforts.
The developer of CozoDB praised our bug-finding abilities, saying, \textit{"you are really good at finding bugs!"}\footnote{\url{https://github.com/cozodb/cozo/issues/\anonymize{99}}}

\subsection{Sensitivity Analysis}
\label{sect:sensitivity}
In this section, we examine the factors that might influence the efficiency and effectiveness of our test case generation method. 
As detailed next, we have introduced five options, which we believed to be potentially relevant for the efficiency and effectiveness of our approach. 


\paragraph{Methodology}
To illustrate the effect of each configuration on our system, we selected default configurations that we determined to work well empirically. We set 
$Max_{rules} = 100$, $Max_{att} = \infty$, $P_{empty} = 0.1$, $P_{head} = 0.02$, $Max_{iter} = 100$. 
We introduce these options below. In each experiment, we modified only the relevant option to assess its impact. 
For each graph depicted, we ran the test case generation for 24 hours. Where metrics required to be summarized, we used the arithmetic mean.
Subsequently, $Ref.$ refers to the reference program, and $Opt.$ to the optimized program.
We refer to every Datalog program to which we add a rule as a \emph{test case}.
We consider the process of incrementally generating rules to form a test case as a \emph{test iteration}, which will be terminated when the number of rules in this test case reaches the threshold $Max_{rules}$, or the attempts for generating a new rule reaches the threshold $Max_{att}$.
Our \textit{throughput} metric refers to the number of test iterations in a specific period, which is 24 hours for our experiments. 
We conducted all experiments on Souffl\'{e}, as µZ and CozoDB have similar performance trends. We also conducted performance-related experiments on DDlog, which is slower compared to the other engines.

\paragraph{Max number of rules ($Max_{rules} > 0$)}
$Max_{rules}$ determines the maximum number of rules to be generated in a test case.
We anticipate a decrease in the \textit{throughput} with an increase in $Max_{rules}$, as larger $Max_{rules}$ requires more reference and optimized programs in a test iteration. 
Thus, we evaluate how varying $Max_{rules}$ affects \tool's performance.

\begin{figure}
    \centering
    \begin{subfigure}{.30\textwidth}
        \begin{tikzpicture}[scale=1, font=\tiny]
        \begin{axis}[
	ylabel={\tiny $\%\ execution\  time\ of\ Ref.$},
        xlabel={\tiny $Max_{rules}$},
        legend style={at={(0.5,-0.4), anchor=east, cells={align=left}},
	anchor=north,legend columns=2},
        width=4cm,height=4cm
        ]
        \addlegendentry{Souffl\'{e}}
        \addplot[mark=star, mark size=1pt, blue] coordinates {(10, 0.6839994610212473) (20, 0.6127681645416388) (40, 0.5007584871715605) (60, 0.4147357227113309) (80, 0.35069034476825417) (100, 0.30212929661119237)};

        \addlegendentry{DDlog}
        \addplot[mark=otimes, mark size=1pt, green] coordinates {(10, 0.6804879852578759) (20, 0.5718245595833311) (40, 0.5842462118674555) (60, 0.43599714981743676) (80, 0.4138092780700251) (100, 0.3037125384867105)};
        \end{axis}
        \end{tikzpicture}
        \caption{Runtime percentage.}
        \label{fig:ruleweight}
    \end{subfigure}
    \begin{subfigure}{.30\textwidth}
        \begin{tikzpicture}[scale=1, font=\tiny]
        \begin{axis}[
	ylabel={\tiny $Ratio$},
        xlabel={\tiny $Max_{rules}$},
        legend style={at={(0.5,-0.4), anchor=east, cells={align=left}},
	anchor=north,legend columns=2},
        width=4cm,height=4cm
        ]
        \addlegendentry{Souffl\'{e}}
        \addplot[mark=star, mark size=1pt, blue] coordinates {(10, 1.1507676861266551) (20, 1.6347429594293095) (40, 2.6393253577166886) (60, 3.7318596499297154) (80, 4.852600426070166) (100, 6.119791217354439)};

        \addlegendentry{DDlog}
        \addplot[mark=otimes, mark size=1pt, green] coordinates {(10, 1.1666607206825208) (20, 1.262242841105534) (40, 1.5822140469618793) (60, 2.6354944016228643) (80, 2.7195156255665194) (100, 3.613117340537328)};
        \end{axis}
        \end{tikzpicture}
        \caption{The cause of results.}
        \label{fig:ruleratio}
    \end{subfigure}
    \vspace{-7pt}
    \caption{The results of $Max_{rules}$. In a, the curves illustrate the percentage of the cumulative execution time of the reference program out of the total execution time of a test iteration. In b, the curves show the ratio of average execution time between a single reference program and an optimized program. 
    }
    \label{fig:rulenum}
\end{figure}
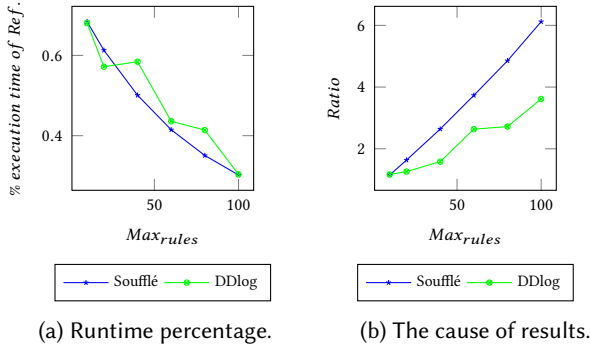

The two graphs in Figure~\ref{fig:rulenum} show the effect of varying $Max_{rules}$. 
The graph in Figure~\ref{fig:ruleweight} illustrates the percentage of the cumulative execution time of the reference program. 
For example, when generating a test case for Souffl\'{e} with 100 rules, the cumulative execution time of the reference programs accounted for 30.2\% of the total execution time of a test iteration, thus we can infer that the optimized programs accounted for 69.8\%. 
As $Max_{rules}$ increases, we can observe a decline in the cumulative execution time percentage of the reference programs. 
The explanation for this can be found in Figure~\ref{fig:ruleratio}. The execution time for a single reference program is constant as it always contains a single rule, whereas the average execution time for the optimized program increases with $Max_{rules}$. 
Moreover, we found that regardless of the size of $Max_{rules}$, the number of evaluations required for reference programs when adding a new rule remains almost constant. Thus, we can conclude that as $Max_{rules}$ increases, the time percentage required to generate the test oracle becomes small, and the execution time of the optimized program becomes the limiting factor for the test performance. 

DDlog was slow, hindering the speed of testing. 
When $Max_{rules}$ was set to 100, we could conduct only 5 test iterations in the 24-hour period. Despite this, we can still observe that it has similar characteristics to Souffl\'{e}. 
Therefore, for $Max_{att}$, $P_{empty}$, $P_{head}$, and $Max_{iter}$, we omitted the results for DDlog, because it has the same trend as Souffl\'{e}, but due to the small number of test iterations, has a higher level of noise in the results.

\begin{figure}
    \centering
    \begin{subfigure}{.3\textwidth}
        \begin{tikzpicture}[scale=1, font=\tiny]
        \begin{axis}[
	ylabel={\tiny $Number\ of\ rules$},
        axis y line*=left,
        xlabel={\tiny $Max_{att}$},
	legend style={at={(0.5,-0.3)},
	anchor=north,legend columns=-1},
        width=4cm,height=4cm
        ]
        \addlegendentry{Size}
        \addplot[mark=star, mark size=1pt, blue]
        coordinates {(10, 73.94749498997996)(20, 96.12779464532156)(30, 98.71420422133485)(40, 99.63258785942492)(50, 99.82206613988879)(60, 99.92974720752498)(70, 99.98029411764706)};\label{Size}
        \legend{}
        \end{axis}
        \end{tikzpicture}
    \end{subfigure}
    \hspace{7pt}
    \begin{subfigure}{.3\textwidth}
        \begin{tikzpicture}[scale=1, font=\tiny]
        \begin{axis}[
	ylabel={\tiny $\#\ Test\ Iterations$},
        axis y line*=left,
        xlabel={\tiny $Max_{att}$},
	legend style={at={(0.5,-0.3)},
	anchor=north,legend columns=-1},
        width=4cm,height=4cm
        ]
        \addlegendentry{Num}
        \addplot[mark=star, mark size=1pt, blue]
        coordinates {(10, 4990)(20, 3623)(30, 3506)(40, 3443)(50, 3417)(60, 3402)(70, 3400)};
        \legend{}
        \end{axis}
        \end{tikzpicture}
    \end{subfigure}
    \vspace{-4pt}
    \caption{The results of $Max_{att}$.}
    \vspace*{-10pt}
    \label{fig:attempts}
\end{figure}
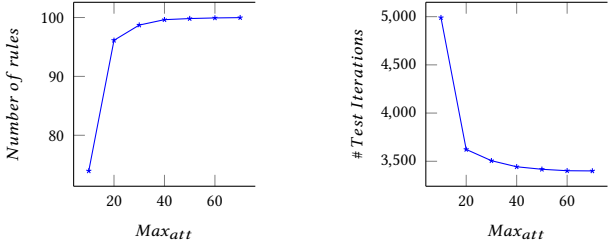

\paragraph{Max attempts to generate new rules ($Max_{att} > 0$ or $Max_{att} = \infty$)}
$Max_{att}$ determines the maximum number of attempts to generate a new rule if the current rule produces an empty result or results in a semantic error; the current number of attempts is set to zero after successfully generating a rule.
$Max_{att}$ requires a positive number or is set to $\infty$ for an infinite number of attempts. 
We introduced $Max_{att}$, because the presence of existing 
relations and their associated facts can make it difficult to generate valid rules that produce non-empty results. For example, the rule \lstinline[style=lstinlinestyle]{a(A) :- b(A), A < 0, A > 0.} produces an empty result, as no facts can satisfy its predicates. 
If the number of failed consecutive rule generations reaches this threshold, we will stop generating more rules in this test iteration.
We measured the impact of $Max_{att}$ on the average size of generated test cases and the \textit{throughput} of testing. 

The results of $Max_{att}$ are shown in Figure~\ref{fig:attempts}. 
When the threshold was lower than 20, the average number of rules in the test cases when the test iteration was terminated increased rapidly, reaching 96 at a threshold of 20. After that, the rate of increase slowed down and this average number gradually approached 100. This indicates that, typically, only a small number of attempts are required to generate a valid rule that maintains the probability of non-empty results.
However, we encountered rare cases where it took a higher number of attempts for a rule. 
For the \textit{throughput}, the opposite seems to hold; when the threshold was lower than 20, the number of test iterations conducted in 24 hours decreased rapidly, and remained stable when the threshold was higher than 20. 
As mentioned in the previous evaluation about $Max_{rules}$, the execution time of the optimized programs accounts for the majority of the total test time, so the decrease in throughput is mainly due to the larger $Max_{att}$ result in the larger test case, which requires more time to execute the optimized programs. 

\begin{figure}
    \centering
    \begin{subfigure}{.3\textwidth}
        \begin{tikzpicture}[scale=1, font=\tiny]
        \begin{axis}[
	ylabel={\tiny $\%\ Opt.\ empty$},
        axis y line*=left,
        xlabel={\tiny $P_{empty}(\%)$},
	legend style={at={(0.5,-0.3)},
	anchor=north,legend columns=-1},
        width=4cm,height=4cm
        ]
        \addlegendentry{Opt.}
        \addplot[mark=star, mark size=1pt, blue]
        coordinates {(0, 0) (20, 35.93763676148797) (40, 68.0870265914585) (60, 83.60516795865633) (80, 89.52472736495054) (100, 92.7251714503429)};\label{Opt}

        \legend{}
        \end{axis}
        \end{tikzpicture}
    \end{subfigure}
    \hspace{7pt}
    \begin{subfigure}{.3\textwidth}
        \begin{tikzpicture}[scale=1, font=\tiny]
        \begin{axis}[
	ylabel={\tiny $\#\ Test\ Iterations$},
        ymin=0,
        axis y line*=left,
        xlabel={\tiny $P_{empty}(\%)$},
	legend style={at={(0.5,-0.3)},
	anchor=north,legend columns=-1},
        width=4cm,height=4cm
        ]
        \addlegendentry{Num}
        \addplot[mark=star, mark size=1pt, blue, smooth]
        coordinates {(0, 3593)(20, 3656)(40, 3723)(60, 3870)(80, 3943)(100, 3937)};
        \legend{}
        \end{axis}
        \end{tikzpicture}
    \end{subfigure}
    \vspace*{-3pt}
    \caption{The results of $P_{empty}$.}
    \vspace*{-7pt}
    \label{fig:empty}
\end{figure}
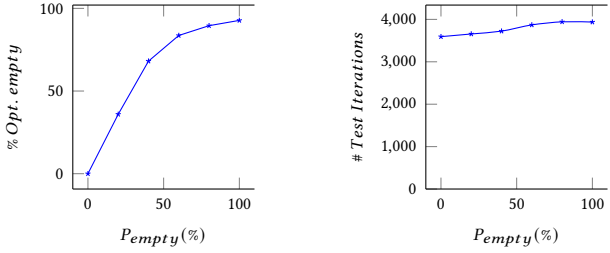

\paragraph{Probability to allow empty results for a rule ($0 \leq P_{empty} \leq 1$)}
$P_{empty}$ determines the probability of generating a new rule that is permitted to produce an empty result. 
Most bugs require relations with non-empty facts to be detected, but relations with empty facts are also necessary to detect some bugs (\eg 2 out of 13 bugs detected by queryFuzz). We set $P_{empty}$ to maintain a certain percentage of relations without any facts. 
For $P_{empty}$, we measured the probability that the optimized program will produce empty results and its impact on the \textit{throughput} of testing.

The results of $P_{empty}$ are shown in Figure~\ref{fig:empty}. 
For example, when we set 60\% for $P_{empty}$, the likelihood of the optimized program (\ie test case) producing empty results is 83.3\%. 
While the results are for all test cases, the trend looks similar for the last test case when test iterations was terminated. 
Even if there is a small number of empty relations in the available relations, the probability of newly generated relations being empty is significantly higher. This is, because, for most rules, we require that all relations in the body are non-empty to ensure that the final result is non-empty. 
Even if most relations are empty, there is a chance to generate rules with non-empty results.
This is why the curve increases quickly at the beginning and then flattens. 
Furthermore, we noticed a slight improvement in testing throughput as $P_{empty}$ increases.
This is because a lower value for $P_{empty}$ requires more attempts to accept a rule, resulting in an increase in the number of reference program executions. 
However, since the execution of the reference program is fast (\ie about 0.03s per execution), this results only in a slight improvement in testing throughput. 


\begin{figure}
    \centering
    \begin{subfigure}{.35\textwidth}
        \begin{tikzpicture}[scale=1, font=\tiny]
        \begin{axis}[
	ylabel={\tiny $Num\ of\ cycles$},
        axis y line*=left,
        xlabel={\tiny $P_{head}(\%)$},
	legend style={at={(0.5,-0.4)},
	anchor=north,legend columns=-1},
        width=4cm,height=4cm
        ]
        \addplot[mark=star, mark size=1pt, blue]
        coordinates {(0, 0.0) (2, 0.8320337881741391) (4, 2.7260900140646975) (6, 6.932001536688436) (8, 19.06288209606987) (10, 112.57482014388489)};\label{Num}
        \legend{}
        \end{axis}
        \begin{axis}[
	ylabel={\tiny $Len\ of\ cycles$},
        axis y line*=right,
        axis x line=none,
	legend style={at={(0.5,-0.4)},
	anchor=north,legend columns=2},
        width=4cm,height=4cm
        ]
        \addlegendimage{/pgfplots/refstyle=Num}\addlegendentry{Num}
        \addlegendentry{Len}
        \addplot[mark=otimes, mark size=1pt, green] coordinates {(0, 0) (2, 3.360796563842249) (4, 4.381787695085773) (6, 6.041010862336511) (8, 7.877926421404682) (10, 11.072897960748726)};
        \end{axis}
        \end{tikzpicture}
        \caption{$P_{head}$}
        \label{fig:rqprob}
    \end{subfigure}
    \hspace{10pt}
    \begin{subfigure}{.35\textwidth}
        \begin{tikzpicture}[scale=1, font=\tiny]
        \begin{axis}[
	ylabel={\tiny $Num\ of\ cycles$},
        axis y line*=left,
        xlabel={\tiny $Max_{iter}$},
	legend style={at={(0.5,-0.4)},
	anchor=north,legend columns=-1},
        width=4cm,height=4cm
        ]
        \addplot[mark=star, mark size=1pt, blue]
        coordinates {(10, 0.4552819183408944) (50, 0.8158415841584158) (100, 0.8320337881741391) (200, 0.9128838451268364) (300, 1.016474464579901)};\label{NumIter}
        \legend{}
        \end{axis}
        \begin{axis}[
	ylabel={\tiny $Len\ of\ cycles$},
        axis y line*=right,
        axis x line=none,
	legend style={at={(0.5,-0.4)},
	anchor=north,legend columns=2},
        width=4cm,height=4cm
        ]
        \addlegendimage{/pgfplots/refstyle=Num}\addlegendentry{NumIter}
        \addlegendentry{Len}
        \addplot[mark=otimes, mark size=1pt, green] coordinates {(10, 2.3437722419928826) (50, 3.2091423948220066) (100, 3.360796563842249) (200, 3.3952468007312615) (300, 4.436628849270664)};
        \end{axis}
        \end{tikzpicture}
        \caption{$Max_{iter}$}
        \label{fig:rqlen}
    \end{subfigure}
    \caption{The effect of $P_{head}$ and $Max_{iter}$ on recursions.}
    \vspace*{-7pt}
    \label{fig:reucrsive}
\end{figure}
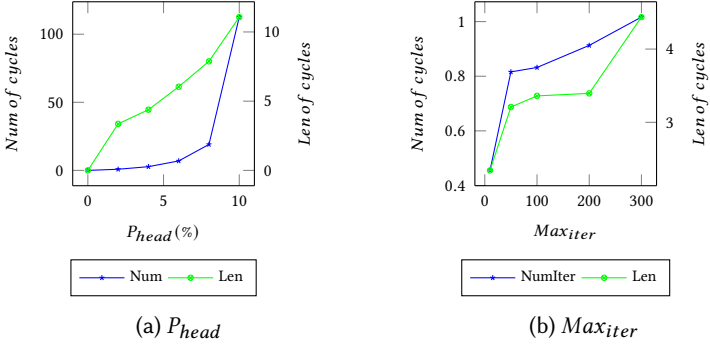

\paragraph{Probability of choosing an existing relation as head ($0 \leq P_{head} \leq 1$)}
$P_{head}$ determines the probability of generating a new rule with an already existing relation as the head.
This option controls the generation of recursive queries (\eg \lstinline[style=lstinlinestyle]{a(A):-b(A). b(A):-a(A).} introduces a recursion because we select an existing relation \lstinline[style=lstinlinestyle]{b} as the head of the second rule). The larger $P_{head}$ is, the more recursions will be generated;
meanwhile, the more time it will take to evaluate the newly generated rule. 
For $P_{head}$, we measured its impact on the number and complexity of recursions in the last test case when the test iteration was terminated.

\paragraph{Max iterations for recursions ($Max_{iter} > 0$)}
$Max_{iter}$ controls up to how many iterations are executed in the fixpoint computation for recursion.
As shown in Section~\ref{sect:oracle}, for recursions without negation, we continually evaluate the rules individually until all of them reach a fixpoint. However, not all cases have a fixpoint (\eg \lstinline[style=lstinlinestyle]{a(A+1):-b(A). b(A):-a(A).}). 
If the iterations exceed this threshold, we need to abandon this rule and generate a new one. Larger thresholds may result in spending more time on invalid recursive queries, while smaller thresholds may cause us to overlook complex recursions that could potentially trigger bugs. For $Max_{iter}$, we also measured its impact on the number and complexity of recursions. 

For $P_{head}$ and $Max_{iter}$, determining the precise number and complexity of recursions in a test case can be challenging as some intersecting recursions are difficult to differentiate. We use the average number and size of the cycles in the precedence graph to approximate this, and treat two intersecting cycles as separate entities. 
As shown in Figure~\ref{fig:reucrsive}, we can observe that an increase in $P_{head}$ results in a significant increase in the number and size of circles, indicating an increase in test case complexity. 
An increase in $Max_{iter}$ leads to an increase in both the number and size of cycles, though the impact is not significant when compared with $P_{head}$.

\subsection{Efficiency}
\label{sect:efficiency}
We compared the efficiency of our approach in generating complex test cases that produce non-empty results with a random test case generation approach. 
We evaluated this comparison under different test case sizes, denoted by $Max_{rules}$, which determines the maximum number of rules generated in a test case.

\paragraph{Methodology}
We implemented a random test case generator, which adds the newly generated rule directly to the test case until reaching the $Max_{rules}$ threshold, and generates test oracles for the final test case by evaluating each rule individually. 
For this experiment, we only gather statistics on the last test case when the test iteration was terminated. 
Finally, we evaluated the total number of test cases, valid test cases, and test cases with non-empty results considering various test case sizes (\ie $Max_{rules}$).
We conducted this experiment on Souffl\'{e} and DDlog, as µZ and CozoDB have similar performance trends with Souffl\'{e}, and DDlog is slower compared to the other engines.

\begin{figure}
\centering
    \begin{tikzpicture}[scale=1, font=\tiny]
    \begin{axis}[
        name=souffle,
        ybar,
	x tick label style={
		/pgf/number format/1000 sep=},
	ylabel={$\#\ Test\ Cases$},
        xlabel={Souffl\'{e} Program Size},
	enlargelimits=0.05,
	legend style={at={(-0.1,0.-0.2)},anchor=north west,legend columns=3},
        symbolic x coords={40, 60, 80},
        xtick=data,
        enlarge x limits=+0.3,
        bar width=0.06cm,
        width=.48\linewidth, height=.48\linewidth,
    ]
    \addlegendentry{Inc-total}
    \addplot coordinates {(40, 12932) (60, 7190) (80, 4535)};
    
    \addlegendentry{Inc-vaild}
    \addplot coordinates {(40, 12932) (60, 7190) (80, 4535)};

    \addlegendentry{Inc-non\_empty}
    \addplot coordinates {(40, 11008) (60, 6110) (80, 3901)};

    \addlegendentry{Ran-total}
    \addplot coordinates {(40, 43503) (60, 31309) (80, 25025)};

    \addlegendentry{Ran-valid}
    \addplot coordinates {(40, 10647) (60, 4412) (80, 2125)};

    \addlegendentry{Ran-non\_empty}
    \addplot coordinates {(40, 658) (60, 197) (80, 89)};
\legend{}
\end{axis}

    \begin{axis}[
        name=ddlog,
        ybar,
        at={(0.5\linewidth,0)},
	x tick label style={
		/pgf/number format/1000 sep=},
	ylabel={$\#\ Test\ Cases$},
        xlabel={DDlog Program Size},
	enlargelimits=0.05,
	legend style={at={(-1.05,0.-0.2)},anchor=north west,legend columns=6},
        symbolic x coords={40, 60, 80},
        xtick=data,
        enlarge x limits=+0.3,
        bar width=0.06cm,
        width=.48\linewidth, height=.48\linewidth
    ]
    \addlegendentry{Inc-total}
    \addplot coordinates {(40, 16) (60, 8) (80, 7)};
    
    \addlegendentry{Inc-vaild}
    \addplot coordinates {(40, 16) (60, 8) (80, 7)};

    \addlegendentry{Inc-non\_empty}
    \addplot coordinates {(40, 14) (60, 7) (80, 7)};

    \addlegendentry{Ran-total}
    \addplot coordinates {(40, 216) (60, 157) (80, 140)};

    \addlegendentry{Ran-valid}
    \addplot coordinates {(40, 168) (60, 112) (80, 78)};

    \addlegendentry{Ran-non\_empty}
    \addplot coordinates {(40, 20) (60, 6) (80, 5)};
\end{axis}
\end{tikzpicture}
    \caption{Comparison with a random test case generation approach. \textit{Inc} refers to the incremental test case generation, and \textit{Ran} refers to the random method.}
    \label{fig:randcompsouffle}
\end{figure}
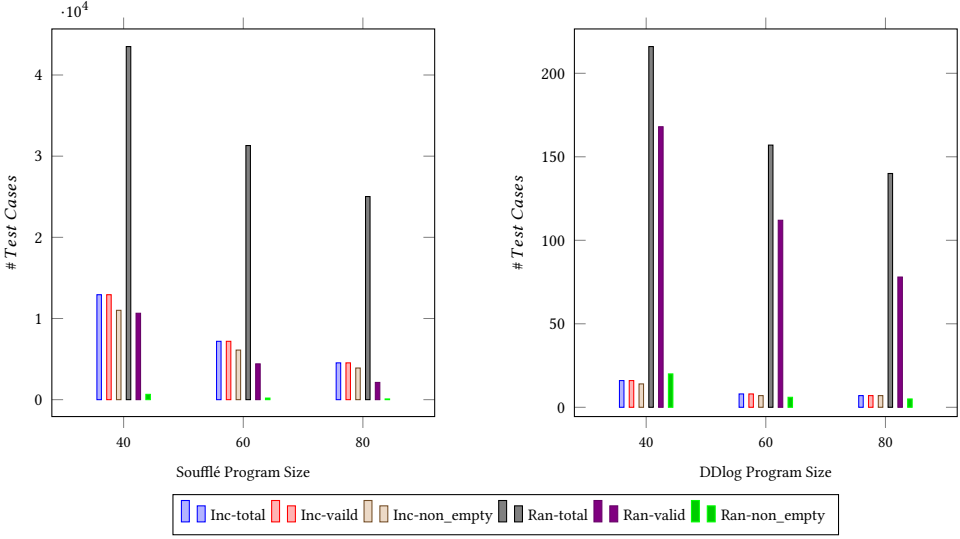

\paragraph{Results.} 
Figure~\ref{fig:randcompsouffle} shows that the random test case generation method consistently produces more test cases over time. 
However, for Souffl\'{e}, as $Max_{rules}$ increases, the number of valid test cases generated by our method surpasses that of the random method (\ie when $Max_{rules}>40$). 
Additionally, the number of test cases with non-empty results generated by our method consistently exceeds that of the random method. For example, when $Max_{rules}=40$, the number of test cases with non-empty results generated by the incremental method is 16.73$\times$ as high as that of the random method for Souffl\'{e}; when $Max_{rules}=60$, this ratio is 31.02 for Souffl\'{e}, and 1.17 for DDlog.
Therefore, incremental test case generation, while taking longer for the execution of the reference programs, allows for generating more complex and, thus, potentially interesting test cases.

\subsection{Comparison with queryFuzz}
\label{sec:compareqf}
queryFuzz is a state-of-the-art approach for testing Datalog engines, 
which is why we compared its effectiveness with \tool.

\paragraph{Challenges}
Fairly comparing the two approaches is difficult.
Our approach primarily aims at finding optimization bugs applied to multiple rules, while queryFuzz finds logic bugs in general, but is restricted to specific transformations; due to the different focus, an apples-to-apples comparison is impossible.
Both queryFuzz and \tool support Souffl\'{e}, µZ, and DDlog, enabling testing these Datalog engines by both tools.
However, while we could test the same version of the Datalog engine that contains the logic bugs with the respectively other tool, we cannot determine whether the two bugs triggered by different test cases have the same root cause, given that many bugs reported by the two tools remain unfixed.

\paragraph{Methodology}
We implemented a best-effort comparison based on manually inspecting and analyzing the bug-inducing test cases and the bugs.
We adopted this methodology from prior testing works~\cite{rigger2020finding, 10.1145/3368089.3409710}.
Specifically, we manually analyzed the bug-inducing test cases generated by the two tools, and attempted to apply the test oracle of the respectively other tools. 
For the logic bugs found by queryFuzz, we evaluated the rules in bug-inducing test cases individually to determine whether our approach could find them. 
However, for the logic bugs found by \tool, we could only reason about which bugs might not be detectable by queryFuzz based on its conceptual limitations. 
It is possible, however, that queryFuzz could trigger the same underlying bug with a different test case, in which case we would incorrectly classify this case as a bug missed by queryFuzz.

\paragraph{Results}
Through manual analysis, we confirmed that all 13 bugs found by queryFuzz could also be detected by \tool. 
For each bug, we included an analysis in our supplementary materials.
Since we could construct bug-inducing test cases using our proposed oracle, we could provide ``witnesses'' that \tool could indeed have found them. Out of the bugs found by \tool, there were five that were not detectable by queryFuzz.

\paragraph{Bugs found by queryFuzz}
\begin{figure}
    \centering
    \parbox{0.7\textwidth}{
        \lstset{style=mystyle, numbers=left, framexleftmargin=7pt}
\begin{lstlisting}[morekeywords={decl, output, symbol, number, magic, no_magic, unsigned}, basicstyle=\scriptsize\ttfamily\bfseries]
.type Word <: symbol
.decl words (i:number, w:Word)
.decl out (w:Word, w1:Word)
.output out
words(0,"a"). words(1, "b").
out(w, w1)  :- words(i, w), words(i+1, w1).
// expected result: out("a", "b").
// buggy result:
// out("a", "a"). out("a", "b"). out("b", "a"). out("b", "b").
\end{lstlisting}
    }
    \caption{A bug in Souffl\'{e} was detected by queryFuzz, which is specific to a single rule and cannot be detected by \approach.}
    \label{fig:buginonerule}
\end{figure}
We divide the 13 logic bugs found by queryFuzz into two categories: those that can be detected by our core test oracle (\ie \approach), and those that can only be detected by an extension of \tool.
12 out of 13 bugs belonged to the first category, and our test oracle was able to detect them by individually evaluating the rules. 
To demonstrate that the correct test oracles were generated by evaluating the rules individually, we used the same keywords and execution options for both the reference programs and the optimized programs for the Souffl\'{e} bugs.
In the second category, the bug was triggered by a single rule, but \approach did not allow us to divide the rule into smaller elements for testing. 
The bug shown in Figure~\ref{fig:buginonerule} was caused by a transformer that identified aliases between the logical variables of subgoals. \approach was unable to identify this bug, because it only occurred in a single rule.  
However, such bugs could have been found also by a naive differential testing approach that would remove or include optimization keywords.\footnote{\url{https://github.com/souffle-lang/souffle/issues/1848}} Integrating this into our method is straightforward by ignoring all keywords in the reference program and maintaining them in the optimized program.
After discovering this, we extended \tool to this end, allowing us to find all 13 bugs found by queryFuzz.
\paragraph{Bugs found by \tool}
We discovered 5 out of \totallogicbug bugs identified by \tool could not be detected by queryFuzz.
queryFuzz's method for detecting bugs in programs beyond conjunctive queries relies on the concept of monotonicity~\cite{10.1145/3468264.3468573};
queryFuzz applies transformations only to conjunctive queries, and the final results should not contradict their oracles, because of monotonicity. 
Four bugs found in Souffl\'{e} do not violate the monotonicity of Datalog. 
\begin{figure}
    \centering
    \parbox{0.45\textwidth}{
        \lstset{style=mystyle, numbers=left, framexleftmargin=7pt}
\begin{lstlisting}[morekeywords={decl, output, symbol, number, magic, no_magic, unsigned}, basicstyle=\scriptsize\ttfamily\bfseries]
.decl a(A:unsigned) magic
.decl b(A:unsigned)
.output b
a(6). a(7). a(3).
b(E) :- a(E).
b(E1) <= b(E2) :- E1<E2.
// expected result: b(7).
// buggy result: b(3). b(6). b(7).
\end{lstlisting}
    }
    \caption{A bug found in Souffl\'{e} that subsumption does not work under the magic sets transformation.}
    \label{fig:subsumption}
\end{figure}
For example, 
the bug shown in Figure~\ref{fig:subsumption} was caused by subsumption, 
which removes facts from a relation based on the conditions specified in its body. 
Since \lstinline[style=lstinlinestyle]{b} is equivalent to \lstinline[style=lstinlinestyle]{a}, we can conclude that there are three facts associated with \lstinline[style=lstinlinestyle]{b}: \lstinline[style=lstinlinestyle]{3}, \lstinline[style=lstinlinestyle]{6}, and \lstinline[style=lstinlinestyle]{7}.
The condition of this subsumption is \lstinline[style=lstinlinestyle]{E1<E2}, which implies that if there are two facts, with one fact being smaller than the other, the smaller fact will be removed. 
Therefore, the correct result is \lstinline[style=lstinlinestyle]{7}.
This bug causes the subsumption to fail to remove any facts, even if they meet the specified conditions. 
Hence, the results of this test case remain unchanged, namely \lstinline[style=lstinlinestyle]{3}, \lstinline[style=lstinlinestyle]{6}, and \lstinline[style=lstinlinestyle]{7}.
This behavior is identical to the case where no facts meet the conditions, and the subsumption will not remove any facts.
Since queryFuzz's test oracle checks a set relationship, it fails to detect this bug.
Suppose that the transformation applied to the test case leads to the derivation of additional facts for the relation, which will undergo subsumption computation. Since queryFuzz is unaware of whether the new facts satisfy the conditions, its test oracle can only rely on comparing the old results to ensure they are either equivalent or a subset of the new results.
queryFuzz could not detect the other three bugs for the same reason. These bugs were related to equivalence relations and the buggy results did not violate queryFuzz's oracle.
More detailed information about these bugs can be found in the supplementary material.
Although subsumption is not implemented in queryFuzz, we believe implementing it in \tool is fair, because it is unclear how queryFuzz could conceptually support such a language feature.

\begin{figure}
    \centering
    \parbox{0.5\textwidth}{
        \lstset{style=mystyle, numbers=left, framexleftmargin=7pt}
\begin{lstlisting}[morekeywords={decl, output, symbol, number, magic, no_magic, unsigned}, basicstyle=\scriptsize\ttfamily\bfseries]
a[A, B] <- [[-0.0, null], [0.0, null]]
b[A, B] := a[A, B], A >= 0
c[A] := a[_, A]
?[A, B] := c[B], b[A, B]
// expected result: [[0.0, null]]
// buggy result: [[0.0, null], [-0.0, null]]
\end{lstlisting}
    }
    \caption{We found a bug in CozoDB resulting from the usage of different comparison functions in the magic sets rewrites.} 
    \label{fig:cozobug2}
\end{figure}

Another bug found in CozoDB is shown in Figure~\ref{fig:cozobug2}, which was caused by the discrepancy between the comparison functions employed in the magic sets rewrite~\cite{DBS-017} and the binary operators utilized in the rules.\footnote{\url{https://github.com/cozodb/cozo/issues/\anonymize{122}}}
The correct result is \lstinline[style=lstinlinestyle]{[[0.0, null]]}, but CozoDB returned \lstinline[style=lstinlinestyle]{[[0.0, null], [-0.0, null]]}. 
After performing the magic sets rewrite, CozoDB employed range scanning on the relation \lstinline[style=lstinlinestyle]{a} instead of testing each individual value. However, the comparison function used in the range scan considers \lstinline[style=lstinlinestyle]{-0.0 >= 0} to be true, while the comparison function used in the binary operators considers it to be false.
This bug can be identified only when the program is evaluated twice, once with the magic sets rewrite and once without it.
We believe that queryFuzz would not be able to detect this bug, because the transformations it applies are designed specifically for conjunctive queries, which consist of single non-recursive function-free Horn rules, and produce conjunctive queries. 
We received feedback from the developer of CozoDB that their magic sets rewrite only applies to conjunctive queries. 
Therefore, the transformations supported by queryFuzz, such as adding an existing subgoal into the rule body, modifying a variable, or removing a subgoal from the rule body, cannot prevent the magic sets rewrite in this particular scenario, since the optimization would be performed on both the original and transformed program.

\section{Discussion}

\paragraph{Summary}
Overall, the results of our evaluation are promising. 
In terms of effectiveness (see Section~\ref{sect:effectiveness}), \tool discovered a total of \totalbugs bugs.
Of these, \totallogicbug were logic bugs.
A key finding of our sensitivity analysis (see Section~\ref{sect:sensitivity}) demonstrates that the execution time of the optimized program is the limiting factor for the test performance. 
Our efficiency comparison (see Section~\ref{sect:efficiency}) indicates that our incremental test case generation method is more efficient than a random test case generation method.
The comparison with queryFuzz (see Section~\ref{sec:compareqf}) reveals that \tool can detect all bugs identified by queryFuzz, a state-of-the-art approach. Out of the bugs identified by \tool, queryFuzz might be unable to detect 5. 

\paragraph{Optimization flags}
An assumption of our work is that Datalog engines do not provide a straightforward way to disable optimizations.
Thus, it could be argued that it is not useful for Datalog engines that provide optimization options that allow disabling all cross-rule optimizations.
However, we believe that the test-case generation approach would still be useful.
Our test-case generation technique incrementally generates and adds rules, which allows it to ensure that the final Datalog program consists of many rules, while also avoiding intermediate steps resulting in empty results, which would make it unlikely to observe any potential discrepancies indicating bugs for the results in the subsequent rules.
As shown in Section~\ref{sect:efficiency}, our incremental test case generation method can produce 1.17$\times$ (for DDlog) to 31.02$\times$ (for Souffl\'{e}) as many valid test cases with non-empty results as a naive random method.
When combining our test-case generation approach with a differential testing approach that uses the same Datalog engine, enabling and disabling all optimization flags, the test oracle could be either applied for every intermediate program---as for IRE---or for the final program. Applying the test oracle for every intermediate program would be costly, as the Datalog engine with optimizations disabled would need to recompute all facts for every intermediate program, while IRE stores the intermediate results as the unoptimized programs. Applying the test oracle for the final Datalog program would be undesirable due to two reasons. First, if a bug is found, reduction techniques are necessary to reduce the bug-inducing test case~\cite{10.1145/3468264.3468573}. Second, once a bug is identified, further expanding the Datalog program after finding a bug is efficient, as the optimizing version of the Datalog engine is called for every intermediate program to prevent subsequent empty results.

\paragraph{Reimplementing an actual reference engine}
Another potential approach to detect logic bugs in optimizations could be to reimplement an actual reference engine,  requiring us to add support for all the data types, operators, and functions, which is not needed for \approach.
A white-box approach could partly reuse these from the existing source code, but that would require access to the source code and a detailed understanding of the Datalog engine under test.

\paragraph{Missed bugs}
\approach cannot detect all logic bugs or optimization bugs.
By directing the Datalog engine's focus to individual rules, \approach provides a test oracle to validate the correctness of the cross-rule optimizations of Datalog engines.
However, while cross-rule optimizations are an important aspect of Datalog engine optimizations, there are also optimizations that are specific to individual rules. These rule-level optimizations pose a challenge for our method in detecting certain bugs, as illustrated by the bug depicted in Figure~\ref{fig:buginonerule}.
We believe that this is not a significant limitation, as an automated testing approach is valuable if it efficiently finds some important bugs.
In fact, this points towards an inherent limitation of testing as coined by
Edsger W. Dijkstra~\cite{dijkstra1970notes}, ``Program testing can be used to show the presence of bugs, but never to show their absence!''

\paragraph{Importance of cross-rule optimization bugs}
The number of fixed bugs is often deemed an important indicator of bug importance.
Regarding the bugs reported in CozoDB, the developer fixed one out of the two bugs we reported. The CozoDB developers expressed the belief that the incorrect results would go unnoticed if someone casually glanced at them.\footnote{\url{https://github.com/cozodb/cozo/issues/\anonymize{101}}} The CozoDB developers attributed the unfixed bug to the inherent complexity of the system.
Although the Datalog engine in µZ is no longer maintained, the developers fixed one of the two reported bugs. 
While the percentage of fixed bugs in Souffl\'{e} may be relatively low, we received feedback from the developer of Souffl\'{e} acknowledging the significance of our findings and expressing the hope that we continue to discover more bugs. However, due to limited resources, they are unable to fix all of the identified bugs at the moment.
Furthermore, as demonstrated in Section~\ref{sec:compareqf}, \approach successfully detects all the bugs identified by queryFuzz that are attributed to cross-rule optimizations.
Overall, we thus believe that \approach can find important optimization bugs.

\section{RELATED WORK}\label{sect:rela}
\paragraph{Automatic testing for Datalog}
queryFuzz~\cite{10.1145/3468264.3468573} is a state-of-the-art technique for detecting logic bugs in Datalog engines. It transforms conjunctive queries with three mutation strategies on a single rule: by adding transformations, modifying transformations, and removing transformations. 
Its test oracle checks one of 
three relationships between the original and transformed program: equivalence, containment, and substitution. queryFuzz detects bugs based on the monotonicity of Datalog programs without negation. 
For a monotonic query $Q$, it holds that $Q(I)\subseteq Q(J)$ when $I\subseteq J$; $I$ and $J$ are the inputs of $Q$. 
DLSmith~\cite{DBLP:conf/issta/MansurWC23} extends queryFuzz by adding an annotated precedence graph, enabling rule-level program transformations such as adding or removing a rule.
DLSmith performs complex transformations by considering the entire program, while queryFuzz only focuses on a single rule.
While both queryFuzz and DLSmith have been successful, they have limitations with respect to \tool. They cannot detect bugs in negation, because they can apply transformations only in lower strata than the rules with negation. 
Their reliance on a set relationship as a test oracle renders them inadequate in detecting bugs that do not violate the monotonicity property. 
Furthermore, it is unclear whether and how these approaches handle recursive queries.
Furthermore, our method incrementally generates test cases, ensuring that they produce valid and non-empty results. Our efficiency evaluation demonstrates that our incremental approach outperforms 
 random generation, which is used also by queryFuzz and DLSmith, in generating non-empty and valid test cases.
We omitted a comparison with DLSmith in our evaluation as we consider it concurrent work.
We reported the first bug four days after the last DLSmith-related bug report. For 9 of the logic bugs that we found in Souffl\'{e}, 8 were introduced before DLSmith's first report, indicating that DLSmith overlooked them.

\paragraph{Debugging for Datalog} 
Datalog debugging helps to identify the underlying cause when encountering incorrect results. \citet{caballero2008theoretical} presented a theoretical framework for identifying bugs based on declarative debugging and computation graphs. 
Later, \citet{caballero2015debugging} proposed another Datalog debugging approach based on the principles of algorithmic debugging, which involves posing questions about the validity or invalidity of intermediate results until the bug is identified. 
Computing provenance reveals the process of Datalog computation in order to assist developers in analyzing the root cause of bugs. \citet{kohler2012declarative} computed provenance by tracking the history of rule derivations, while \citet{10.1145/3379446} computed provenance based on partial proof trees, enabling the computation of provenance for large-scale Datalog programs.

\paragraph{Optimizations of Datalog engines} Bottom-up and top-down are two evaluation strategies for Datalog interpreters~\cite{bancilhon1986amateur}. Bottom-up methods evaluate all rules in a program based on the EDB and generate all possible tuples until no more facts can be derived. The semi-naïve method, a bottom-up approach, seeks to minimize redundant derivations. On the other hand, top-down methods start from the output relation and only evaluate the necessary rules that are related to the output. 
Magic set transformation~\cite{10.1145/6012.15399,     alviano2012magic, 10.1145/1376616.1376673} is an optimization technique for the bottom-up approach that involves rewriting rules for improved performance or reduced memory usage, such as through merging or splitting rules. 
Approaches to join ordering problems~\cite{10.1007/978-3-031-16767-6_5} seek to find the most effective way to incorporate relations into rules in order to improve performance. 
Index selection~\cite{subotic2018automatic, ryzhyk2019differential} is employed to reduce the time required to look up relations during the evaluation.
The above-mentioned optimizations aim to improve the efficiency of executing rules.
Many other optimization techniques exist that leverage specific language features for optimization, such as, subsumption~\cite{kiessling1994database}, equivalence relation~\cite{nappa2019fast}, choice~\cite{hu2021choice}, and specialized data structures~\cite{jordan2019brie, jordan2019specialized, jordan2022specializing}.

\paragraph{Metamorphic testing}
Metamorphic testing~\cite{chen2020metamorphic} is a methodology for detecting bugs based on so-called metamorphic relations between a test case and its follow-up version. 
Both queryFuzz, as well as our test oracle, are metamorphic testing approaches, as they define a relation between the results of two Datalog programs. 
For database engines, NoREC~\cite{10.1145/3368089.3409710} rewrites queries to prevent their optimization; while the high-level idea of NoREC and IRE are similar, their actual approaches differ significantly.
NoREC transforms the predicate in the \lstinline[style=lstinlinestyle]{WHERE} clause into a query that evaluates the predicate's result for each row of the table. This transformation exploits the difference in optimization strategies between the fetch clause and the \texttt{WHERE} clause.
IRE leverages the compositional nature of Datalog programs, where rules can be evaluated individually, enabling the prevention of cross-rule optimizations.
Many other metamorphic testing approaches have been proposed for various domains, which have been surveyed in the literature~\cite{segura2016survey, chen2018metamorphic}.

\section{conclusion}\label{sect:con}

We have presented a general, highly-effective method, called \emph{Incremental Rule Evaluation (\approach)} for detecting bugs in Datalog engines.  Our key insight is that by evaluating the rules of an optimized Datalog program individually as reference programs, we can disable the optimizations applied to multiple rules in a black-box manner, enabling a test oracle that can detect optimization bugs by comparing the result sets of the two programs. 
We also proposed an incremental test case generation method that efficiently generates valid test cases with a controlled likelihood of producing non-empty results. 
We have implemented this approach as a tool called \tool.
We evaluated \tool on four mature Datalog engines, and discovered a total of \totalbugs bugs. Of these, \totallogicbug were logic bugs. 
Moreover, our evaluation illustrates that our incremental test case generation approach is more efficient than a random method.
We believe that our incremental approach for test case and test oracle generation can be extended to other Datalog engines and data-centric systems to detect bugs.
Furthermore, we hope that it will be widely adopted as a simple, general testing approach.

\section*{Data-Availability Statement}
We include \tool, the tool that we developed, as well as all experimental results as an artifact at \url{https://github.com/nus-test/Deopt}, and archived it for long-term storage at \url{https://zenodo.org/records/10609061}.


\begin{acks}
 This work was partially supported by the National Natural Science Foundation of China under Grant No.62032010, No.62232001 and No.62202220. 
 We would like to thank the support from the Collaborative Innovation Center of Novel Software Technology and Industrialization, Jiangsu, China. 
 We also thank for the financial support from the program of China Scholarships Council (No.202106190065).
\end{acks}

\bibliography{reference}
\end{document}


\title[Supplementary Material]{Supplementary Material}

\begin{center}
    \huge\textbf{Supplementary Material}
\end{center}

\section{Q1}
In this section, we illustrate all the logic bugs detected by \tool.

\subsection{Bug 1}

Bug 1 was found in Souffl\'{e}. In interpreter mode of Souffl\'{e}, \texttt{0.0} not equals to \texttt{-0.0}, because they use bitwise comparison for float type. 
This bug was found due to an inconsistency, as the presence of \lstinline[morekeywords={inline}, style=lstinlinestyle]{inline} keyword changed the result from \texttt{-0} to \texttt{0}. 
And we checked that when we remove the \lstinline[morekeywords={inline}, style=lstinlinestyle]{inline} keyword or replaced it with \lstinline[style=lstinlinestyle]{magic} keyword, the final result equaled to \texttt{-0}, which means the \lstinline[morekeywords={inline}, style=lstinlinestyle]{inline} keyword result in this inconsistent behavior.
Here, we omit the \lstinline[morekeywords={inline}, style=lstinlinestyle]{inline} keyword for the reference program, as Souffle does not support inline optimization for input and output relations. 
However, we think that our test oracle can block the inline optimization if we can add \lstinline[morekeywords={inline}, style=lstinlinestyle]{inline} keyword for reference programs.

\begin{table}[H]\footnotesize
    \centering
    \caption{Bug 1 found by \tool.}
    \begin{tabular}{c l l l l} \toprule
         iteration & $P^{ref}_{n}$ & $Facts^{ref}_{output\_rel}$ & $P^{opt}_{n}$ & $Facts^{opt}_{output\_rel}$ \\\midrule
0
&
\begin{minipage}[c]{.3\linewidth}
\lstset{style=mystyle, frame=0}
\begin{lstlisting}[morekeywords={decl, output, symbol, number, magic}, basicstyle=\scriptsize\ttfamily\bfseries]
.decl olkw(A:float)
.decl kkcb(A:float)
.output kkcb
olkw(-0).
kkcb(A) :- olkw(A).
\end{lstlisting}
\end{minipage} 
&  
\begin{minipage}[c]{.1\linewidth}
\lstset{style=mystyle, frame=0}
\begin{lstlisting}[morekeywords={decl, output, symbol, number, magic}, basicstyle=\scriptsize\ttfamily\bfseries]
kkcb(-0).
\end{lstlisting}
\end{minipage} 
&
\begin{minipage}[c]{.3\linewidth}
\lstset{style=mystyle, frame=0}
\begin{lstlisting}[morekeywords={decl, output, symbol, number, magic, inline}, basicstyle=\scriptsize\ttfamily\bfseries]
.decl olkw(A:float)
.decl kkcb(A:float)
.output kkcb
olkw(-0).
kkcb(A) :- olkw(A).
\end{lstlisting}
\end{minipage} 
&
\begin{minipage}[c]{.1\linewidth}
\lstset{style=mystyle, frame=0}
\begin{lstlisting}[morekeywords={decl, output, symbol, number, magic}, basicstyle=\scriptsize\ttfamily\bfseries]
kkcb(-0).
\end{lstlisting}
\end{minipage} 
\\ 
1
&
\begin{minipage}[c]{.3\linewidth}
\lstset{style=mystyle, frame=0}
\begin{lstlisting}[morekeywords={decl, output, symbol, number, magic}, basicstyle=\scriptsize\ttfamily\bfseries]
.decl kkcb(A:float)
.decl nvuk(A:float)
.output nvuk
kkcb(-0).
nvuk(A) :- kkcb(A), 0.0 = A.
\end{lstlisting}
\end{minipage} 
&  
\begin{minipage}[c]{.1\linewidth}
\lstset{style=mystyle, frame=0}
\begin{lstlisting}[morekeywords={decl, output, symbol, number, magic}, basicstyle=\scriptsize\ttfamily\bfseries]
nvuk(-0).
\end{lstlisting}
\end{minipage} 
&
\begin{minipage}[c]{.3\linewidth}
\lstset{style=mystyle, frame=0}
\begin{lstlisting}[morekeywords={decl, output, symbol, number, magic, inline}, basicstyle=\scriptsize\ttfamily\bfseries]
.decl olkw(A:float)
.decl kkcb(A:float) inline
.decl nvuk(A:float)
.output nvuk
olkw(-0).
kkcb(A) :- olkw(A).
nvuk(A) :- kkcb(A), 0.0 = A.
\end{lstlisting}
\end{minipage} 
&
\begin{minipage}[c]{.1\linewidth}
\lstset{style=mystyle, frame=0}
\begin{lstlisting}[morekeywords={decl, output, symbol, number, magic}, basicstyle=\scriptsize\ttfamily\bfseries]
nvuk(0).
\end{lstlisting}
\end{minipage} 
\\\bottomrule
    \end{tabular}
\end{table}

\subsection{Bug 2}

Bug 2 is much similar with bug 1. 
In this test case, the inconsistency was only caused by \lstinline[style=lstinlinestyle]{magic} transformation. 
Based on this, we determined they were different unique bugs, and also different from bug 5 found by queryFuzz shown in Section~\ref{sec:qb5}. To demonstrate the effectiveness of our method, we did not remove the \lstinline[style=lstinlinestyle]{magic} keyword in the reference program.

\begin{table}[H]\footnotesize
    \centering
    \caption{Bug 2 found by \tool.}
    \begin{tabular}{c l l l l} \toprule
         iteration & $P^{ref}_{n}$ & $Facts^{ref}_{output\_rel}$ & $P^{opt}_{n}$ & $Facts^{opt}_{output\_rel}$ \\\midrule
0
&
\begin{minipage}[c]{.3\linewidth}
\lstset{style=mystyle, frame=0}
\begin{lstlisting}[morekeywords={decl, output, symbol, number, magic}, basicstyle=\scriptsize\ttfamily\bfseries]
.decl yona(A:symbol, C:float)
.decl idck(B:float) magic
.output idck
yona("DrdskfIQ9A", 0).
idck(A) :- yona(D, A).
\end{lstlisting}
\end{minipage} 
&  
\begin{minipage}[c]{.1\linewidth}
\lstset{style=mystyle, frame=0}
\begin{lstlisting}[morekeywords={decl, output, symbol, number, magic}, basicstyle=\scriptsize\ttfamily\bfseries]
idck(0).
\end{lstlisting}
\end{minipage} 
&
\begin{minipage}[c]{.3\linewidth}
\lstset{style=mystyle, frame=0}
\begin{lstlisting}[morekeywords={decl, output, symbol, number, magic}, basicstyle=\scriptsize\ttfamily\bfseries]
.decl yona(A:symbol, C:float)
.decl idck(B:float) magic
.output idck
yona("DrdskfIQ9A", 0).
idck(A) :- yona(D, A).
\end{lstlisting}
\end{minipage} 
&
\begin{minipage}[c]{.1\linewidth}
\lstset{style=mystyle, frame=0}
\begin{lstlisting}[morekeywords={decl, output, symbol, number, magic}, basicstyle=\scriptsize\ttfamily\bfseries]
idck(0).
\end{lstlisting}
\end{minipage} 
\\
1
&
\begin{minipage}[c]{.27\linewidth}
\lstset{style=mystyle, frame=0}
\begin{lstlisting}[morekeywords={decl, output, symbol, number, magic}, basicstyle=\scriptsize\ttfamily\bfseries]
.decl idck(B:float) magic
.decl zsjo(A:float) 
.output zsjo
idck(0).
zsjo(D) :- idck(D), idck(min(-D,D)).
\end{lstlisting}
\end{minipage} 
&  
\begin{minipage}[c]{.13\linewidth}
\lstset{style=mystyle, frame=0}
\begin{lstlisting}[morekeywords={decl, output, symbol, number, magic}, basicstyle=\scriptsize\ttfamily\bfseries]
zsjo(0).
\end{lstlisting}
\end{minipage} 
&
\begin{minipage}[c]{.27\linewidth}
\lstset{style=mystyle, frame=0}
\begin{lstlisting}[morekeywords={decl, output, symbol, number, magic}, basicstyle=\scriptsize\ttfamily\bfseries]
.decl yona(A:symbol, C:float)
.decl idck(B:float) magic
.decl zsjo(A:float) 
.output zsjo
yona("DrdskfIQ9A", 0).
idck(A) :- yona(D, A).
zsjo(D) :- idck(D), idck(min(-D,D)).
\end{lstlisting}
\end{minipage} 
&
\begin{minipage}[c]{.13\linewidth}
\lstset{style=mystyle, frame=0}
\begin{lstlisting}[morekeywords={decl, output, symbol, number, magic}, basicstyle=\scriptsize\ttfamily\bfseries]
-
\end{lstlisting}
\end{minipage} 
\\\bottomrule
    \end{tabular}
\end{table}

\clearpage
\subsection{Bug 3}
\label{sec:debug3}

Bug 3 was a bug in Souffl\'{e}, related to subsumption and \lstinline[style=lstinlinestyle]{magic} transformation. For this bug, we thought that queryFuzz was difficult to detect. queryFuzz tests queries which beyond conjunctive queries based on the monotonicity of conjunctive queries. Subsumption is used to remove some facts of a relation, based on the conditions in its body. If no facts meet these conditions, the number of facts will not change. So it is difficult to predict how the number of facts will change after subsumption. Adding or removing facts to or from the relation in subsumption may leave the results of this relation unchanged after the subsumption. Because we cannot tell whether the added or deleted facts are the ones to be removed by the subsumption. Such as we want to add facts to the EDB, we can only assume that the old results are a subset of the new results, and cannot tell if they should be equal. But in this bug, subsumption not works under \lstinline[style=lstinlinestyle]{magic} transformation, which means it will remove nothing from the relation. This does not contradict the old result is the subset of the new result. So queryFuzz can not detect this bug. 

\begin{table}[H]\footnotesize
    \centering
    \caption{Bug 3 found by \tool.}
    \begin{tabular}{c l l l l} \toprule
         iteration & $P^{ref}_{n}$ & $Facts^{ref}_{output\_rel}$ & $P^{opt}_{n}$ & $Facts^{opt}_{output\_rel}$ \\\midrule
0
&
\begin{minipage}[c]{.3\linewidth}
\lstset{style=mystyle, frame=0}
\begin{lstlisting}[morekeywords={decl, output, symbol, number, magic, unsigned}, basicstyle=\scriptsize\ttfamily\bfseries]
.decl buyh(A:unsigned) magic
.decl eozn(A:unsigned)
.output eozn
buyh(6). buyh(7). buyh(3).
eozn(E) :- buyh(E).
\end{lstlisting}
\end{minipage} 
&  
\begin{minipage}[c]{.1\linewidth}
\lstset{style=mystyle, frame=0}
\begin{lstlisting}[morekeywords={decl, output, symbol, number, magic}, basicstyle=\scriptsize\ttfamily\bfseries]
eozn(6). 
eozn(7). 
eozn(3).
\end{lstlisting}
\end{minipage} 
&
\begin{minipage}[c]{.3\linewidth}
\lstset{style=mystyle, frame=0}
\begin{lstlisting}[morekeywords={decl, output, symbol, number, magic, unsigned}, basicstyle=\scriptsize\ttfamily\bfseries]
.decl buyh(A:unsigned) magic
.decl eozn(A:unsigned)
.output eozn
buyh(6). buyh(7). buyh(3).
eozn(E) :- buyh(E).
\end{lstlisting}
\end{minipage} 
&
\begin{minipage}[c]{.1\linewidth}
\lstset{style=mystyle, frame=0}
\begin{lstlisting}[morekeywords={decl, output, symbol, number, magic}, basicstyle=\scriptsize\ttfamily\bfseries]
eozn(6). 
eozn(7). 
eozn(3).
\end{lstlisting}
\end{minipage} 
\\
1
&
\begin{minipage}[c]{.27\linewidth}
\lstset{style=mystyle, frame=0}
\begin{lstlisting}[morekeywords={decl, output, symbol, number, magic, unsigned}, basicstyle=\scriptsize\ttfamily\bfseries]
.decl eozn(A:unsigned)
.output eozn
eozn(6). eozn(7). eozn(3).
eozn(E1) <= eozn(E2) :- E1<E2.
\end{lstlisting}
\end{minipage} 
&  
\begin{minipage}[c]{.13\linewidth}
\lstset{style=mystyle, frame=0}
\begin{lstlisting}[morekeywords={decl, output, symbol, number, magic}, basicstyle=\scriptsize\ttfamily\bfseries]
eozn(7).
\end{lstlisting}
\end{minipage} 
&
\begin{minipage}[c]{.27\linewidth}
\lstset{style=mystyle, frame=0}
\begin{lstlisting}[morekeywords={decl, output, symbol, number, magic, unsigned}, basicstyle=\scriptsize\ttfamily\bfseries]
.decl buyh(A:unsigned) magic
.decl eozn(A:unsigned)
.output eozn
buyh(6). buyh(7). buyh(3).
eozn(E) :- buyh(E).
eozn(E1) <= eozn(E2) :- E1<E2.
\end{lstlisting}
\end{minipage} 
&
\begin{minipage}[c]{.13\linewidth}
\lstset{style=mystyle, frame=0}
\begin{lstlisting}[morekeywords={decl, output, symbol, number, magic}, basicstyle=\scriptsize\ttfamily\bfseries]
eozn(6). 
eozn(7). 
eozn(3).
\end{lstlisting}
\end{minipage} 
\\\bottomrule
    \end{tabular}
\end{table}

\subsection{Bug 4}
\label{sec:debug4}

Bug 4 was a bug in Souffl\'{e}, related to subsumption and \lstinline[style=lstinlinestyle]{magic} transformation. The difference from bug 3 was that in this test case, the results of subsumption were correct. Similar with bug 3, queryFuzz was also unable to detect this bug due to the same underlying reason.

\begin{table}[H]\footnotesize
    \centering
    \caption{Bug 4 found by \tool.}
    \begin{tabular}{c l l l l} \toprule
         iteration & $P^{ref}_{n}$ & $Facts^{ref}_{output\_rel}$ & $P^{opt}_{n}$ & $Facts^{opt}_{output\_rel}$ \\\midrule
0
&
\begin{minipage}[c]{.3\linewidth}
\lstset{style=mystyle, frame=0}
\begin{lstlisting}[morekeywords={decl, output, symbol, number, magic, unsigned}, basicstyle=\scriptsize\ttfamily\bfseries]
.decl kdof(A:unsigned, B:number)
.output kdof
kdof(1, -2). kdof(1, 9).
kdof(B, A1) <= kdof(B, A2) :- A1<A2.
\end{lstlisting}
\end{minipage} 
&  
\begin{minipage}[c]{.1\linewidth}
\lstset{style=mystyle, frame=0}
\begin{lstlisting}[morekeywords={decl, output, symbol, number, magic}, basicstyle=\scriptsize\ttfamily\bfseries]
kdof(1, 9).
\end{lstlisting}
\end{minipage} 
&
\begin{minipage}[c]{.3\linewidth}
\lstset{style=mystyle, frame=0}
\begin{lstlisting}[morekeywords={decl, output, symbol, number, magic, unsigned}, basicstyle=\scriptsize\ttfamily\bfseries]
.decl kdof(A:unsigned, B:number)
.output kdof
kdof(1, -2). kdof(1, 9).
kdof(B, A1) <= kdof(B, A2) :- A1<A2.
\end{lstlisting}
\end{minipage} 
&
\begin{minipage}[c]{.1\linewidth}
\lstset{style=mystyle, frame=0}
\begin{lstlisting}[morekeywords={decl, output, symbol, number, magic}, basicstyle=\scriptsize\ttfamily\bfseries]
kdof(1, 9).
\end{lstlisting}
\end{minipage} 
\\
1
&
\begin{minipage}[c]{.27\linewidth}
\lstset{style=mystyle, frame=0}
\begin{lstlisting}[morekeywords={decl, output, symbol, number, magic, unsigned}, basicstyle=\scriptsize\ttfamily\bfseries]
.decl kdof(A:unsigned, B:number)
.decl ovqb(A:number, B:unsigned) magic
.output khkw
kdof(1, 9).
ovqb(A^-3, B) :- kdof(B, A).
\end{lstlisting}
\end{minipage} 
&  
\begin{minipage}[c]{.13\linewidth}
\lstset{style=mystyle, frame=0}
\begin{lstlisting}[morekeywords={decl, output, symbol, number, magic}, basicstyle=\scriptsize\ttfamily\bfseries]
ovqb(0, 1).
\end{lstlisting}
\end{minipage} 
&
\begin{minipage}[c]{.27\linewidth}
\lstset{style=mystyle, frame=0}
\begin{lstlisting}[morekeywords={decl, output, symbol, number, magic, unsigned}, basicstyle=\scriptsize\ttfamily\bfseries]
.decl kdof(A:unsigned, B:number)
.decl ovqb(A:number, B:unsigned) magic
.decl khkw(A:number, B:unsigned)
.output ovqb
kdof(1, -2).
kdof(1, 9).
kdof(B, A1) <= kdof(B, A2) :- A1<A2.
ovqb(A^-3, B) :- kdof(B, A).
\end{lstlisting}
\end{minipage} 
&
\begin{minipage}[c]{.13\linewidth}
\lstset{style=mystyle, frame=0}
\begin{lstlisting}[morekeywords={decl, output, symbol, number, magic}, basicstyle=\scriptsize\ttfamily\bfseries]
ovqb(0, 1).
\end{lstlisting}
\end{minipage} 
\\
2
&
\begin{minipage}[c]{.27\linewidth}
\lstset{style=mystyle, frame=0}
\begin{lstlisting}[morekeywords={decl, output, symbol, number, magic, unsigned}, basicstyle=\scriptsize\ttfamily\bfseries]
.decl kdof(A:unsigned, B:number)
.decl ovqb(A:number, B:unsigned) magic
.decl khkw(A:number, B:unsigned)
.output khkw
kdof(1, 9). ovqb(0, 1).
khkw(A, B) :- kdof(B, A), !ovqb(A, B).
\end{lstlisting}
\end{minipage} 
&  
\begin{minipage}[c]{.13\linewidth}
\lstset{style=mystyle, frame=0}
\begin{lstlisting}[morekeywords={decl, output, symbol, number, magic}, basicstyle=\scriptsize\ttfamily\bfseries]
khkw(9, 1).
\end{lstlisting}
\end{minipage} 
&
\begin{minipage}[c]{.27\linewidth}
\lstset{style=mystyle, frame=0}
\begin{lstlisting}[morekeywords={decl, output, symbol, number, magic, unsigned}, basicstyle=\scriptsize\ttfamily\bfseries]
.decl kdof(A:unsigned, B:number)
.decl ovqb(A:number, B:unsigned) magic
.decl khkw(A:number, B:unsigned)
.output khkw
kdof(1, -2). kdof(1, 9).
kdof(B, A1) <= kdof(B, A2) :- A1<A2.
ovqb(A^-3, B) :- kdof(B, A).
khkw(A, B) :- kdof(B, A), !ovqb(A, B).
\end{lstlisting}
\end{minipage} 
&
\begin{minipage}[c]{.13\linewidth}
\lstset{style=mystyle, frame=0}
\begin{lstlisting}[morekeywords={decl, output, symbol, number, magic}, basicstyle=\scriptsize\ttfamily\bfseries]
khkw(-2, 1).
khkw(9, 1).
\end{lstlisting}
\end{minipage} 
\\\bottomrule
    \end{tabular}
\end{table}

\clearpage
\subsection{Bug 5}
\label{sec:debug5}

Bug 5 was a bug in Souffl\'{e}, related to equivalence relations and \lstinline[style=lstinlinestyle]{magic} transformation. Equivalence relation is a binary relation in Souffl\'{e} and exhibits three properties: reflexivity, symmetry, and transitivity. So when given an input to the equivalence relation, that input should be a subset of the final result, but there is no way to tell if they are equal. This bug caused by that equivalence relation not works under \lstinline[style=lstinlinestyle]{magic} transformation. So similar with bug 3, queryFuzz can not detect this bug. This bug also could be triggered with \lstinline[style=lstinlinestyle]{inline} transformation.

\begin{table}[H]\footnotesize
    \centering
    \caption{Bug 5 found by \tool.}
    \begin{tabular}{c l l l l} \toprule
         iteration & $P^{ref}_{n}$ & $Facts^{ref}_{output\_rel}$ & $P^{opt}_{n}$ & $Facts^{opt}_{output\_rel}$ \\\midrule
0
&
\begin{minipage}[c]{.3\linewidth}
\lstset{style=mystyle, frame=0}
\begin{lstlisting}[morekeywords={decl, output, symbol, number, magic, eqrel, unsigned}, basicstyle=\scriptsize\ttfamily\bfseries]
.decl a(c:unsigned)
.decl d(b:unsigned, c:unsigned) magic eqrel
.output d
a(8). a(4).
d(f+4, f) :- a(f).
\end{lstlisting}
\end{minipage} 
&  
\begin{minipage}[c]{.1\linewidth}
\lstset{style=mystyle, frame=0}
\begin{lstlisting}[morekeywords={decl, output, symbol, number, magic}, basicstyle=\scriptsize\ttfamily\bfseries]
d(4, 4).
d(4, 8).
d(4, 12).
d(8, 4).
d(8, 8).
d(8, 12).
d(12, 4).
d(12, 8).
d(12, 12).
\end{lstlisting}
\end{minipage} 
&
\begin{minipage}[c]{.3\linewidth}
\lstset{style=mystyle, frame=0}
\begin{lstlisting}[morekeywords={decl, output, symbol, number, magic, eqrel, unsigned}, basicstyle=\scriptsize\ttfamily\bfseries]
.decl a(c:unsigned)
.decl d(b:unsigned, c:unsigned) magic eqrel
.output d
a(8). a(4).
d(f+4, f) :- a(f).
\end{lstlisting}
\end{minipage} 
&
\begin{minipage}[c]{.1\linewidth}
\lstset{style=mystyle, frame=0}
\begin{lstlisting}[morekeywords={decl, output, symbol, number, magic}, basicstyle=\scriptsize\ttfamily\bfseries]
d(4, 4).
d(4, 8).
d(4, 12).
d(8, 4).
d(8, 8).
d(8, 12).
d(12, 4).
d(12, 8).
d(12, 12).
\end{lstlisting}
\end{minipage} 
\\
1
&
\begin{minipage}[c]{.27\linewidth}
\lstset{style=mystyle, frame=0}
\begin{lstlisting}[morekeywords={decl, output, symbol, number, magic, eqrel, unsigned}, basicstyle=\scriptsize\ttfamily\bfseries]
.decl a(c:unsigned)
.decl d(b:unsigned, c:unsigned) magic eqrel
.decl bxax(e:unsigned)
.output bxax
a(8). a(4). d(4, 4). d(4, 8). d(4, 12). 
d(8, 4). d(8, 8). d(8, 12). d(12, 4). 
d(12, 8). d(12, 12).
bxax(c) :- d(c, e), a(c).
\end{lstlisting}
\end{minipage} 
&  
\begin{minipage}[c]{.13\linewidth}
\lstset{style=mystyle, frame=0}
\begin{lstlisting}[morekeywords={decl, output, symbol, number, magic}, basicstyle=\scriptsize\ttfamily\bfseries]
bxax(4).
bxax(8).
\end{lstlisting}
\end{minipage} 
&
\begin{minipage}[c]{.27\linewidth}
\lstset{style=mystyle, frame=0}
\begin{lstlisting}[morekeywords={decl, output, symbol, number, magic, inline, eqrel, unsigned}, basicstyle=\scriptsize\ttfamily\bfseries]
.decl a(c:unsigned)
.decl d(b:unsigned, c:unsigned) magic eqrel
.decl bxax(e:unsigned)
.output bxax
a(8). a(4).
d(f+4, f) :- a(f).
bxax(c) :- d(c, e), a(c).
\end{lstlisting}
\end{minipage} 
&
\begin{minipage}[c]{.13\linewidth}
\lstset{style=mystyle, frame=0}
\begin{lstlisting}[morekeywords={decl, output, symbol, number, magic}, basicstyle=\scriptsize\ttfamily\bfseries]
bxax(8).
\end{lstlisting}
\end{minipage} 
\\\bottomrule
    \end{tabular}
\end{table}

\subsection{Bug 6}

Bug 6 was a bug in Souffl\'{e}, related to the conflict of \lstinline[style=lstinlinestyle]{inline} and \lstinline[style=lstinlinestyle]{no_magic} keywords.

\begin{table}[H]\footnotesize
    \centering
    \caption{Bug 6 found by \tool.}
    \begin{tabular}{c l l l l} \toprule
         iteration & $P^{ref}_{n}$ & $Facts^{ref}_{output\_rel}$ & $P^{opt}_{n}$ & $Facts^{opt}_{output\_rel}$ \\\midrule
0
&
\begin{minipage}[c]{.3\linewidth}
\lstset{style=mystyle, frame=0}
\begin{lstlisting}[morekeywords={decl, output, symbol, number, magic}, basicstyle=\scriptsize\ttfamily\bfseries]
.decl a(A:number)
.decl b(A:number)
.output b
a(-1). a(1). a(2).
b(A) :- a(A), a(-A).
\end{lstlisting}
\end{minipage} 
&  
\begin{minipage}[c]{.1\linewidth}
\lstset{style=mystyle, frame=0}
\begin{lstlisting}[morekeywords={decl, output, symbol, number, magic}, basicstyle=\scriptsize\ttfamily\bfseries]
b(1).
b(-1).
\end{lstlisting}
\end{minipage} 
&
\begin{minipage}[c]{.3\linewidth}
\lstset{style=mystyle, frame=0}
\begin{lstlisting}[morekeywords={decl, output, symbol, number, magic}, basicstyle=\scriptsize\ttfamily\bfseries]
.decl a(A:number)
.decl b(A:number)
.output b
a(-1). a(1). a(2).
b(A) :- a(A), a(-A).
\end{lstlisting}
\end{minipage} 
&
\begin{minipage}[c]{.1\linewidth}
\lstset{style=mystyle, frame=0}
\begin{lstlisting}[morekeywords={decl, output, symbol, number, magic}, basicstyle=\scriptsize\ttfamily\bfseries]
b(1).
b(-1).
\end{lstlisting}
\end{minipage} 
\\
1
&
\begin{minipage}[c]{.27\linewidth}
\lstset{style=mystyle, frame=0}
\begin{lstlisting}[morekeywords={decl, output, symbol, number, magic, no_magic}, basicstyle=\scriptsize\ttfamily\bfseries]
.decl b(A:number)
.decl c(A:number) no_magic
.output c
b(1). b(-1).
c(A) :- b(A).
\end{lstlisting}
\end{minipage} 
&  
\begin{minipage}[c]{.13\linewidth}
\lstset{style=mystyle, frame=0}
\begin{lstlisting}[morekeywords={decl, output, symbol, number, magic}, basicstyle=\scriptsize\ttfamily\bfseries]
c(1).
c(-1).
\end{lstlisting}
\end{minipage} 
&
\begin{minipage}[c]{.27\linewidth}
\lstset{style=mystyle, frame=0}
\begin{lstlisting}[morekeywords={decl, output, symbol, number, magic, inline, no_magic}, basicstyle=\scriptsize\ttfamily\bfseries]
.decl a(A:number)
.decl b(A:number) inline
.decl c(A:number) no_magic
.output c
a(-1). a(1). a(2).
b(A) :- a(A), a(-A).
c(A) :- b(A).
\end{lstlisting}
\end{minipage} 
&
\begin{minipage}[c]{.13\linewidth}
\lstset{style=mystyle, frame=0}
\begin{lstlisting}[morekeywords={decl, output, symbol, number, magic}, basicstyle=\scriptsize\ttfamily\bfseries]
c(-1).
c(1).
c(2).
\end{lstlisting}
\end{minipage} 
\\\bottomrule
    \end{tabular}
\end{table}

\clearpage
\subsection{Bug 7}
\label{sect:modebug}
Bug 7 was a bug in Souffl\'{e}, it produced different results under interpreter mode and synthesizer mode. In interpreter mode, it could produce the correct results, but in synthesizer mode, it produced empty results. Only when execute the test case with \lstinline[style=lstinlinestyle]{-c} option, we can trigger this bug.

\begin{figure}[H]
    \centering
    \lstset{style=mystyle}
    \parbox{.35\linewidth}{
\begin{lstlisting}[morekeywords={decl, output, symbol, number, magic, eqrel},basicstyle=\scriptsize\ttfamily\bfseries]
.decl payb(A:float, B:float)
.decl uuyd(A:float)
.output uuyd

payb(0.80300000000000082, 0.80300000000000082).
uuyd(F) :- payb(0.80300000000000082, F).

// expected result:
// uuyd(0.80300000000000082).
// buggy result:
// -
\end{lstlisting}}
    \vspace{-7pt}
    \caption{Bug 7 found by \tool.}
    \label{fig:qfbug9}
\end{figure}

\subsection{Bug 8}
\label{sec:debug8}

Bug 8 was a bug in Souffl\'{e}, equivalence relation not worked when we applied provenance to the program, which was controlled by execution option \lstinline[style=lstinlinestyle]{-t}. Provenance is utilized to furnish insight for troubleshooting Datalog programs. This bug was triggered under two specific execution options: \lstinline[style=lstinlinestyle]{-t none} and \lstinline[style=lstinlinestyle]{--disable-transformers=ExpandEqrelsTransformer}. 
Similar with bug 3, queryFuzz can not detect this bug.

\begin{figure}[H]
    \centering
    \lstset{style=mystyle}
    \parbox{.35\linewidth}{
\begin{lstlisting}[morekeywords={decl, output, symbol, number, magic, eqrel},basicstyle=\scriptsize\ttfamily\bfseries]
.decl ytfh(A:float)
.decl bkbi(A:float, B:float) eqrel
.output bkbi

ytfh(1).

bkbi(E, E+1) :- ytfh(E).
// expected result: 
// bkbi(1, 2). bkbi(1, 1). bkbi(2, 2). bkbi(2, 1).
// buggy result:
// bkbi(1, 2).
\end{lstlisting}}
    \vspace{-7pt}
    \caption{Bug 8 found by \tool.}
    \label{fig:qfbug9}
\end{figure}

\subsection{Bug 9}

Bug 9 was a bug in Souffl\'{e}, related to \lstinline[style=lstinlinestyle]{subsumption}.

\begin{table}[H]\footnotesize
    \centering
    \caption{Bug 9 found by \tool.}
    \begin{tabular}{c l l l l} \toprule
         iteration & $P^{ref}_{n}$ & $Facts^{ref}_{output\_rel}$ & $P^{opt}_{n}$ & $Facts^{opt}_{output\_rel}$ \\\midrule
0
&
\begin{minipage}[c]{.3\linewidth}
\lstset{style=mystyle, frame=0}
\begin{lstlisting}[morekeywords={decl, output, symbol, number, magic}, basicstyle=\scriptsize\ttfamily\bfseries]
.decl a(A:number)
.output a
a(1). a(2).
a(A1)<=a(A2):-A1<A2.
\end{lstlisting}
\end{minipage} 
&  
\begin{minipage}[c]{.1\linewidth}
\lstset{style=mystyle, frame=0}
\begin{lstlisting}[morekeywords={decl, output, symbol, number, magic}, basicstyle=\scriptsize\ttfamily\bfseries]
a(2).
\end{lstlisting}
\end{minipage} 
&
\begin{minipage}[c]{.3\linewidth}
\lstset{style=mystyle, frame=0}
\begin{lstlisting}[morekeywords={decl, output, symbol, number, magic}, basicstyle=\scriptsize\ttfamily\bfseries]
.decl a(A:number)
.output a
a(1). a(2).
a(A1)<=a(A2):-A1<A2.
\end{lstlisting}
\end{minipage} 
&
\begin{minipage}[c]{.1\linewidth}
\lstset{style=mystyle, frame=0}
\begin{lstlisting}[morekeywords={decl, output, symbol, number, magic}, basicstyle=\scriptsize\ttfamily\bfseries]
a(2).
\end{lstlisting}
\end{minipage} 
\\
1
&
\begin{minipage}[c]{.27\linewidth}
\lstset{style=mystyle, frame=0}
\begin{lstlisting}[morekeywords={decl, output, symbol, number, magic}, basicstyle=\scriptsize\ttfamily\bfseries]
.decl a(A:number)
.output a
a(2).
a(A1)<=a(A2):-A1>A2.
\end{lstlisting}
\end{minipage} 
&  
\begin{minipage}[c]{.13\linewidth}
\lstset{style=mystyle, frame=0}
\begin{lstlisting}[morekeywords={decl, output, symbol, number, magic}, basicstyle=\scriptsize\ttfamily\bfseries]
a(2).
\end{lstlisting}
\end{minipage} 
&
\begin{minipage}[c]{.27\linewidth}
\lstset{style=mystyle, frame=0}
\begin{lstlisting}[morekeywords={decl, output, symbol, number, magic}, basicstyle=\scriptsize\ttfamily\bfseries]
.decl a(A:number)
.output a
a(1). a(2).
a(A1)<=a(A2):-A1<A2.
a(A1)<=a(A2):-A1>A2.
\end{lstlisting}
\end{minipage} 
&
\begin{minipage}[c]{.13\linewidth}
\lstset{style=mystyle, frame=0}
\begin{lstlisting}[morekeywords={decl, output, symbol, number, magic}, basicstyle=\scriptsize\ttfamily\bfseries]
-
\end{lstlisting}
\end{minipage} 
\\\bottomrule
    \end{tabular}
\end{table}

\clearpage
\subsection{Bug 10}

Bug 10 was a bug in µZ. The developer fixed this bug after we report it, but did not give us any feedback about this bug. This was an interesting bug that all the four rules in this test case did not depend on each other, but each of the first three rules could have an effect on the results of this bug-inducing rule. Also, both of the results of test oracle and test case were wrong.

\begin{table}[H]\footnotesize
    \centering
    \caption{Bug 10 found by \tool.}
    \begin{tabular}{c l l l l} \toprule
         iteration & $P^{ref}_{n}$ & $Facts^{ref}_{output\_rel}$ & $P^{opt}_{n}$ & $Facts^{opt}_{output\_rel}$ \\\midrule
0
&
\begin{minipage}[c]{.3\linewidth}
\lstset{style=mystyle, frame=0}
\begin{lstlisting}[morekeywords={input, printtuples}, basicstyle=\scriptsize\ttfamily\bfseries]
Z 128
mxsr(A:Z) input
ebbj(A:Z) printtuples
mxsr(1).
ebbj(A) :- mxsr(A), 43 != A.
\end{lstlisting}
\end{minipage} 
&  
\begin{minipage}[c]{.1\linewidth}
\lstset{style=mystyle, frame=0}
\begin{lstlisting}[morekeywords={input, printtuples}, basicstyle=\scriptsize\ttfamily\bfseries]
ebbj(1).
\end{lstlisting}
\end{minipage} 
&
\begin{minipage}[c]{.3\linewidth}
\lstset{style=mystyle, frame=0}
\begin{lstlisting}[morekeywords={input, printtuples}, basicstyle=\scriptsize\ttfamily\bfseries]
Z 128
mxsr(A:Z) input
qjfp(A:Z) input
jrkr(A:Z, B:Z) input
rtkv(A:Z) input
ebbj(A:Z) printtuples
jrkr(80, 80). jrkr(29, 29).
jrkr(4, 4). mxsr(1).
qjfp(1). rtkv(1).
ebbj(A) :- mxsr(A), 43 != A.
\end{lstlisting}
\end{minipage} 
&
\begin{minipage}[c]{.1\linewidth}
\lstset{style=mystyle, frame=0}
\begin{lstlisting}[morekeywords={input, printtuples}, basicstyle=\scriptsize\ttfamily\bfseries]
ebbj(1).
\end{lstlisting}
\end{minipage} 
\\
1
&
\begin{minipage}[c]{.27\linewidth}
\lstset{style=mystyle, frame=0}
\begin{lstlisting}[morekeywords={input, printtuples}, basicstyle=\scriptsize\ttfamily\bfseries]
Z 128
qjfp(A:Z) input
oxyx(A:Z) printtuples
qjfp(1).
oxyx(C) :- qjfp(C), 76 != C.
\end{lstlisting}
\end{minipage} 
&  
\begin{minipage}[c]{.13\linewidth}
\lstset{style=mystyle, frame=0}
\begin{lstlisting}[morekeywords={input, printtuples}, basicstyle=\scriptsize\ttfamily\bfseries]
oxyx(1).
\end{lstlisting}
\end{minipage} 
&
\begin{minipage}[c]{.27\linewidth}
\lstset{style=mystyle, frame=0}
\begin{lstlisting}[morekeywords={input, printtuples}, basicstyle=\scriptsize\ttfamily\bfseries]
Z 128
mxsr(A:Z) input
qjfp(A:Z) input
jrkr(A:Z, B:Z) input
rtkv(A:Z) input
ebbj(A:Z)
oxyx(A:Z) printtuples
jrkr(80, 80). jrkr(29, 29).
jrkr(4, 4). mxsr(1).
qjfp(1). rtkv(1).
ebbj(A) :- mxsr(A), 43 != A.
oxyx(C) :- qjfp(C), 76 != C.
\end{lstlisting}
\end{minipage} 
&
\begin{minipage}[c]{.13\linewidth}
\lstset{style=mystyle, frame=0}
\begin{lstlisting}[morekeywords={input, printtuples}, basicstyle=\scriptsize\ttfamily\bfseries]
oxyx(1).
\end{lstlisting}
\end{minipage} 
\\
2
&
\begin{minipage}[c]{.3\linewidth}
\lstset{style=mystyle, frame=0}
\begin{lstlisting}[morekeywords={input, printtuples}, basicstyle=\scriptsize\ttfamily\bfseries]
Z 128
rtkv(A:Z) input
iypi(A:Z) printtuples
rtkv(1).
iypi(A) :- rtkv(A), 77 < A.
\end{lstlisting}
\end{minipage} 
&  
\begin{minipage}[c]{.1\linewidth}
\lstset{style=mystyle, frame=0}
\begin{lstlisting}[morekeywords={input, printtuples}, basicstyle=\scriptsize\ttfamily\bfseries]
iypi(1).
\end{lstlisting}
\end{minipage} 
&
\begin{minipage}[c]{.3\linewidth}
\lstset{style=mystyle, frame=0}
\begin{lstlisting}[morekeywords={input, printtuples}, basicstyle=\scriptsize\ttfamily\bfseries]
Z 128
mxsr(A:Z) input
qjfp(A:Z) input
jrkr(A:Z, B:Z) input
rtkv(A:Z) input
ebbj(A:Z)
oxyx(A:Z)
iypi(A:Z) printtuples
jrkr(80, 80). jrkr(29, 29).
jrkr(4, 4). mxsr(1).
qjfp(1). rtkv(1).
ebbj(A) :- mxsr(A), 43 != A.
oxyx(C) :- qjfp(C), 76 != C.
iypi(A) :- rtkv(A), 77 < A.
\end{lstlisting}
\end{minipage} 
&
\begin{minipage}[c]{.1\linewidth}
\lstset{style=mystyle, frame=0}
\begin{lstlisting}[morekeywords={input, printtuples}, basicstyle=\scriptsize\ttfamily\bfseries]
iypi(1).
\end{lstlisting}
\end{minipage} 
\\
3
&
\begin{minipage}[c]{.27\linewidth}
\lstset{style=mystyle, frame=0}
\begin{lstlisting}[morekeywords={input, printtuples}, basicstyle=\scriptsize\ttfamily\bfseries]
Z 128
jrkr(A:Z, B:Z) input
fvof(A:Z) printtuples
jrkr(80, 80).
jrkr(29, 29).
jrkr(4, 4).
fvof(E) :- jrkr(D, E), 8 != E, 71 < D.
\end{lstlisting}
\end{minipage} 
&  
\begin{minipage}[c]{.13\linewidth}
\lstset{style=mystyle, frame=0}
\begin{lstlisting}[morekeywords={input, printtuples}, basicstyle=\scriptsize\ttfamily\bfseries]
fvof(29).
fvof(80).
fvof(4).
\end{lstlisting}
\end{minipage} 
&
\begin{minipage}[c]{.27\linewidth}
\lstset{style=mystyle, frame=0}
\begin{lstlisting}[morekeywords={input, printtuples}, basicstyle=\scriptsize\ttfamily\bfseries]
Z 128
mxsr(A:Z) input
qjfp(A:Z) input
jrkr(A:Z, B:Z) input
rtkv(A:Z) input
ebbj(A:Z)
oxyx(A:Z)
iypi(A:Z)
fvof(A:Z) printtuples
jrkr(80, 80). jrkr(29, 29).
jrkr(4, 4). mxsr(1).
qjfp(1). rtkv(1).
ebbj(A) :- mxsr(A), 43 != A.
oxyx(C) :- qjfp(C), 76 != C.
iypi(A) :- rtkv(A), 77 < A.
fvof(E) :- jrkr(D, E), 8 != E, 71 < D.
\end{lstlisting}
\end{minipage} 
&
\begin{minipage}[c]{.13\linewidth}
\lstset{style=mystyle, frame=0}
\begin{lstlisting}[morekeywords={input, printtuples}, basicstyle=\scriptsize\ttfamily\bfseries]
fvof(29).
fvof(80).
\end{lstlisting}
\end{minipage} 
\\\bottomrule
    \end{tabular}
\end{table}

\clearpage
\subsection{Bug 11}

Bug 11 was a bug in µZ. Similar with the bug 10, the rules in this test case did not depend on each other. The reference program in this example trigger the bug.

\begin{table}[H]\footnotesize
    \centering
    \caption{Bug 11 found by \tool.}
    \begin{tabular}{c l l l l} \toprule
         iteration & $P^{ref}_{n}$ & $Facts^{ref}_{output\_rel}$ & $P^{opt}_{n}$ & $Facts^{opt}_{output\_rel}$ \\\midrule
0
&
\begin{minipage}[c]{.3\linewidth}
\lstset{style=mystyle, frame=0}
\begin{lstlisting}[morekeywords={input, printtuples}, basicstyle=\scriptsize\ttfamily\bfseries]
Z 128
hmyo(A:Z) input
cdan(A:Z) printtuples
hmyo(70).
cdan(C) :- hmyo(C), 65 < C.
\end{lstlisting}
\end{minipage} 
&  
\begin{minipage}[c]{.1\linewidth}
\lstset{style=mystyle, frame=0}
\begin{lstlisting}[morekeywords={input, printtuples}, basicstyle=\scriptsize\ttfamily\bfseries]
cdan(70).
\end{lstlisting}
\end{minipage} 
&
\begin{minipage}[c]{.3\linewidth}
\lstset{style=mystyle, frame=0}
\begin{lstlisting}[morekeywords={input, printtuples}, basicstyle=\scriptsize\ttfamily\bfseries]
Z 128
szra(A:Z) input
hmyo(A:Z) input
cdan(A:Z) printtuples
szra(1).
hmyo(70).
cdan(C) :- hmyo(C), 65 < C.
\end{lstlisting}
\end{minipage} 
&
\begin{minipage}[c]{.1\linewidth}
\lstset{style=mystyle, frame=0}
\begin{lstlisting}[morekeywords={input, printtuples}, basicstyle=\scriptsize\ttfamily\bfseries]
cdan(70).
\end{lstlisting}
\end{minipage} 
\\
1
&
\begin{minipage}[c]{.27\linewidth}
\lstset{style=mystyle, frame=0}
\begin{lstlisting}[morekeywords={input, printtuples}, basicstyle=\scriptsize\ttfamily\bfseries]
Z 128
szra(A:Z) input
fbnd(A:Z) printtuples
szra(1).
fbnd(F) :- szra(F), 72 != F, 97 = F.
\end{lstlisting}
\end{minipage} 
&  
\begin{minipage}[c]{.13\linewidth}
\lstset{style=mystyle, frame=0}
\begin{lstlisting}[morekeywords={input, printtuples}, basicstyle=\scriptsize\ttfamily\bfseries]
fbnd(1).
\end{lstlisting}
\end{minipage} 
&
\begin{minipage}[c]{.27\linewidth}
\lstset{style=mystyle, frame=0}
\begin{lstlisting}[morekeywords={input, printtuples}, basicstyle=\scriptsize\ttfamily\bfseries]
Z 128
szra(A:Z) input
hmyo(A:Z) input
cdan(A:Z)
fbnd(A:Z) printtuples
szra(1).
hmyo(70).
cdan(C) :- hmyo(C), 65 < C.
fbnd(F) :- szra(F), 72 != F, 97 = F.
\end{lstlisting}
\end{minipage} 
&
\begin{minipage}[c]{.13\linewidth}
\lstset{style=mystyle, frame=0}
\begin{lstlisting}[morekeywords={input, printtuples}, basicstyle=\scriptsize\ttfamily\bfseries]
-
\end{lstlisting}
\end{minipage} 
\\\bottomrule
    \end{tabular}
\end{table}

\subsection{Bug 12}

Bug 12 was a bug in CozoDB, which was caused by an error in the implementation of semi-naïve algorithm.

\begin{table}[H]\footnotesize
    \centering
    \caption{Bug 12 found by \tool.}
    \begin{tabular}{c l l l l} \toprule
         iteration & $P^{ref}_{n}$ & $Facts^{ref}_{output\_rel}$ & $P^{opt}_{n}$ & $Facts^{opt}_{output\_rel}$ \\\midrule
0
&
\begin{minipage}[c]{.3\linewidth}
\lstset{style=mystyle, frame=0}
\begin{lstlisting}[morekeywords={input, printtuples}, basicstyle=\scriptsize\ttfamily\bfseries]
x[A] := A = 1
?[A] := x[A]
\end{lstlisting}
\end{minipage} 
&  
\begin{minipage}[c]{.1\linewidth}
\lstset{style=mystyle, frame=0}
\begin{lstlisting}[morekeywords={input, printtuples}, basicstyle=\scriptsize\ttfamily\bfseries]
[[1]]
\end{lstlisting}
\end{minipage} 
&
\begin{minipage}[c]{.3\linewidth}
\lstset{style=mystyle, frame=0}
\begin{lstlisting}[morekeywords={input, printtuples}, basicstyle=\scriptsize\ttfamily\bfseries]
x[A] := A = 1
?[A] := x[A]
\end{lstlisting}
\end{minipage} 
&
\begin{minipage}[c]{.1\linewidth}
\lstset{style=mystyle, frame=0}
\begin{lstlisting}[morekeywords={input, printtuples}, basicstyle=\scriptsize\ttfamily\bfseries]
[[1]]
\end{lstlisting}
\end{minipage} 
\\
1
&
\begin{minipage}[c]{.27\linewidth}
\lstset{style=mystyle, frame=0}
\begin{lstlisting}[morekeywords={input, printtuples}, basicstyle=\scriptsize\ttfamily\bfseries]
y[A, A] := A = 1
?[A, B] := y[A, B]
\end{lstlisting}
\end{minipage} 
&  
\begin{minipage}[c]{.13\linewidth}
\lstset{style=mystyle, frame=0}
\begin{lstlisting}[morekeywords={input, printtuples}, basicstyle=\scriptsize\ttfamily\bfseries]
[[1, 1]]
\end{lstlisting}
\end{minipage} 
&
\begin{minipage}[c]{.27\linewidth}
\lstset{style=mystyle, frame=0}
\begin{lstlisting}[morekeywords={input, printtuples}, basicstyle=\scriptsize\ttfamily\bfseries]
x[A] := A = 1
y[A, A] := A = 1
?[A, B] := y[A, B]
\end{lstlisting}
\end{minipage} 
&
\begin{minipage}[c]{.13\linewidth}
\lstset{style=mystyle, frame=0}
\begin{lstlisting}[morekeywords={input, printtuples}, basicstyle=\scriptsize\ttfamily\bfseries]
[[1, 1]]
\end{lstlisting}
\end{minipage} 
\\
2
&
\begin{minipage}[c]{.27\linewidth}
\lstset{style=mystyle, frame=0}
\begin{lstlisting}[morekeywords={input, printtuples}, basicstyle=\scriptsize\ttfamily\bfseries]
x[A] <- [[1]]
y[A, B] := A = 0, B = 1, x[B]
?[A, B] := y[A, B]
\end{lstlisting}
\end{minipage} 
&  
\begin{minipage}[c]{.13\linewidth}
\lstset{style=mystyle, frame=0}
\begin{lstlisting}[morekeywords={input, printtuples}, basicstyle=\scriptsize\ttfamily\bfseries]
[[0, 1]]
\end{lstlisting}
\end{minipage} 
&
\begin{minipage}[c]{.27\linewidth}
\lstset{style=mystyle, frame=0}
\begin{lstlisting}[morekeywords={input, printtuples}, basicstyle=\scriptsize\ttfamily\bfseries]
x[A] := A = 1
y[A, A] := A = 1
y[A, B] := A = 0, B = 1, x[B]
?[A, B] := y[A, B]
\end{lstlisting}
\end{minipage} 
&
\begin{minipage}[c]{.13\linewidth}
\lstset{style=mystyle, frame=0}
\begin{lstlisting}[morekeywords={input, printtuples}, basicstyle=\scriptsize\ttfamily\bfseries]
[[0, 1]]
\end{lstlisting}
\end{minipage} 
\\
3
&
\begin{minipage}[c]{.27\linewidth}
\lstset{style=mystyle, frame=0}
\begin{lstlisting}[morekeywords={input, printtuples}, basicstyle=\scriptsize\ttfamily\bfseries]
y[A, B] <- [[0, 1], [1, 1]]
?[C] := y[A, _], y[C, A]
\end{lstlisting}
\end{minipage} 
&  
\begin{minipage}[c]{.13\linewidth}
\lstset{style=mystyle, frame=0}
\begin{lstlisting}[morekeywords={input, printtuples}, basicstyle=\scriptsize\ttfamily\bfseries]
[[0], [1]]
\end{lstlisting}
\end{minipage} 
&
\begin{minipage}[c]{.27\linewidth}
\lstset{style=mystyle, frame=0}
\begin{lstlisting}[morekeywords={input, printtuples}, basicstyle=\scriptsize\ttfamily\bfseries]
x[A] := A = 1
y[A, A] := A = 1
y[A, B] := A = 0, B = 1, x[B]
?[C] := y[A, _], y[C, A]
\end{lstlisting}
\end{minipage} 
&
\begin{minipage}[c]{.13\linewidth}
\lstset{style=mystyle, frame=0}
\begin{lstlisting}[morekeywords={input, printtuples}, basicstyle=\scriptsize\ttfamily\bfseries]
[[1]]
\end{lstlisting}
\end{minipage} 
\\\bottomrule
    \end{tabular}
\end{table}

\clearpage
\subsection{Bug 13}
\label{sec:debug13}

Bug 13 was a bug in CozoDB, which was caused by the discrepancy between the comparison functions employed in the magic sets rewrite and the binary operators utilized in the rules. 
We anticipated the results to be \lstinline[style=lstinlinestyle]{[[0.0, null]]}, but CozoDB returned \lstinline[style=lstinlinestyle]{[[0.0, null], [-0.0, null]]}, where the second result violates the constraint \lstinline[style=lstinlinestyle]{A >= 0}.
During the magic sets rewrite, CozoDB employs range scanning on the relation \lstinline[style=lstinlinestyle]{a} instead of testing each individual value. However, the comparison function used in the range scan considers \lstinline[style=lstinlinestyle]{-0.0 >= 0} to be true, while the comparison function used in the binary operators considers it to be false.
This bug can only be identified when the program is evaluated twice, once with the magic sets rewrite and once without it.
We believe that queryFuzz would not be able to detect this bug because the transformations it applies are designed specifically for conjunctive queries, which consist of single non-recursive function-free Horn rules. 
We got feedback from the developer of CozoDB that their magic sets rewrite only applied on conjunctive queries. 
Therefore, the transformations supported by queryFuzz, such as adding an existing subgoal into the rule body, modifying a variable, or removing a subgoal from the rule body, are inadequate for preventing the magic sets rewrite in this particular scenario.

\begin{table}[H]\footnotesize
    \centering
    \caption{Bug 13 found by \tool.}
    \begin{tabular}{c l l l l} \toprule
         iteration & $P^{ref}_{n}$ & $Facts^{ref}_{output\_rel}$ & $P^{opt}_{n}$ & $Facts^{opt}_{output\_rel}$ \\\midrule
0
&
\begin{minipage}[c]{.3\linewidth}
\lstset{style=mystyle, frame=0}
\begin{lstlisting}[morekeywords={input, printtuples}, basicstyle=\scriptsize\ttfamily\bfseries]
phve[A, B] <- [[-0.0, null], [0.0, null]]
xukw[A, B] := phve[A, B], ge(A, 0)
?[A, B] := xukw[A, B]
\end{lstlisting}
\end{minipage} 
&  
\begin{minipage}[c]{.1\linewidth}
\lstset{style=mystyle, frame=0}
\begin{lstlisting}[morekeywords={input, printtuples}, basicstyle=\scriptsize\ttfamily\bfseries]
[[0.0, null]]
\end{lstlisting}
\end{minipage} 
&
\begin{minipage}[c]{.3\linewidth}
\lstset{style=mystyle, frame=0}
\begin{lstlisting}[morekeywords={input, printtuples}, basicstyle=\scriptsize\ttfamily\bfseries]
phve[A, B] <- [[-0.0, null], [0.0, null]]
xukw[A, B] := phve[A, B], ge(A, 0)
?[A, B] := xukw[A, B]
\end{lstlisting}
\end{minipage} 
&
\begin{minipage}[c]{.1\linewidth}
\lstset{style=mystyle, frame=0}
\begin{lstlisting}[morekeywords={input, printtuples}, basicstyle=\scriptsize\ttfamily\bfseries]
[[0.0, null]]
\end{lstlisting}
\end{minipage} 
\\
1
&
\begin{minipage}[c]{.27\linewidth}
\lstset{style=mystyle, frame=0}
\begin{lstlisting}[morekeywords={input, printtuples}, basicstyle=\scriptsize\ttfamily\bfseries]
phve[A, B] <- [[-0.0, null], [0.0, null]]
nwku[F] := phve[_, F]
?[A] := nwku[A]
\end{lstlisting}
\end{minipage} 
&  
\begin{minipage}[c]{.13\linewidth}
\lstset{style=mystyle, frame=0}
\begin{lstlisting}[morekeywords={input, printtuples}, basicstyle=\scriptsize\ttfamily\bfseries]
[[null]]
\end{lstlisting}
\end{minipage} 
&
\begin{minipage}[c]{.27\linewidth}
\lstset{style=mystyle, frame=0}
\begin{lstlisting}[morekeywords={input, printtuples}, basicstyle=\scriptsize\ttfamily\bfseries]
phve[A, B] <- [[-0.0, null], [0.0, null]]
xukw[A, B] := phve[A, B], ge(A, 0)
nwku[F] := phve[_, F]
?[A] := nwku[A]
\end{lstlisting}
\end{minipage} 
&
\begin{minipage}[c]{.13\linewidth}
\lstset{style=mystyle, frame=0}
\begin{lstlisting}[morekeywords={input, printtuples}, basicstyle=\scriptsize\ttfamily\bfseries]
[[null]]
\end{lstlisting}
\end{minipage} 
\\
2
&
\begin{minipage}[c]{.27\linewidth}
\lstset{style=mystyle, frame=0}
\begin{lstlisting}[morekeywords={input, printtuples}, basicstyle=\scriptsize\ttfamily\bfseries]
xukw[A, B] <- [[0.0, null]]
nwku[A] <- [[null]]
ssie[F, C] := nwku[C], xukw[F, C]
?[A, B] := ssie[A, B]
\end{lstlisting}
\end{minipage} 
&  
\begin{minipage}[c]{.13\linewidth}
\lstset{style=mystyle, frame=0}
\begin{lstlisting}[morekeywords={input, printtuples}, basicstyle=\scriptsize\ttfamily\bfseries]
[[0.0, null]]
\end{lstlisting}
\end{minipage} 
&
\begin{minipage}[c]{.27\linewidth}
\lstset{style=mystyle, frame=0}
\begin{lstlisting}[morekeywords={input, printtuples}, basicstyle=\scriptsize\ttfamily\bfseries]
phve[A, B] <- [[-0.0, null], [0.0, null]]
xukw[A, B] := phve[A, B], ge(A, 0)
nwku[F] := phve[_, F]
ssie[F, C] := nwku[C], xukw[F, C]
?[A, B] := ssie[A, B]
\end{lstlisting}
\end{minipage} 
&
\begin{minipage}[c]{.13\linewidth}
\lstset{style=mystyle, frame=0}
\begin{lstlisting}[morekeywords={input, printtuples}, basicstyle=\scriptsize\ttfamily\bfseries]
[[0.0, null], 
[-0.0, null]]
\end{lstlisting}
\end{minipage} 
\\\bottomrule
    \end{tabular}
\end{table}

\clearpage
\section{Q2}
In this section, we demonstrate the best-effort comparison based on manually inspecting and analyzing the bug-inducing test cases and the bugs. 
We first checkout to the version of Datalog engine which contains the bug to confirm that the bug can be reproduced. Then we emulated our process of generating test oracles on all the logic bugs found by queryFuzz. For the bugs found by \tool, we analyze the reasons why it might not be detected by queryFuzz.

\subsection{Bugs found by queryFuzz}
\subsubsection{Bug 1}

Through the description in the GitHub issue of this bug we found that the bug 1 was related to magic transformation, which was used to minimize the program. But queryFuzz provided too much inputs, we picked the minimum inputs that triggered the bug. We can see that in the $iteration_4$, the result of $P^{opt}_{4}$ was different from that of $P^{ref}_{4}$.

\begin{table}[H]\footnotesize
    \centering
    \caption{Bug 1 found by queryFuzz.}
    \begin{tabular}{c l l l l} \toprule
         iteration & $P^{ref}_{n}$ & $Facts^{ref}_{output\_rel}$ & $P^{opt}_{n}$ & $Facts^{opt}_{output\_rel}$ \\\midrule
0
&
\begin{minipage}[c]{.3\linewidth}
\lstset{style=mystyle, frame=0}
\begin{lstlisting}[morekeywords={decl, output, symbol, number, magic}, basicstyle=\scriptsize\ttfamily\bfseries]
.decl MZV(U:number, V:number)
.decl HqV(U:number)
.output HqV
MZV(9, 28). MZV(18, 18).
HqV(a) :- MZV(a,b).
\end{lstlisting}
\end{minipage} 
&  
\begin{minipage}[c]{.1\linewidth}
\lstset{style=mystyle, frame=0}
\begin{lstlisting}[morekeywords={decl, output, symbol, number, magic}, basicstyle=\scriptsize\ttfamily\bfseries]
HqV(9).
HqV(18).
\end{lstlisting}
\end{minipage} 
&
\begin{minipage}[c]{.3\linewidth}
\lstset{style=mystyle, frame=0}
\begin{lstlisting}[morekeywords={decl, output, symbol, number, magic}, basicstyle=\scriptsize\ttfamily\bfseries]
.decl MZV(U:number, V:number)
.decl HqV(U:number)
.output HqV
MZV(9, 28). MZV(18, 18).
HqV(a) :- MZV(a,b).
\end{lstlisting}
\end{minipage} 
&
\begin{minipage}[c]{.1\linewidth}
\lstset{style=mystyle, frame=0}
\begin{lstlisting}[morekeywords={decl, output, symbol, number, magic}, basicstyle=\scriptsize\ttfamily\bfseries]
HqV(9).
HqV(18).
\end{lstlisting}
\end{minipage} 
\\
1
&
\begin{minipage}[c]{.3\linewidth}
\lstset{style=mystyle, frame=0}
\begin{lstlisting}[morekeywords={decl, output, symbol, number, magic}, basicstyle=\scriptsize\ttfamily\bfseries]
.decl MZV(U:number, V:number)
.decl gQk(U:number)
.output gQk
MZV(9, 28). MZV(18, 18).
gQk(jW) :- MZV(jW,jW).
\end{lstlisting}
\end{minipage} 
&  
\begin{minipage}[c]{.1\linewidth}
\lstset{style=mystyle, frame=0}
\begin{lstlisting}[morekeywords={decl, output, symbol, number, magic}, basicstyle=\scriptsize\ttfamily\bfseries]
gQk(18).
\end{lstlisting}
\end{minipage} 
&
\begin{minipage}[c]{.3\linewidth}
\lstset{style=mystyle, frame=0}
\begin{lstlisting}[morekeywords={decl, output, symbol, number, magic}, basicstyle=\scriptsize\ttfamily\bfseries]
.decl MZV(U:number, V:number)
.decl HqV(U:number)
.decl gQk(U:number)
.output gQk
MZV(9, 28). MZV(18, 18).
HqV(a) :- MZV(a,b).
gQk(jW) :- MZV(jW,jW).
\end{lstlisting}
\end{minipage} 
&
\begin{minipage}[c]{.1\linewidth}
\lstset{style=mystyle, frame=0}
\begin{lstlisting}[morekeywords={decl, output, symbol, number, magic}, basicstyle=\scriptsize\ttfamily\bfseries]
gQk(18).
\end{lstlisting}
\end{minipage} 
\\
2
&
\begin{minipage}[c]{.3\linewidth}
\lstset{style=mystyle, frame=0}
\begin{lstlisting}[morekeywords={decl, output, symbol, number, magic}, basicstyle=\scriptsize\ttfamily\bfseries]
.decl MZV(U:number, V:number)
.decl HqV(U:number)
.decl gQk(U:number)
.decl QOq(U:number, V:number)
MZV(9, 28). MZV(18, 18).
HqV(9). HqV(18). gQk(18).
QOq(aS,GF) :- MZV(GF,GF), gQk(M), 
              HqV(aS), MZV(aS,M).
.output QOq
\end{lstlisting}
\end{minipage} 
&  
\begin{minipage}[c]{.1\linewidth}
\lstset{style=mystyle, frame=0}
\begin{lstlisting}[morekeywords={decl, output, symbol, number, magic}, basicstyle=\scriptsize\ttfamily\bfseries]
QOq(18, 18).
\end{lstlisting}
\end{minipage} 
&
\begin{minipage}[c]{.3\linewidth}
\lstset{style=mystyle, frame=0}
\begin{lstlisting}[morekeywords={decl, output, symbol, number, magic}, basicstyle=\scriptsize\ttfamily\bfseries]
.decl MZV(U:number, V:number)
.decl HqV(U:number)
.decl gQk(U:number)
.decl QOq(U:number, V:number)
.output QOq
MZV(9, 28). MZV(18, 18).
HqV(a) :- MZV(a,b).
gQk(jW) :- MZV(jW,jW).
QOq(aS,GF) :- MZV(GF,GF), gQk(M), 
              HqV(aS), MZV(aS,M).
\end{lstlisting}
\end{minipage} 
&
\begin{minipage}[c]{.1\linewidth}
\lstset{style=mystyle, frame=0}
\begin{lstlisting}[morekeywords={decl, output, symbol, number, magic}, basicstyle=\scriptsize\ttfamily\bfseries]
QOq(18, 18).
\end{lstlisting}
\end{minipage} 
\\
3
&
\begin{minipage}[c]{.3\linewidth}
\lstset{style=mystyle, frame=0}
\begin{lstlisting}[morekeywords={decl, output, symbol, number, magic}, basicstyle=\scriptsize\ttfamily\bfseries]
.decl gQk(U:number)
.decl QOq(U:number, V:number)
.decl RwL(U:number)
.output RwL
gQk(18). QOq(18, 18).
RwL(qr) :- QOq(u,qr), gQk(u), gQk(u).
\end{lstlisting}
\end{minipage} 
&  
\begin{minipage}[c]{.1\linewidth}
\lstset{style=mystyle, frame=0}
\begin{lstlisting}[morekeywords={decl, output, symbol, number, magic}, basicstyle=\scriptsize\ttfamily\bfseries]
RwL(18).
\end{lstlisting}
\end{minipage} 
&
\begin{minipage}[c]{.3\linewidth}
\lstset{style=mystyle, frame=0}
\begin{lstlisting}[morekeywords={decl, output, symbol, number, magic}, basicstyle=\scriptsize\ttfamily\bfseries]
.decl MZV(U:number, V:number)
.decl HqV(U:number)
.decl gQk(U:number)
.decl QOq(U:number, V:number)
.decl RwL(U:number)
.output RwL
MZV(9, 28). MZV(18, 18).
HqV(a) :- MZV(a,b).
gQk(jW) :- MZV(jW,jW).
QOq(aS,GF) :- MZV(GF,GF), gQk(M), 
              HqV(aS), MZV(aS,M).
RwL(qr) :- QOq(u,qr), gQk(u), gQk(u).
\end{lstlisting}
\end{minipage} 
&
\begin{minipage}[c]{.1\linewidth}
\lstset{style=mystyle, frame=0}
\begin{lstlisting}[morekeywords={decl, output, symbol, number, magic}, basicstyle=\scriptsize\ttfamily\bfseries]
RwL(18).
\end{lstlisting}
\end{minipage} 
\\
4
&
\begin{minipage}[c]{.3\linewidth}
\lstset{style=mystyle, frame=0}
\begin{lstlisting}[morekeywords={decl, output, symbol, number, magic}, basicstyle=\scriptsize\ttfamily\bfseries]
.decl MZV(U:number, V:number)
.decl gQk(U:number)
.decl RwL(U:number)
.decl out(U:number, V:number)
.output out
MZV(9, 28). MZV(18, 18).
gQk(18). RwL(18).
out(jB,ym) :- gQk(h), RwL(ym), MZV(h,jB).
\end{lstlisting}
\end{minipage} 
&  
\begin{minipage}[c]{.1\linewidth}
\lstset{style=mystyle, frame=0}
\begin{lstlisting}[morekeywords={decl, output, symbol, number, magic}, basicstyle=\scriptsize\ttfamily\bfseries]
out(18, 18).
\end{lstlisting}
\end{minipage} 
&
\begin{minipage}[c]{.3\linewidth}
\lstset{style=mystyle, frame=0}
\begin{lstlisting}[morekeywords={decl, output, symbol, number, magic}, basicstyle=\scriptsize\ttfamily\bfseries]
.decl MZV(U:number, V:number)
.decl HqV(U:number)
.decl gQk(U:number)
.decl QOq(U:number, V:number)
.decl RwL(U:number)
.decl out(U:number, V:number)
.output out
MZV(9, 28). MZV(18, 18).
HqV(a) :- MZV(a,b).
gQk(jW) :- MZV(jW,jW).
QOq(aS,GF) :- MZV(GF,GF), gQk(M), 
              HqV(aS), MZV(aS,M).
RwL(qr) :- QOq(u,qr), gQk(u), gQk(u).
out(jB,ym) :- gQk(h), RwL(ym), MZV(h,jB).
\end{lstlisting}
\end{minipage} 
&
\begin{minipage}[c]{.1\linewidth}
\lstset{style=mystyle, frame=0}
\begin{lstlisting}[morekeywords={decl, output, symbol, number, magic}, basicstyle=\scriptsize\ttfamily\bfseries]
out(18, 18).
out(28, 18).
\end{lstlisting}
\end{minipage} 
\\\bottomrule
    \end{tabular}
\end{table}

\clearpage
\subsubsection{Bug 2}

The bug 2 was caused by magic transformation, in the same time, the inputs of Datalog program must come from an input file. 
So we needed two options, which were the necessary conditions for triggering the bug. The first one was \lstinline[style=lstinlinestyle]{-F}, which means to specify the path where the input file is located, we also created an input file named \lstinline[style=lstinlinestyle]{IBf.facts} which only contains one entry \lstinline[style=lstinlinestyle]{1}. The second was \lstinline[style=lstinlinestyle]{--magic-transform=*}, which means to perform magic transformation on all the relations in the program. We added these two options to both the reference programs and the optimized programs to demonstrate the accuracy of our approach in providing test oracles, without relying on changing the execution options. We can see that in the $iteration_2$, $P^{opt}_{2}$ produced different results from that of $P^{ref}_{2}$.

\begin{table}[H]\footnotesize
    \centering
    \caption{Bug 2 found by queryFuzz.}
    \begin{tabular}{c l l l l} \toprule
         iteration & $P^{ref}_{n}$ & $Facts^{ref}_{output\_rel}$ & $P^{opt}_{n}$ & $Facts^{opt}_{output\_rel}$ \\\midrule
0
&
\begin{minipage}[c]{.3\linewidth}
\lstset{style=mystyle, frame=0}
\begin{lstlisting}[morekeywords={decl, output, symbol, number, magic}, basicstyle=\scriptsize\ttfamily\bfseries]
.decl IBf(U:unsigned)
.decl dfm(U:unsigned)
.output dfm
IBf(1).
dfm(a) :- IBf(a).
\end{lstlisting}
\end{minipage} 
&  
\begin{minipage}[c]{.1\linewidth}
\lstset{style=mystyle, frame=0}
\begin{lstlisting}[morekeywords={decl, output, symbol, number, magic}, basicstyle=\scriptsize\ttfamily\bfseries]
dfm(1).
\end{lstlisting}
\end{minipage} 
&
\begin{minipage}[c]{.3\linewidth}
\lstset{style=mystyle, frame=0}
\begin{lstlisting}[morekeywords={decl, output, symbol, number, magic, input}, basicstyle=\scriptsize\ttfamily\bfseries]
.decl IBf(U:unsigned)
.decl dfm(U:unsigned)
.input IBf
.output dfm
dfm(a) :- IBf(a).
\end{lstlisting}
\end{minipage} 
&
\begin{minipage}[c]{.1\linewidth}
\lstset{style=mystyle, frame=0}
\begin{lstlisting}[morekeywords={decl, output, symbol, number, magic}, basicstyle=\scriptsize\ttfamily\bfseries]
dfm(1).
\end{lstlisting}
\end{minipage} 
\\
1
&
\begin{minipage}[c]{.3\linewidth}
\lstset{style=mystyle, frame=0}
\begin{lstlisting}[morekeywords={decl, output, symbol, number, magic}, basicstyle=\scriptsize\ttfamily\bfseries]
.decl IBf(U:unsigned)
.decl dfm(U:unsigned)
.decl onG(U:unsigned, V:unsigned)
.output onG
IBf(1). dfm(1).
onG(a,b) :- IBf(a), dfm(b), IBf(a).
\end{lstlisting}
\end{minipage} 
&  
\begin{minipage}[c]{.1\linewidth}
\lstset{style=mystyle, frame=0}
\begin{lstlisting}[morekeywords={decl, output, symbol, number, magic}, basicstyle=\scriptsize\ttfamily\bfseries]
onG(1, 1).
\end{lstlisting}
\end{minipage} 
&
\begin{minipage}[c]{.3\linewidth}
\lstset{style=mystyle, frame=0}
\begin{lstlisting}[morekeywords={decl, output, symbol, number, magic, input}, basicstyle=\scriptsize\ttfamily\bfseries]
.decl IBf(U:unsigned)
.decl dfm(U:unsigned)
.decl onG(U:unsigned, V:unsigned)
.input IBf
.output onG
dfm(a) :- IBf(a).
onG(a,b) :- IBf(a), dfm(b), IBf(a).
\end{lstlisting}
\end{minipage} 
&
\begin{minipage}[c]{.1\linewidth}
\lstset{style=mystyle, frame=0}
\begin{lstlisting}[morekeywords={decl, output, symbol, number, magic}, basicstyle=\scriptsize\ttfamily\bfseries]
onG(1, 1).
\end{lstlisting}
\end{minipage} 
\\
2
&
\begin{minipage}[c]{.3\linewidth}
\lstset{style=mystyle, frame=0}
\begin{lstlisting}[morekeywords={decl, output, symbol, number, magic}, basicstyle=\scriptsize\ttfamily\bfseries]
.decl onG(U:unsigned, V:unsigned)
.decl YVz(U:unsigned)
.output YVz
onG(1, 1).
YVz(a) :- onG(a,a), onG(b,a).
\end{lstlisting}
\end{minipage} 
&  
\begin{minipage}[c]{.1\linewidth}
\lstset{style=mystyle, frame=0}
\begin{lstlisting}[morekeywords={decl, output, symbol, number, magic}, basicstyle=\scriptsize\ttfamily\bfseries]
YVz(1).
\end{lstlisting}
\end{minipage} 
&
\begin{minipage}[c]{.3\linewidth}
\lstset{style=mystyle, frame=0}
\begin{lstlisting}[morekeywords={decl, output, symbol, number, magic, input}, basicstyle=\scriptsize\ttfamily\bfseries]
.decl IBf(U:unsigned)
.decl dfm(U:unsigned)
.decl onG(U:unsigned, V:unsigned)
.decl YVz(U:unsigned)
.input IBf
.output YVz
dfm(a) :- IBf(a).
onG(a,b) :- IBf(a), dfm(b), IBf(a).
YVz(a) :- onG(a,a), onG(b,a).
\end{lstlisting}
\end{minipage} 
&
\begin{minipage}[c]{.1\linewidth}
\lstset{style=mystyle, frame=0}
\begin{lstlisting}[morekeywords={decl, output, symbol, number, magic}, basicstyle=\scriptsize\ttfamily\bfseries]
-
\end{lstlisting}
\end{minipage} 
\\\bottomrule
    \end{tabular}
\end{table}

\subsubsection{Bug 3}

The bug 3 was caused by the different behaviors of the Datalog engine on a corner case under an optimization. In first rule, there will be \lstinline[style=lstinlinestyle]{-nan} in the facts of \lstinline[style=lstinlinestyle]{A}, because there is a negative base (\ie \lstinline[style=lstinlinestyle]{-1.5}) and a floating point exponent (\ie \lstinline[style=lstinlinestyle]{1.1}). The bug occurred in the second rule. In the default execution, Souffl\'{e} thought \lstinline[style=lstinlinestyle]{-nan} equals to \lstinline[style=lstinlinestyle]{-nan}, so \lstinline[style=lstinlinestyle]{-nan} would appear in the \lstinline[style=lstinlinestyle]{B}'s facts. But when added the \lstinline[style=lstinlinestyle]{--disable-transformers=ResolveAliasesTransformer} option, Souffl\'{e} thought \lstinline[style=lstinlinestyle]{-nan} not equals to \lstinline[style=lstinlinestyle]{-nan}, \lstinline[style=lstinlinestyle]{-nan} would not appear in the \lstinline[style=lstinlinestyle]{B}'s facts. In Souffl\'{e}, we can not take \lstinline[style=lstinlinestyle]{-nan} and \lstinline[style=lstinlinestyle]{inf} as input, so we construct \lstinline[style=lstinlinestyle]{-nan} with $(-1)\string^{0.5}$, construct \lstinline[style=lstinlinestyle]{inf} with $1/0$. We added the \lstinline[style=lstinlinestyle]{--disable-transformers=ResolveAliasesTransformer} option for both reference programs and optimized programs and find this bug in two iterations.

\begin{table}[H]\footnotesize
    \centering
    \caption{Bug 3 found by queryFuzz.}
    \begin{tabular}{c l l l l} \toprule
         iteration & $P^{ref}_{n}$ & $Facts^{ref}_{output\_rel}$ & $P^{opt}_{n}$ & $Facts^{opt}_{output\_rel}$ \\\midrule
0
&
\begin{minipage}[c]{.3\linewidth}
\lstset{style=mystyle, frame=0}
\begin{lstlisting}[morekeywords={decl, output, symbol, number, magic}, basicstyle=\scriptsize\ttfamily\bfseries]
.decl fac(U:float, V:float)
.decl A(U:float)
.output A
fac(2.1, -1.5). fac(1.1, 1.1).
A(n^m) :- fac(v,n), fac(m,m).
\end{lstlisting}
\end{minipage} 
&  
\begin{minipage}[c]{.1\linewidth}
\lstset{style=mystyle, frame=0}
\begin{lstlisting}[morekeywords={decl, output, symbol, number, magic}, basicstyle=\scriptsize\ttfamily\bfseries]
A(-nan).
A(1.11053).
\end{lstlisting}
\end{minipage} 
&
\begin{minipage}[c]{.3\linewidth}
\lstset{style=mystyle, frame=0}
\begin{lstlisting}[morekeywords={decl, output, symbol, number, magic}, basicstyle=\scriptsize\ttfamily\bfseries]
.decl fac(U:float, V:float)
.decl A(U:float)
.output A
fac(2.1, -1.5). fac(1.1, 1.1).
A(n^m) :- fac(v,n), fac(m,m).
\end{lstlisting}
\end{minipage} 
&
\begin{minipage}[c]{.1\linewidth}
\lstset{style=mystyle, frame=0}
\begin{lstlisting}[morekeywords={decl, output, symbol, number, magic}, basicstyle=\scriptsize\ttfamily\bfseries]
A(-nan).
A(1.11053).
\end{lstlisting}
\end{minipage} 
\\
1
&
\begin{minipage}[c]{.3\linewidth}
\lstset{style=mystyle, frame=0}
\begin{lstlisting}[morekeywords={decl, output, symbol, number, magic}, basicstyle=\scriptsize\ttfamily\bfseries]
.decl A(U:float)
.decl B(U:float)
.output B
A((-1)^0.5). A(1.11053).
B(m/m) :- A(m), m = m.
\end{lstlisting}
\end{minipage} 
&  
\begin{minipage}[c]{.1\linewidth}
\lstset{style=mystyle, frame=0}
\begin{lstlisting}[morekeywords={decl, output, symbol, number, magic}, basicstyle=\scriptsize\ttfamily\bfseries]
B(-nan).
B(1).
\end{lstlisting}
\end{minipage} 
&
\begin{minipage}[c]{.3\linewidth}
\lstset{style=mystyle, frame=0}
\begin{lstlisting}[morekeywords={decl, output, symbol, number, magic}, basicstyle=\scriptsize\ttfamily\bfseries]
.decl fac(U:float, V:float)
.decl A(U:float)
.decl B(U:float)
.output B
fac(2.1, -1.5). fac(1.1, 1.1).
A(n^m) :- fac(v,n), fac(m,m).
B(m/m) :- A(m), m = m.
\end{lstlisting}
\end{minipage} 
&
\begin{minipage}[c]{.1\linewidth}
\lstset{style=mystyle, frame=0}
\begin{lstlisting}[morekeywords={decl, output, symbol, number, magic}, basicstyle=\scriptsize\ttfamily\bfseries]
B(1).
\end{lstlisting}
\end{minipage} 
\\\bottomrule
    \end{tabular}
\end{table}

\clearpage
\subsubsection{Bug 4}

The bug 4 was caused by the Souffl\'{e} because it missed transforming some operator (\ie float-operator) in magic transformation. In this program, there is a relation \lstinline[style=lstinlinestyle]{Fail} with no attributes. This is a language feature in Souffl\'{e}, and it is usually used as a boolean variable. When it equals to \texttt{true}, it takes ``()'' as results, and it has empty results when it equals to \texttt{false}. We do not ignore the empty result of this type of relations. We cannot directly construct inputs for this type of relation. In our implementation, we take \lstinline[style=lstinlinestyle]{Fail():-1=1.} as its input for \texttt{true} value, and don't need inputs for \texttt{false} value. We show the process that detect this bug with our test oracle generation approach, \ie evaluate the rules one-by-one. In order to generate this test case, we set a probability for \tool so that it allowed the rule with empty results.

\begin{table}[H]\footnotesize
    \centering
    \caption{Bug 4 found by queryFuzz.}
    \begin{tabular}{c l l l l} \toprule
         iteration & $P^{ref}_{n}$ & $Facts^{ref}_{output\_rel}$ & $P^{opt}_{n}$ & $Facts^{opt}_{output\_rel}$ \\\midrule
0
&
\begin{minipage}[c]{.3\linewidth}
\lstset{style=mystyle, frame=0}
\begin{lstlisting}[morekeywords={decl, output, symbol, number, magic}, basicstyle=\scriptsize\ttfamily\bfseries]
.decl Pairs(A:float, B:float)
.decl First(A:float)
.output First
Pairs(0,0).
First(x) :- Pairs(x,_).
\end{lstlisting}
\end{minipage} 
&  
\begin{minipage}[c]{.1\linewidth}
\lstset{style=mystyle, frame=0}
\begin{lstlisting}[morekeywords={decl, output, symbol, number, magic}, basicstyle=\scriptsize\ttfamily\bfseries]
First(0).
\end{lstlisting}
\end{minipage} 
&
\begin{minipage}[c]{.3\linewidth}
\lstset{style=mystyle, frame=0}
\begin{lstlisting}[morekeywords={decl, output, symbol, number, magic}, basicstyle=\scriptsize\ttfamily\bfseries]
.decl Pairs(A:float, B:float)
.decl First(A:float)
.output First
Pairs(0,0).
First(x) :- Pairs(x,_).
\end{lstlisting}
\end{minipage} 
&
\begin{minipage}[c]{.1\linewidth}
\lstset{style=mystyle, frame=0}
\begin{lstlisting}[morekeywords={decl, output, symbol, number, magic}, basicstyle=\scriptsize\ttfamily\bfseries]
First(0).
\end{lstlisting}
\end{minipage} 
\\
1
&
\begin{minipage}[c]{.3\linewidth}
\lstset{style=mystyle, frame=0}
\begin{lstlisting}[morekeywords={decl, output, symbol, number, magic}, basicstyle=\scriptsize\ttfamily\bfseries]
.decl First(A:float)
.decl DupFirst(A:float, B:float)
.output DupFirst
First(0).
DupFirst(x,x) :- First(x), x < 100.
\end{lstlisting}
\end{minipage} 
&  
\begin{minipage}[c]{.1\linewidth}
\lstset{style=mystyle, frame=0}
\begin{lstlisting}[morekeywords={decl, output, symbol, number, magic}, basicstyle=\scriptsize\ttfamily\bfseries]
DupFirst(0, 0).
\end{lstlisting}
\end{minipage} 
&
\begin{minipage}[c]{.3\linewidth}
\lstset{style=mystyle, frame=0}
\begin{lstlisting}[morekeywords={decl, output, symbol, number, magic}, basicstyle=\scriptsize\ttfamily\bfseries]
.decl Pairs(A:float, B:float)
.decl First(A:float)
.decl DupFirst(A:float, B:float)
.output DupFirst
Pairs(0,0).
First(x) :- Pairs(x,_).
DupFirst(x,x) :- First(x), x < 100.
\end{lstlisting}
\end{minipage} 
&
\begin{minipage}[c]{.1\linewidth}
\lstset{style=mystyle, frame=0}
\begin{lstlisting}[morekeywords={decl, output, symbol, number, magic}, basicstyle=\scriptsize\ttfamily\bfseries]
DupFirst(0, 0).
\end{lstlisting}
\end{minipage} 
\\
2
&
\begin{minipage}[c]{.3\linewidth}
\lstset{style=mystyle, frame=0}
\begin{lstlisting}[morekeywords={decl, output, symbol, number, magic}, basicstyle=\scriptsize\ttfamily\bfseries]
.decl DupFirst(A:float, B:float)
.decl FirstAgain(A:float)
.output FirstAgain
DupFirst(0, 0).
FirstAgain(x) :- DupFirst(x,_).
\end{lstlisting}
\end{minipage} 
&  
\begin{minipage}[c]{.1\linewidth}
\lstset{style=mystyle, frame=0}
\begin{lstlisting}[morekeywords={decl, output, symbol, number, magic}, basicstyle=\scriptsize\ttfamily\bfseries]
FirstAgain(0).
\end{lstlisting}
\end{minipage} 
&
\begin{minipage}[c]{.3\linewidth}
\lstset{style=mystyle, frame=0}
\begin{lstlisting}[morekeywords={decl, output, symbol, number, magic}, basicstyle=\scriptsize\ttfamily\bfseries]
.decl Pairs(A:float, B:float)
.decl First(A:float)
.decl DupFirst(A:float, B:float)
.decl FirstAgain(A:float)
.output FirstAgain
Pairs(0,0).
First(x) :- Pairs(x,_).
DupFirst(x,x) :- First(x), x < 100.
FirstAgain(x) :- DupFirst(x,_).
\end{lstlisting}
\end{minipage} 
&
\begin{minipage}[c]{.1\linewidth}
\lstset{style=mystyle, frame=0}
\begin{lstlisting}[morekeywords={decl, output, symbol, number, magic}, basicstyle=\scriptsize\ttfamily\bfseries]
FirstAgain(0).
\end{lstlisting}
\end{minipage} 
\\
3
&
\begin{minipage}[c]{.3\linewidth}
\lstset{style=mystyle, frame=0}
\begin{lstlisting}[morekeywords={decl, output, symbol, number, magic}, basicstyle=\scriptsize\ttfamily\bfseries]
.decl FirstAgain(A:float)
.decl Fail()
.output Fail
FirstAgain(0).
Fail() :- FirstAgain(x), !FirstAgain(x).
\end{lstlisting}
\end{minipage} 
&  
\begin{minipage}[c]{.1\linewidth}
\lstset{style=mystyle, frame=0}
\begin{lstlisting}[morekeywords={decl, output, symbol, number, magic}, basicstyle=\scriptsize\ttfamily\bfseries]
Fail(False).
\end{lstlisting}
\end{minipage} 
&
\begin{minipage}[c]{.3\linewidth}
\lstset{style=mystyle, frame=0}
\begin{lstlisting}[morekeywords={decl, output, symbol, number, magic}, basicstyle=\scriptsize\ttfamily\bfseries]
.decl Pairs(A:float, B:float)
.decl First(A:float)
.decl DupFirst(A:float, B:float)
.decl FirstAgain(A:float)
.decl Fail()
.output Fail
Pairs(0,0).
First(x) :- Pairs(x,_).
DupFirst(x,x) :- First(x), x < 100.
FirstAgain(x) :- DupFirst(x,_).
Fail() :- FirstAgain(x), !FirstAgain(x).
\end{lstlisting}
\end{minipage} 
&
\begin{minipage}[c]{.1\linewidth}
\lstset{style=mystyle, frame=0}
\begin{lstlisting}[morekeywords={decl, output, symbol, number, magic}, basicstyle=\scriptsize\ttfamily\bfseries]
Fail(False).
\end{lstlisting}
\end{minipage} 
\\
4
&
\begin{minipage}[c]{.3\linewidth}
\lstset{style=mystyle, frame=0}
\begin{lstlisting}[morekeywords={decl, output, symbol, number, magic}, basicstyle=\scriptsize\ttfamily\bfseries]
.decl First(A:float)
.decl Fail()
.decl Out(A:float)
.output Out
First(0).
Out(x) :- Fail(), First(x).
\end{lstlisting}
\end{minipage} 
&  
\begin{minipage}[c]{.1\linewidth}
\lstset{style=mystyle, frame=0}
\begin{lstlisting}[morekeywords={decl, output, symbol, number, magic}, basicstyle=\scriptsize\ttfamily\bfseries]
-
\end{lstlisting}
\end{minipage} 
&
\begin{minipage}[c]{.3\linewidth}
\lstset{style=mystyle, frame=0}
\begin{lstlisting}[morekeywords={decl, output, symbol, number, magic}, basicstyle=\scriptsize\ttfamily\bfseries]
.decl Pairs(A:float, B:float)
.decl First(A:float)
.decl DupFirst(A:float, B:float)
.decl FirstAgain(A:float)
.decl Fail()
.decl Out(A:float)
.output Out
Pairs(0,0).
First(x) :- Pairs(x,_).
DupFirst(x,x) :- First(x), x < 100.
FirstAgain(x) :- DupFirst(x,_).
Fail() :- FirstAgain(x), !FirstAgain(x).
Out(x) :- Fail(), First(x).
\end{lstlisting}
\end{minipage} 
&
\begin{minipage}[c]{.1\linewidth}
\lstset{style=mystyle, frame=0}
\begin{lstlisting}[morekeywords={decl, output, symbol, number, magic}, basicstyle=\scriptsize\ttfamily\bfseries]
Out(0).
\end{lstlisting}
\end{minipage} 
\\\bottomrule
    \end{tabular}
\end{table}

\clearpage
\subsubsection{Bug 5}
\label{sec:qb5}
The bug 5 was related to the \lstinline[style=lstinlinestyle]{inline} keyword. Here, we were unable to add inline optimization to the reference program because Souffl\'{e} does not allow adding inline optimization for input and output relations.

\begin{table}[H]\footnotesize
    \centering
    \caption{Bug 5 found by queryFuzz.}
    \begin{tabular}{c l l l l} \toprule
         iteration & $P^{ref}_{n}$ & $Facts^{ref}_{output\_rel}$ & $P^{opt}_{n}$ & $Facts^{opt}_{output\_rel}$ \\\midrule
0
&
\begin{minipage}[c]{.3\linewidth}
\lstset{style=mystyle, frame=0}
\begin{lstlisting}[morekeywords={decl, output, symbol, number, magic}, basicstyle=\scriptsize\ttfamily\bfseries]
.decl A(x:float)
.output A
A(0.0).
A(x+1)  :- A(x), x=-0.0.
\end{lstlisting}
\end{minipage} 
&  
\begin{minipage}[c]{.1\linewidth}
\lstset{style=mystyle, frame=0}
\begin{lstlisting}[morekeywords={decl, output, symbol, number, magic}, basicstyle=\scriptsize\ttfamily\bfseries]
A(0).
A(1).
\end{lstlisting}
\end{minipage} 
&
\begin{minipage}[c]{.3\linewidth}
\lstset{style=mystyle, frame=0}
\begin{lstlisting}[morekeywords={decl, output, symbol, number, magic}, basicstyle=\scriptsize\ttfamily\bfseries]
.decl A(x:float)
.output A
A(0.0).
A(x+1)  :- A(x), x=-0.0.
\end{lstlisting}
\end{minipage} 
&
\begin{minipage}[c]{.1\linewidth}
\lstset{style=mystyle, frame=0}
\begin{lstlisting}[morekeywords={decl, output, symbol, number, magic}, basicstyle=\scriptsize\ttfamily\bfseries]
A(0).
A(1).
\end{lstlisting}
\end{minipage} 
\\
1
&
\begin{minipage}[c]{.3\linewidth}
\lstset{style=mystyle, frame=0}
\begin{lstlisting}[morekeywords={decl, output, symbol, number, magic}, basicstyle=\scriptsize\ttfamily\bfseries]
.decl A(x:float)
.decl B(A:float, B:float, C:float)
.output B
A(0). A(1).
B(e,e,max(-D,e)) :- A(D), A(e).
\end{lstlisting}
\end{minipage} 
&  
\begin{minipage}[c]{.1\linewidth}
\lstset{style=mystyle, frame=0}
\begin{lstlisting}[morekeywords={decl, output, symbol, number, magic}, basicstyle=\scriptsize\ttfamily\bfseries]
B(0, 0, -0).
B(0, 0, 0).
B(1, 1, 1).
\end{lstlisting}
\end{minipage} 
&
\begin{minipage}[c]{.3\linewidth}
\lstset{style=mystyle, frame=0}
\begin{lstlisting}[morekeywords={decl, output, symbol, number, magic}, basicstyle=\scriptsize\ttfamily\bfseries]
.decl A(x:float)
.decl B(A:float, B:float, C:float)
.output B
A(0.0).
A(x+1)  :- A(x), x=-0.0.
B(e,e,max(-D,e)) :- A(D), A(e).
\end{lstlisting}
\end{minipage} 
&
\begin{minipage}[c]{.1\linewidth}
\lstset{style=mystyle, frame=0}
\begin{lstlisting}[morekeywords={decl, output, symbol, number, magic}, basicstyle=\scriptsize\ttfamily\bfseries]
B(0, 0, -0).
B(0, 0, 0).
B(1, 1, 1).
\end{lstlisting}
\end{minipage} 
\\
2
&
\begin{minipage}[c]{.3\linewidth}
\lstset{style=mystyle, frame=0}
\begin{lstlisting}[morekeywords={decl, output, symbol, number, magic}, basicstyle=\scriptsize\ttfamily\bfseries]
.decl A(x:float)
.decl B(A:float, B:float, C:float)
.decl out(A:float)
.output out
A(0). A(1).
B(0, 0, -0). B(0, 0, 0). B(1, 1, 1).
out(R) :- A(R), B(a,b,a), B(a,R,R), B(R,R,R).
\end{lstlisting}
\end{minipage} 
&  
\begin{minipage}[c]{.1\linewidth}
\lstset{style=mystyle, frame=0}
\begin{lstlisting}[morekeywords={decl, output, symbol, number, magic}, basicstyle=\scriptsize\ttfamily\bfseries]
out(0).
out(1).
\end{lstlisting}
\end{minipage} 
&
\begin{minipage}[c]{.3\linewidth}
\lstset{style=mystyle, frame=0}
\begin{lstlisting}[morekeywords={decl, output, symbol, number, magic, inline}, basicstyle=\scriptsize\ttfamily\bfseries]
.decl A(x:float)
.decl B(A:float, B:float, C:float) inline
.decl out(A:float)
.output out
A(0.0).
A(x+1)  :- A(x), x=-0.0.
B(e,e,max(-D,e)) :- A(D), A(e).
out(R) :- A(R), B(a,b,a), B(a,R,R), B(R,R,R).
\end{lstlisting}
\end{minipage} 
&
\begin{minipage}[c]{.1\linewidth}
\lstset{style=mystyle, frame=0}
\begin{lstlisting}[morekeywords={decl, output, symbol, number, magic}, basicstyle=\scriptsize\ttfamily\bfseries]
out(-0).
out(0).
out(1).
\end{lstlisting}
\end{minipage} 
\\\bottomrule
    \end{tabular}
\end{table}

\clearpage
\subsubsection{Bug 6}

The bug 6 was caused by the data structure \lstinline[style=lstinlinestyle]{brie} and only occurred in synthesizer mode. \lstinline[style=lstinlinestyle]{brie} is also an optimization method. In this example, we evaluate the optimized program with the synthesizer mode of Souffl\'{e} and found this bug with our approach. To demonstrate that our test oracles generation method did mitigate the bugs caused by optimization, we evaluated the reference program with synthesizer mode and add \lstinline[style=lstinlinestyle]{brie} to $P^{ref}_{4}$, and the result showed that our reference program can be executed correctly.

\begin{table}[H]\footnotesize
    \centering
    \caption{Bug 6 found by queryFuzz.}
    \begin{tabular}{c l l l l} \toprule
         iteration & $P^{ref}_{n}$ & $Facts^{ref}_{output\_rel}$ & $P^{opt}_{n}$ & $Facts^{opt}_{output\_rel}$ \\\midrule
0
&
\begin{minipage}[c]{.3\linewidth}
\lstset{style=mystyle, frame=0}
\begin{lstlisting}[morekeywords={decl, output, symbol, number, magic}, basicstyle=\scriptsize\ttfamily\bfseries]
.decl BCCS(A:number, B:number)
.decl Bkft(A:number)
.output Bkft
BCCS(12, 4). BCCS(1, 15).
Bkft(max(--Q,-9)) :- BCCS(IAV,Q).
\end{lstlisting}
\end{minipage} 
&  
\begin{minipage}[c]{.1\linewidth}
\lstset{style=mystyle, frame=0}
\begin{lstlisting}[morekeywords={decl, output, symbol, number, magic}, basicstyle=\scriptsize\ttfamily\bfseries]
Bkft(4).
Bkft(15).
\end{lstlisting}
\end{minipage} 
&
\begin{minipage}[c]{.3\linewidth}
\lstset{style=mystyle, frame=0}
\begin{lstlisting}[morekeywords={decl, output, symbol, number, magic}, basicstyle=\scriptsize\ttfamily\bfseries]
.decl UteE(A:number, B:number)
.decl BCCS(A:number, B:number)
.decl Bkft(A:number)
.output Bkft
UteE(-6, 15). UteE(5, 4).	
UteE(-10, 4). UteE(-10, -29).
BCCS(12, 4). BCCS(1, 15).
Bkft(max(--Q,-9)) :- BCCS(IAV,Q).
\end{lstlisting}
\end{minipage} 
&
\begin{minipage}[c]{.1\linewidth}
\lstset{style=mystyle, frame=0}
\begin{lstlisting}[morekeywords={decl, output, symbol, number, magic}, basicstyle=\scriptsize\ttfamily\bfseries]
Bkft(4).
Bkft(15).
\end{lstlisting}
\end{minipage} 
\\
1
&
\begin{minipage}[c]{.3\linewidth}
\lstset{style=mystyle, frame=0}
\begin{lstlisting}[morekeywords={decl, output, symbol, number, magic}, basicstyle=\scriptsize\ttfamily\bfseries]
.decl BCCS(A:number, B:number)
.decl HXxe(A:number)
.output HXxe
BCCS(12, 4). BCCS(1, 15).
HXxe(S-5-9*Oiz) :- BCCS(S,Oiz).
\end{lstlisting}
\end{minipage} 
&  
\begin{minipage}[c]{.1\linewidth}
\lstset{style=mystyle, frame=0}
\begin{lstlisting}[morekeywords={decl, output, symbol, number, magic}, basicstyle=\scriptsize\ttfamily\bfseries]
HXxe(-139).
HXxe(-29).
\end{lstlisting}
\end{minipage} 
&
\begin{minipage}[c]{.3\linewidth}
\lstset{style=mystyle, frame=0}
\begin{lstlisting}[morekeywords={decl, output, symbol, number, magic}, basicstyle=\scriptsize\ttfamily\bfseries]
.decl UteE(A:number, B:number)
.decl BCCS(A:number, B:number)
.decl Bkft(A:number)
.decl HXxe(A:number)
.output HXxe
UteE(-6, 15). UteE(5, 4).	
UteE(-10, 4). UteE(-10, -29).
BCCS(12, 4). BCCS(1, 15).
Bkft(max(--Q,-9)) :- BCCS(IAV,Q).
HXxe(S-5-9*Oiz) :- BCCS(S,Oiz).
\end{lstlisting}
\end{minipage} 
&
\begin{minipage}[c]{.1\linewidth}
\lstset{style=mystyle, frame=0}
\begin{lstlisting}[morekeywords={decl, output, symbol, number, magic}, basicstyle=\scriptsize\ttfamily\bfseries]
HXxe(-139).
HXxe(-29).
\end{lstlisting}
\end{minipage} 
\\
2
&
\begin{minipage}[c]{.3\linewidth}
\lstset{style=mystyle, frame=0}
\begin{lstlisting}[morekeywords={decl, output, symbol, number, magic}, basicstyle=\scriptsize\ttfamily\bfseries]
.decl UteE(A:number, B:number)
.decl HXxe(A:number)
.decl Appy(A:number)
.output Appy
UteE(-6, 15). UteE(5, 4).	
UteE(-10, 4). UteE(-10, -29).
HXxe(-139). HXxe(-29).
Appy(D) :- HXxe(D), UteE(SNu,D).
\end{lstlisting}
\end{minipage} 
&  
\begin{minipage}[c]{.1\linewidth}
\lstset{style=mystyle, frame=0}
\begin{lstlisting}[morekeywords={decl, output, symbol, number, magic}, basicstyle=\scriptsize\ttfamily\bfseries]
Appy(-29).
\end{lstlisting}
\end{minipage} 
&
\begin{minipage}[c]{.3\linewidth}
\lstset{style=mystyle, frame=0}
\begin{lstlisting}[morekeywords={decl, output, symbol, number, magic}, basicstyle=\scriptsize\ttfamily\bfseries]
.decl UteE(A:number, B:number)
.decl BCCS(A:number, B:number)
.decl Bkft(A:number)
.decl HXxe(A:number)
.decl Appy(A:number)
.output Appy
UteE(-6, 15). UteE(5, 4).	
UteE(-10, 4). UteE(-10, -29).
BCCS(12, 4). BCCS(1, 15).
Bkft(max(--Q,-9)) :- BCCS(IAV,Q).
HXxe(S-5-9*Oiz) :- BCCS(S,Oiz).
Appy(D) :- HXxe(D), UteE(SNu,D).
\end{lstlisting}
\end{minipage} 
&
\begin{minipage}[c]{.1\linewidth}
\lstset{style=mystyle, frame=0}
\begin{lstlisting}[morekeywords={decl, output, symbol, number, magic}, basicstyle=\scriptsize\ttfamily\bfseries]
Appy(-29).
\end{lstlisting}
\end{minipage} 
\\
3
&
\begin{minipage}[c]{.3\linewidth}
\lstset{style=mystyle, frame=0}
\begin{lstlisting}[morekeywords={decl, output, symbol, number, magic}, basicstyle=\scriptsize\ttfamily\bfseries]
.decl BCCS(A:number, B:number)
.decl Bkft(A:number)
.decl Appy(A:number)
.output Bkft
BCCS(12, 4). BCCS(1, 15).
Appy(-29).
Bkft(Y) :- BCCS(rzG,Y), !Appy(Y).
\end{lstlisting}
\end{minipage} 
&  
\begin{minipage}[c]{.1\linewidth}
\lstset{style=mystyle, frame=0}
\begin{lstlisting}[morekeywords={decl, output, symbol, number, magic}, basicstyle=\scriptsize\ttfamily\bfseries]
Bkft(4).
Bkft(15).
\end{lstlisting}
\end{minipage} 
&
\begin{minipage}[c]{.3\linewidth}
\lstset{style=mystyle, frame=0}
\begin{lstlisting}[morekeywords={decl, output, symbol, number, magic}, basicstyle=\scriptsize\ttfamily\bfseries]
.decl UteE(A:number, B:number)
.decl BCCS(A:number, B:number)
.decl Bkft(A:number)
.decl HXxe(A:number)
.decl Appy(A:number)
.output Bkft
UteE(-6, 15). UteE(5, 4).	
UteE(-10, 4). UteE(-10, -29).
BCCS(12, 4). BCCS(1, 15).
Bkft(max(--Q,-9)) :- BCCS(IAV,Q).
HXxe(S-5-9*Oiz) :- BCCS(S,Oiz).
Appy(D) :- HXxe(D), UteE(SNu,D).
Bkft(Y) :- BCCS(rzG,Y), !Appy(Y).
\end{lstlisting}
\end{minipage} 
&
\begin{minipage}[c]{.1\linewidth}
\lstset{style=mystyle, frame=0}
\begin{lstlisting}[morekeywords={decl, output, symbol, number, magic}, basicstyle=\scriptsize\ttfamily\bfseries]
Bkft(4, 15).
\end{lstlisting}
\end{minipage} 
\\
4
&
\begin{minipage}[c]{.3\linewidth}
\lstset{style=mystyle, frame=0}
\begin{lstlisting}[morekeywords={decl, output, symbol, number, magic, brie}, basicstyle=\scriptsize\ttfamily\bfseries]
.decl UteE(A:number, B:number)
.decl Bkft(A:number)
.decl bNtR(A:number) brie
.output bNtR
UteE(-6, 15). UteE(5, 4).	
UteE(-10, 4). UteE(-10, -29).
Bkft(4). Bkft(15).
bNtR(J) :- UteE(J,VHk), Bkft(VHk).
\end{lstlisting}
\end{minipage} 
&  
\begin{minipage}[c]{.1\linewidth}
\lstset{style=mystyle, frame=0}
\begin{lstlisting}[morekeywords={decl, output, symbol, number, magic}, basicstyle=\scriptsize\ttfamily\bfseries]
bNtR(5).
bNtR(-6).
bNtR(-10).
\end{lstlisting}
\end{minipage} 
&
\begin{minipage}[c]{.3\linewidth}
\lstset{style=mystyle, frame=0}
\begin{lstlisting}[morekeywords={decl, output, symbol, number, magic, brie}, basicstyle=\scriptsize\ttfamily\bfseries]
.decl UteE(A:number, B:number)
.decl BCCS(A:number, B:number)
.decl Bkft(A:number)
.decl HXxe(A:number)
.decl Appy(A:number)
.decl bNtR(A:number) brie
.output bNtR
UteE(-6, 15). UteE(5, 4).	
UteE(-10, 4). UteE(-10, -29).
BCCS(12, 4). BCCS(1, 15).
Bkft(max(--Q,-9)) :- BCCS(IAV,Q).
HXxe(S-5-9*Oiz) :- BCCS(S,Oiz).
Appy(D) :- HXxe(D), UteE(SNu,D).
Bkft(Y) :- BCCS(rzG,Y), !Appy(Y).
bNtR(J) :- UteE(J,VHk), Bkft(VHk).
\end{lstlisting}
\end{minipage} 
&
\begin{minipage}[c]{.1\linewidth}
\lstset{style=mystyle, frame=0}
\begin{lstlisting}[morekeywords={decl, output, symbol, number, magic}, basicstyle=\scriptsize\ttfamily\bfseries]
bNtR(5).
bNtR(-10).
\end{lstlisting}
\end{minipage} 
\\\bottomrule
    \end{tabular}
\end{table}

\clearpage
\subsubsection{Bug 7}

Similar with the bug 6, the bug 7 was also caused by the data structure \lstinline[style=lstinlinestyle]{brie} and only occurred in synthesizer mode. It was challenging for us to generate test oracles for this test case because it only contained a single rule. To overcome this issue, we made an equivalent transformation to add a new rule without modifying its semantics, as shown in Figure~\ref{fig:qfbug7transfor}. Unlike the previous test cases, we were able to trigger this bug using the reference program, which allowed us to identify it.

\begin{figure}[H]
    \centering
    \begin{subfigure}{0.40\textwidth}
        \lstset{style=mystyle, numbers=left, framexleftmargin=7pt}
\begin{lstlisting}[morekeywords={decl, output, symbol, number, magic, brie, inline},basicstyle=\scriptsize\ttfamily\bfseries]
.decl wRnH(A:float, B:float)
.decl out(A:float) brie

.output out
wRnH(-21.360, -21.360). wRnH(-20.607, 2.486).
wRnH(2.486, 2.486). wRnH(-13.652, -13.652).
wRnH(-18.915, -18.915).

out(r) :- wRnH(a, r), wRnH(r, r).
\end{lstlisting}
        \vspace{-7pt}
        \caption{The original test case.}
        \vspace{-7pt}
        \label{fig:qfbug7old}
    \end{subfigure}
    \hspace{15pt}
    \begin{subfigure}{0.40\textwidth}
        \lstset{style=mystyle, numbers=left, framexleftmargin=7pt}
\begin{lstlisting}[morekeywords={decl, output, symbol, number, magic, brie, inline},basicstyle=\scriptsize\ttfamily\bfseries]
.decl wRnH(A:float, B:float)
.decl out(A:float) brie inline
.decl res(A:float)
.output res
wRnH(-21.360, -21.360). wRnH(-20.607, 2.486).
wRnH(2.486, 2.486). wRnH(-13.652, -13.652).
wRnH(-18.915, -18.915).
out(r) :- wRnH(a, r), wRnH(r, r).
res(a) :- out(a).
\end{lstlisting}
        \vspace{-7pt}
        \caption{The transformed test case.}
        \vspace{-7pt}
        \label{fig:qfbug7new}
    \end{subfigure}
    \caption{An equivalent transformation for bug 7.}
    \label{fig:qfbug7transfor}
\end{figure}

\begin{table}[H]\footnotesize
    \centering
    \caption{Bug 7 found by queryFuzz.}
    \begin{tabular}{c l l l l} \toprule
         iteration & $P^{ref}_{n}$ & $Facts^{ref}_{output\_rel}$ & $P^{opt}_{n}$ & $Facts^{opt}_{output\_rel}$ \\\midrule
0
&
\begin{minipage}[c]{.27\linewidth}
\lstset{style=mystyle, frame=0}
\begin{lstlisting}[morekeywords={decl, output, symbol, number, magic, brie, inline}, basicstyle=\scriptsize\ttfamily\bfseries]
.decl wRnH(A:float, B:float)
.decl out(A:float) brie
.output out
wRnH(-21.360, -21.360).
wRnH(-20.607, 2.486).
wRnH(2.486, 2.486).
wRnH(-13.652, -13.652).
wRnH(-18.915, -18.915).
out(r) :- wRnH(a,r), wRnH(r,r).
\end{lstlisting}
\end{minipage} 
&  
\begin{minipage}[c]{.13\linewidth}
\lstset{style=mystyle, frame=0}
\begin{lstlisting}[morekeywords={decl, output, symbol, number, magic}, basicstyle=\scriptsize\ttfamily\bfseries]
out(-18.9150009).
out(-13.6520004).
out(2.48600006).
\end{lstlisting}
\end{minipage} 
&
\begin{minipage}[c]{.27\linewidth}
\lstset{style=mystyle, frame=0}
\begin{lstlisting}[morekeywords={decl, output, symbol, number, magic, brie, inline}, basicstyle=\scriptsize\ttfamily\bfseries]
.decl wRnH(A:float, B:float)
.decl out(A:float) brie
.output out
wRnH(-21.360, -21.360).
wRnH(-20.607, 2.486).
wRnH(2.486, 2.486).
wRnH(-13.652, -13.652).
wRnH(-18.915, -18.915).
out(r) :- wRnH(a,r), wRnH(r,r).
\end{lstlisting}
\end{minipage} 
&
\begin{minipage}[c]{.13\linewidth}
\lstset{style=mystyle, frame=0}
\begin{lstlisting}[morekeywords={decl, output, symbol, number, magic}, basicstyle=\scriptsize\ttfamily\bfseries]
out(-18.9150009).
out(-13.6520004).
out(2.48600006).
\end{lstlisting}
\end{minipage} 
\\
1
&
\begin{minipage}[c]{.27\linewidth}
\lstset{style=mystyle, frame=0}
\begin{lstlisting}[morekeywords={decl, output, symbol, number, magic, brie, inline}, basicstyle=\scriptsize\ttfamily\bfseries]
.decl out(A:float) brie 
.decl res(A:float)
.output res
out(-18.9150009).
out(-13.6520004).
out(2.48600006).
res(a) :- out(a).
\end{lstlisting}
\end{minipage} 
&  
\begin{minipage}[c]{.13\linewidth}
\lstset{style=mystyle, frame=0}
\begin{lstlisting}[morekeywords={decl, output, symbol, number, magic}, basicstyle=\scriptsize\ttfamily\bfseries]
res(-18.9150009).
res(-13.6520004).
res(2.48600006).
\end{lstlisting}
\end{minipage} 
&
\begin{minipage}[c]{.27\linewidth}
\lstset{style=mystyle, frame=0}
\begin{lstlisting}[morekeywords={decl, output, symbol, number, magic, brie, inline}, basicstyle=\scriptsize\ttfamily\bfseries]
.decl wRnH(A:float, B:float)
.decl out(A:float) brie inline
.decl res(A:float)
.output res
wRnH(-21.360, -21.360).
wRnH(-20.607, 2.486).
wRnH(2.486, 2.486).
wRnH(-13.652, -13.652).
wRnH(-18.915, -18.915).
out(r) :- wRnH(a,r), wRnH(r,r).
res(a) :- out(a).
\end{lstlisting}
\end{minipage} 
&
\begin{minipage}[c]{.13\linewidth}
\lstset{style=mystyle, frame=0}
\begin{lstlisting}[morekeywords={decl, output, symbol, number, magic}, basicstyle=\scriptsize\ttfamily\bfseries]
res(-21.3600006).
res(-18.9150009).
res(-13.6520004).
res(2.48600006).
\end{lstlisting}
\end{minipage} 
\\\bottomrule
    \end{tabular}
\end{table}

\subsubsection{Bug 8}

Similar to bug 7, bug 8 also occurred within a single rule, but the difference was that the rule in bug 8 was not influenced by other rules. This bug is introduced by the optimization between two subgoals in a rule. It was difficult for us to generate test oracles for this bug because it did not align with our expectations, which were based on the assumption that the bug was introduced by the optimization between different rules. Therefore, it was not possible to detect this bug by using the same options in the reference program and the optimized program. However, we were able to identify it by removing the execution options in the reference program. This bug was only triggered when execute the program with the option \lstinline[style=lstinlinestyle]{--disable-transformers=ResolveAliasesTransformer}.

\begin{figure}[H]
    \centering
    \lstset{style=mystyle}
    \parbox{.45\linewidth}{
\begin{lstlisting}[morekeywords={decl, output, symbol, number, magic, eqrel},basicstyle=\scriptsize\ttfamily\bfseries]
.type Word <: symbol
.decl words (i:number, w:Word)
.decl out (w:Word, w1:Word)
.output out
words(0,"a").
words(1, "b").
out(w, w1)  :- words(i, w), words(i+1, w1).

// expected result:
// out("a", "b").
// buggy result:
// out("a", "a"). out("a", "b"). out("b", "a"). out("b", "b").
\end{lstlisting}}
    \vspace{-7pt}
    \caption{Bug 8 found by queryFuzz.}
    \label{fig:qfbug8}
\end{figure}

\subsubsection{Bug 9}

Bug 9 was a bug in µZ and also occurred in only one rule. We made an equivalent transformation for this program and were able to trigger the bug in the reference program.

\begin{figure}[H]
    \centering
    \begin{subfigure}{0.40\textwidth}
        \lstset{style=mystyle, numbers=left, framexleftmargin=7pt}
\begin{lstlisting}[morekeywords={decl, output, symbol, number, magic, brie, inline, input, printtuples},basicstyle=\scriptsize\ttfamily\bfseries]
in(A:Z) input
out(A:Z) printtuples

in(43). in(33).

out(a) :- in(a), in(a), in(a).
\end{lstlisting}
        \vspace{-7pt}
        \caption{The original test case.}
        \vspace{-7pt}
        \label{fig:qfbug9old}
    \end{subfigure}
    \hspace{15pt}
    \begin{subfigure}{0.40\textwidth}
        \lstset{style=mystyle, numbers=left, framexleftmargin=7pt}
\begin{lstlisting}[morekeywords={decl, output, symbol, number, magic, brie, input, printtuples, inline},basicstyle=\scriptsize\ttfamily\bfseries]
aicl(A:Z) input
in(A:Z)
out(A:Z) printtuples
aicl(43). aicl(33).
in(a) :- aicl(a).
out(a) :- in(a), in(a), in(a).
\end{lstlisting}
        \vspace{-7pt}
        \caption{The transformed test case.}
        \vspace{-7pt}
        \label{fig:qfbug9new}
    \end{subfigure}
    \caption{An equivalent transformation for bug 9.}
    \label{fig:qfbug9transfor}
\end{figure}

\begin{table}[H]\footnotesize
    \centering
    \caption{Bug 9 found by queryFuzz.}
    \begin{tabular}{c l l l l} \toprule
         iteration & $P^{ref}_{n}$ & $Facts^{ref}_{output\_rel}$ & $P^{opt}_{n}$ & $Facts^{opt}_{output\_rel}$ \\\midrule
0
&
\begin{minipage}[c]{.3\linewidth}
\lstset{style=mystyle, frame=0}
\begin{lstlisting}[morekeywords={input, printtuples}, basicstyle=\scriptsize\ttfamily\bfseries]
Z 64
aicl(A:Z) input
in(A:Z) printtuples
aicl(43). aicl(33).
in(a) :- aicl(a).
\end{lstlisting}
\end{minipage} 
&  
\begin{minipage}[c]{.1\linewidth}
\lstset{style=mystyle, frame=0}
\begin{lstlisting}[morekeywords={decl, output, symbol, number, magic}, basicstyle=\scriptsize\ttfamily\bfseries]
in(33).
in(43).
\end{lstlisting}
\end{minipage} 
&
\begin{minipage}[c]{.3\linewidth}
\lstset{style=mystyle, frame=0}
\begin{lstlisting}[morekeywords={input, printtuples}, basicstyle=\scriptsize\ttfamily\bfseries]
Z 64
aicl(A:Z) input
in(A:Z) printtuples
aicl(43). aicl(33).
in(a) :- aicl(a).
\end{lstlisting}
\end{minipage} 
&
\begin{minipage}[c]{.1\linewidth}
\lstset{style=mystyle, frame=0}
\begin{lstlisting}[morekeywords={decl, output, symbol, number, magic}, basicstyle=\scriptsize\ttfamily\bfseries]
in(33).
in(43).
\end{lstlisting}
\end{minipage} 
\\
1
&
\begin{minipage}[c]{.3\linewidth}
\lstset{style=mystyle, frame=0}
\begin{lstlisting}[morekeywords={input, printtuples}, basicstyle=\scriptsize\ttfamily\bfseries]
Z 64
in(A:Z) input
out(A:Z) printtuples
in(43). in(33).
out(a) :- in(a), in(a), in(a).
\end{lstlisting}
\end{minipage} 
&  
\begin{minipage}[c]{.1\linewidth}
\lstset{style=mystyle, frame=0}
\begin{lstlisting}[morekeywords={decl, output, symbol, number, magic}, basicstyle=\scriptsize\ttfamily\bfseries]
out(43).
out(33).
out(2).
...
out(63).
\end{lstlisting}
\end{minipage} 
&
\begin{minipage}[c]{.3\linewidth}
\lstset{style=mystyle, frame=0}
\begin{lstlisting}[morekeywords={input, printtuples}, basicstyle=\scriptsize\ttfamily\bfseries]
Z 64
aicl(A:Z) input
in(A:Z)
out(A:Z) printtuples
aicl(43). aicl(33).
in(a) :- aicl(a).
out(a) :- in(a), in(a), in(a).
\end{lstlisting}
\end{minipage} 
&
\begin{minipage}[c]{.1\linewidth}
\lstset{style=mystyle, frame=0}
\begin{lstlisting}[morekeywords={decl, output, symbol, number, magic}, basicstyle=\scriptsize\ttfamily\bfseries]
out(43).
out(33).
\end{lstlisting}
\end{minipage} 
\\\bottomrule
    \end{tabular}
\end{table}

\subsubsection{Bug 10}

Bug 10 was a bug in µZ, our method could generate test oracle for this bug.

\begin{table}[H]\footnotesize
    \centering
    \caption{Bug 10 found by queryFuzz.}
    \begin{tabular}{c l l l l} \toprule
         iteration & $P^{ref}_{n}$ & $Facts^{ref}_{output\_rel}$ & $P^{opt}_{n}$ & $Facts^{opt}_{output\_rel}$ \\\midrule
0
&
\begin{minipage}[c]{.3\linewidth}
\lstset{style=mystyle, frame=0}
\begin{lstlisting}[morekeywords={input, printtuples}, basicstyle=\scriptsize\ttfamily\bfseries]
Z 64
in1(A:Z) input
in2(A:Z, B:Z) input
R(A:Z, B:Z) printtuples
in1(49). in1(10). in2(25, 10). 
in2(16, 13). in2(24, 22).
R(V,M) :- in2(V,M), in1(M).
\end{lstlisting}
\end{minipage} 
&  
\begin{minipage}[c]{.1\linewidth}
\lstset{style=mystyle, frame=0}
\begin{lstlisting}[morekeywords={decl, output, symbol, number, magic}, basicstyle=\scriptsize\ttfamily\bfseries]
R(25, 10).
\end{lstlisting}
\end{minipage} 
&
\begin{minipage}[c]{.3\linewidth}
\lstset{style=mystyle, frame=0}
\begin{lstlisting}[morekeywords={input, printtuples}, basicstyle=\scriptsize\ttfamily\bfseries]
Z 64
in1(A:Z) input
in2(A:Z, B:Z) input
R(A:Z, B:Z) printtuples
in1(49). in1(10). in2(25, 10). 
in2(16, 13). in2(24, 22).
R(V,M) :- in2(V,M), in1(M).
\end{lstlisting}
\end{minipage} 
&
\begin{minipage}[c]{.1\linewidth}
\lstset{style=mystyle, frame=0}
\begin{lstlisting}[morekeywords={decl, output, symbol, number, magic}, basicstyle=\scriptsize\ttfamily\bfseries]
R(25, 10).
\end{lstlisting}
\end{minipage} 
\\
1
&
\begin{minipage}[c]{.3\linewidth}
\lstset{style=mystyle, frame=0}
\begin{lstlisting}[morekeywords={input, printtuples}, basicstyle=\scriptsize\ttfamily\bfseries]
Z 64
R(A:Z, B:Z) input
out(A:Z) printtuples
R(25,10).
out(f) :- R(f,c), R(f,a), R(f,b).
\end{lstlisting}
\end{minipage} 
&  
\begin{minipage}[c]{.1\linewidth}
\lstset{style=mystyle, frame=0}
\begin{lstlisting}[morekeywords={decl, output, symbol, number, magic}, basicstyle=\scriptsize\ttfamily\bfseries]
out(25).
\end{lstlisting}
\end{minipage} 
&
\begin{minipage}[c]{.3\linewidth}
\lstset{style=mystyle, frame=0}
\begin{lstlisting}[morekeywords={input, printtuples}, basicstyle=\scriptsize\ttfamily\bfseries]
Z 64
in1(A:Z) input
in2(A:Z, B:Z) input
R(A:Z, B:Z) 
out(A:Z) printtuples
in1(49). in1(10). in2(25, 10).
in2(16, 13). in2(24, 22).
R(V,M) :- in2(V,M), in1(M).
out(f) :- R(f,c), R(f,a), R(f,b).
\end{lstlisting}
\end{minipage} 
&
\begin{minipage}[c]{.1\linewidth}
\lstset{style=mystyle, frame=0}
\begin{lstlisting}[morekeywords={decl, output, symbol, number, magic}, basicstyle=\scriptsize\ttfamily\bfseries]
out(49).
out(10).
out(25).
...
out(63).
\end{lstlisting}
\end{minipage} 
\\\bottomrule
    \end{tabular}
\end{table}

\clearpage
\subsubsection{Bug 11}

Bug 11 was a bug in µZ and much similar with bug 10, our method could generate test oracle for this bug.

\begin{table}[H]\footnotesize
    \centering
    \caption{Bug 10 found by queryFuzz.}
    \begin{tabular}{c l l l l} \toprule
         iteration & $P^{ref}_{n}$ & $Facts^{ref}_{output\_rel}$ & $P^{opt}_{n}$ & $Facts^{opt}_{output\_rel}$ \\\midrule
0
&
\begin{minipage}[c]{.3\linewidth}
\lstset{style=mystyle, frame=0}
\begin{lstlisting}[morekeywords={input, printtuples}, basicstyle=\scriptsize\ttfamily\bfseries]
Z 64
in(A:Z, B:Z) input
cond(A:Z, B:Z) printtuples
in(45, 40).
cond(c,Z) :- in(c,Z), c > Z.
\end{lstlisting}
\end{minipage} 
&  
\begin{minipage}[c]{.1\linewidth}
\lstset{style=mystyle, frame=0}
\begin{lstlisting}[morekeywords={decl, output, symbol, number, magic}, basicstyle=\scriptsize\ttfamily\bfseries]
cond(45, 40).
\end{lstlisting}
\end{minipage} 
&
\begin{minipage}[c]{.3\linewidth}
\lstset{style=mystyle, frame=0}
\begin{lstlisting}[morekeywords={input, printtuples}, basicstyle=\scriptsize\ttfamily\bfseries]
Z 64
in(A:Z, B:Z) input
cond(A:Z, B:Z) printtuples
in(45, 40).
cond(c,Z) :- in(c,Z), c > Z.
\end{lstlisting}
\end{minipage} 
&
\begin{minipage}[c]{.1\linewidth}
\lstset{style=mystyle, frame=0}
\begin{lstlisting}[morekeywords={decl, output, symbol, number, magic}, basicstyle=\scriptsize\ttfamily\bfseries]
cond(45, 40).
\end{lstlisting}
\end{minipage} 
\\
1
&
\begin{minipage}[c]{.3\linewidth}
\lstset{style=mystyle, frame=0}
\begin{lstlisting}[morekeywords={input, printtuples}, basicstyle=\scriptsize\ttfamily\bfseries]
Z 64
in(A:Z, B:Z) input
cond(A:Z, B:Z) input
out(A:Z) printtuples
in(45, 40). cond(45, 40).
out(y) :- in(A,y), cond(B,y).
\end{lstlisting}
\end{minipage} 
&  
\begin{minipage}[c]{.1\linewidth}
\lstset{style=mystyle, frame=0}
\begin{lstlisting}[morekeywords={decl, output, symbol, number, magic}, basicstyle=\scriptsize\ttfamily\bfseries]
out(40).
\end{lstlisting}
\end{minipage} 
&
\begin{minipage}[c]{.3\linewidth}
\lstset{style=mystyle, frame=0}
\begin{lstlisting}[morekeywords={input, printtuples}, basicstyle=\scriptsize\ttfamily\bfseries]
Z 64
in(A:Z, B:Z) input
cond(A:Z, B:Z)
out(A:Z) printtuples
in(45, 40).
cond(c,Z) :- in(c,Z), c > Z.
out(y) :- in(A,y), cond(B,y).
\end{lstlisting}
\end{minipage} 
&
\begin{minipage}[c]{.1\linewidth}
\lstset{style=mystyle, frame=0}
\begin{lstlisting}[morekeywords={decl, output, symbol, number, magic}, basicstyle=\scriptsize\ttfamily\bfseries]
-
\end{lstlisting}
\end{minipage} 
\\\bottomrule
    \end{tabular}
\end{table}

\subsubsection{Bug 12}

Bug 12 was a bug in µZ, and simialr with bug 4, it depended on an empty relation (\ie \lstinline[style=lstinlinestyle]{y}). We could generate this test case and generate test oracles fot it. In this test case, the last rule of \texttt{out} gave the error result under the influence of unsafe negation in the second rule (\ie \lstinline[style=lstinlinestyle]{!WWW (_)}). Our test oracle generation method can block the propagation of this error.

\begin{table}[H]\footnotesize
    \centering
    \caption{Bug 12 found by queryFuzz.}
    \begin{tabular}{c l l l l} \toprule
         iteration & $P^{ref}_{n}$ & $Facts^{ref}_{output\_rel}$ & $P^{opt}_{n}$ & $Facts^{opt}_{output\_rel}$ \\\midrule
0
&
\begin{minipage}[c]{.3\linewidth}
\lstset{style=mystyle, frame=0}
\begin{lstlisting}[morekeywords={input, printtuples}, basicstyle=\scriptsize\ttfamily\bfseries]
Z 64
WWW(A:Z) input
mAqE(A:Z) input
p(A:Z) printtuples
WWW(18). WWW(15). WWW(25). WWW(16).
mAqE(29). mAqE(39).
p(g) :- mAqE(x), WWW(g), x < g.
\end{lstlisting}
\end{minipage} 
&  
\begin{minipage}[c]{.1\linewidth}
\lstset{style=mystyle, frame=0}
\begin{lstlisting}[morekeywords={decl, output, symbol, number, magic}, basicstyle=\scriptsize\ttfamily\bfseries]
-
\end{lstlisting}
\end{minipage} 
&
\begin{minipage}[c]{.3\linewidth}
\lstset{style=mystyle, frame=0}
\begin{lstlisting}[morekeywords={input, printtuples}, basicstyle=\scriptsize\ttfamily\bfseries]
Z 64
WWW(A:Z) input
mAqE(A:Z) input
p(A:Z) printtuples
WWW(18). WWW(15). WWW(25). WWW(16).
mAqE(29). mAqE(39).
p(g) :- mAqE(x), WWW(g), x < g.
\end{lstlisting}
\end{minipage} 
&
\begin{minipage}[c]{.1\linewidth}
\lstset{style=mystyle, frame=0}
\begin{lstlisting}[morekeywords={decl, output, symbol, number, magic}, basicstyle=\scriptsize\ttfamily\bfseries]
-
\end{lstlisting}
\end{minipage} 
\\
1
&
\begin{minipage}[c]{.3\linewidth}
\lstset{style=mystyle, frame=0}
\begin{lstlisting}[morekeywords={input, printtuples}, basicstyle=\scriptsize\ttfamily\bfseries]
Z 64
WWW(A:Z) input
mAqE(A:Z) input
p(A:Z) input
y(A:Z) printtuples
WWW(18). WWW(15). WWW(25). WWW(16). 
mAqE(29). mAqE(39).
y(d) :- mAqE(a), p(d), !WWW(_).
\end{lstlisting}
\end{minipage} 
&  
\begin{minipage}[c]{.1\linewidth}
\lstset{style=mystyle, frame=0}
\begin{lstlisting}[morekeywords={decl, output, symbol, number, magic}, basicstyle=\scriptsize\ttfamily\bfseries]
-
\end{lstlisting}
\end{minipage} 
&
\begin{minipage}[c]{.3\linewidth}
\lstset{style=mystyle, frame=0}
\begin{lstlisting}[morekeywords={input, printtuples}, basicstyle=\scriptsize\ttfamily\bfseries]
Z 64
WWW(A:Z) input
mAqE(A:Z) input
p(A:Z)
y(A:Z) printtuples
WWW(18). WWW(15). WWW(25). WWW(16).
mAqE(29). mAqE(39).
p(g) :- mAqE(x), WWW(g), x < g.
y(d) :- mAqE(a), p(d), !WWW(_).
\end{lstlisting}
\end{minipage} 
&
\begin{minipage}[c]{.1\linewidth}
\lstset{style=mystyle, frame=0}
\begin{lstlisting}[morekeywords={decl, output, symbol, number, magic}, basicstyle=\scriptsize\ttfamily\bfseries]
-
\end{lstlisting}
\end{minipage} 
\\
2
&
\begin{minipage}[c]{.3\linewidth}
\lstset{style=mystyle, frame=0}
\begin{lstlisting}[morekeywords={input, printtuples}, basicstyle=\scriptsize\ttfamily\bfseries]
Z 64
WWW(A:Z) input
y(A:Z) input
out(A:Z) printtuples
WWW(18). WWW(15). WWW(25). WWW(16).
out(c) :- y(a), WWW(b), WWW(c).
\end{lstlisting}
\end{minipage} 
&  
\begin{minipage}[c]{.1\linewidth}
\lstset{style=mystyle, frame=0}
\begin{lstlisting}[morekeywords={decl, output, symbol, number, magic}, basicstyle=\scriptsize\ttfamily\bfseries]
-
\end{lstlisting}
\end{minipage} 
&
\begin{minipage}[c]{.3\linewidth}
\lstset{style=mystyle, frame=0}
\begin{lstlisting}[morekeywords={input, printtuples}, basicstyle=\scriptsize\ttfamily\bfseries]
Z 64
WWW(A:Z) input
mAqE(A:Z) input
p(A:Z)
y(A:Z)
out(A:Z) printtuples
WWW(18). WWW(15). WWW(25). WWW(16).
mAqE(29). mAqE(39).
p(g) :- mAqE(x), WWW(g), x < g.
y(d) :- mAqE(a), p(d), !WWW(_).
out(c) :- y(a), WWW(b), WWW(c).
\end{lstlisting}
\end{minipage} 
&
\begin{minipage}[c]{.1\linewidth}
\lstset{style=mystyle, frame=0}
\begin{lstlisting}[morekeywords={decl, output, symbol, number, magic}, basicstyle=\scriptsize\ttfamily\bfseries]
out(18).
out(15).
out(25).
out(16).
\end{lstlisting}
\end{minipage} 
\\\bottomrule
    \end{tabular}
\end{table}

\clearpage
\subsubsection{Bug 13}

Bug 13 was a bug in DDlog,  this bug has not been fixed until now and we have detected this bug with our approach. We show the process of generating the oracle for this bug-inducing test case reported by queryFuzz.

\begin{table}[H]\footnotesize
    \centering
    \caption{Bug 13 found by queryFuzz.}
    \begin{tabular}{c l l l l} \toprule
         iteration & $P^{ref}_{n}$ & $Facts^{ref}_{output\_rel}$ & $P^{opt}_{n}$ & $Facts^{opt}_{output\_rel}$ \\\midrule
0
&
\begin{minipage}[c]{.27\linewidth}
\lstset{style=mystyle, frame=0}
\begin{lstlisting}[morekeywords={input, output}, basicstyle=\scriptsize\ttfamily\bfseries]
input relation Z(a:string)
output relation PQRI(a:string)
Z("3u37ezDdBg"). Z("uwsUc").
Z("zDdBgEbgMd"). Z("EbgMd2DV3K").
PQRI(v) :- Z(v), Z(nbj).
\end{lstlisting}
\end{minipage} 
&  
\begin{minipage}[c]{.13\linewidth}
\lstset{style=mystyle, frame=0}
\begin{lstlisting}[morekeywords={decl, output, symbol, number, magic}, basicstyle=\scriptsize\ttfamily\bfseries]
PQRI("3u37ezDdBg"). 
PQRI("uwsUc").
PQRI("zDdBgEbgMd"). 
PQRI("EbgMd2DV3K").
\end{lstlisting}
\end{minipage} 
&
\begin{minipage}[c]{.27\linewidth}
\lstset{style=mystyle, frame=0}
\begin{lstlisting}[morekeywords={input, printtuples}, basicstyle=\scriptsize\ttfamily\bfseries]
input relation Z(a:string)
output relation PQRI(a:string)
Z("3u37ezDdBg"). Z("uwsUc").
Z("zDdBgEbgMd"). Z("EbgMd2DV3K").
PQRI(v) :- Z(v), Z(nbj).
\end{lstlisting}
\end{minipage} 
&
\begin{minipage}[c]{.13\linewidth}
\lstset{style=mystyle, frame=0}
\begin{lstlisting}[morekeywords={decl, output, symbol, number, magic}, basicstyle=\scriptsize\ttfamily\bfseries]
PQRI("3u37ezDdBg"). 
PQRI("uwsUc").
PQRI("zDdBgEbgMd"). 
PQRI("EbgMd2DV3K").
\end{lstlisting}
\end{minipage} 
\\
1
&
\begin{minipage}[c]{.27\linewidth}
\lstset{style=mystyle, frame=0}
\begin{lstlisting}[morekeywords={input, printtuples}, basicstyle=\scriptsize\ttfamily\bfseries]
input relation Z(a:string)
input relation PQRI(a:string)
output relation PLEY(a:string)
Z("3u37ezDdBg"). Z("uwsUc").
Z("zDdBgEbgMd"). Z("EbgMd2DV3K").
PQRI("3u37ezDdBg"). PQRI("uwsUc").
PQRI("zDdBgEbgMd"). PQRI("EbgMd2DV3K").
PLEY(o) :- PQRI(x), Z(o), Z(x), 
           PQRI(z), PQRI(x).
\end{lstlisting}
\end{minipage} 
&  
\begin{minipage}[c]{.13\linewidth}
\lstset{style=mystyle, frame=0}
\begin{lstlisting}[morekeywords={decl, output, symbol, number, magic}, basicstyle=\scriptsize\ttfamily\bfseries]
PLEY("3u37ezDdBg"). 
PLEY("uwsUc").
PLEY("zDdBgEbgMd"). 
PLEY("EbgMd2DV3K").
\end{lstlisting}
\end{minipage} 
&
\begin{minipage}[c]{.27\linewidth}
\lstset{style=mystyle, frame=0}
\begin{lstlisting}[morekeywords={input, printtuples}, basicstyle=\scriptsize\ttfamily\bfseries]
input relation Z(a:string)
relation PQRI(a:string)
output relation PLEY(a:string)
Z("3u37ezDdBg"). Z("uwsUc").
Z("zDdBgEbgMd"). Z("EbgMd2DV3K").
PQRI(v) :- Z(v), Z(nbj).
PLEY(o) :- PQRI(x), Z(o), Z(x), 
            PQRI(z), PQRI(x).
\end{lstlisting}
\end{minipage} 
&
\begin{minipage}[c]{.13\linewidth}
\lstset{style=mystyle, frame=0}
\begin{lstlisting}[morekeywords={decl, output, symbol, number, magic}, basicstyle=\scriptsize\ttfamily\bfseries]
PLEY("3u37ezDdBg"). 
PLEY("uwsUc").
PLEY("zDdBgEbgMd"). 
PLEY("EbgMd2DV3K").
\end{lstlisting}
\end{minipage} 
\\
2
&
\begin{minipage}[c]{.27\linewidth}
\lstset{style=mystyle, frame=0}
\begin{lstlisting}[morekeywords={input, printtuples}, basicstyle=\scriptsize\ttfamily\bfseries]
input relation Z(a:string)
input relation PLEY(a:string)
output relation NFUV(a:string)
Z("3u37ezDdBg"). Z("uwsUc").
Z("zDdBgEbgMd"). Z("EbgMd2DV3K").
PLEY("3u37ezDdBg"). PLEY("uwsUc").
PLEY("zDdBgEbgMd"). PLEY("EbgMd2DV3K").
NFUV(q) :- Z(fym), PLEY(q).
\end{lstlisting}
\end{minipage} 
&  
\begin{minipage}[c]{.13\linewidth}
\lstset{style=mystyle, frame=0}
\begin{lstlisting}[morekeywords={decl, output, symbol, number, magic}, basicstyle=\scriptsize\ttfamily\bfseries]
NFUV("3u37ezDdBg"). 
NFUV("uwsUc").
NFUV("zDdBgEbgMd"). 
NFUV("EbgMd2DV3K").
\end{lstlisting}
\end{minipage} 
&
\begin{minipage}[c]{.27\linewidth}
\lstset{style=mystyle, frame=0}
\begin{lstlisting}[morekeywords={input, printtuples}, basicstyle=\scriptsize\ttfamily\bfseries]
input relation Z(a:string)
relation PQRI(a:string)
relation PLEY(a:string)
output relation NFUV(a:string)
Z("3u37ezDdBg"). Z("uwsUc").
Z("zDdBgEbgMd"). Z("EbgMd2DV3K").
PQRI(v) :- Z(v), Z(nbj).
PLEY(o) :- PQRI(x), Z(o), Z(x), 
            PQRI(z), PQRI(x).
NFUV(q) :- Z(fym), PLEY(q).
\end{lstlisting}
\end{minipage} 
&
\begin{minipage}[c]{.13\linewidth}
\lstset{style=mystyle, frame=0}
\begin{lstlisting}[morekeywords={decl, output, symbol, number, magic}, basicstyle=\scriptsize\ttfamily\bfseries]
NFUV("3u37ezDdBg"). 
NFUV("uwsUc").
NFUV("zDdBgEbgMd"). 
NFUV("EbgMd2DV3K").
\end{lstlisting}
\end{minipage} 
\\
3
&
\begin{minipage}[c]{.27\linewidth}
\lstset{style=mystyle, frame=0}
\begin{lstlisting}[morekeywords={input, printtuples}, basicstyle=\scriptsize\ttfamily\bfseries]
input relation PLEY(a:string)
input relation NFUV(a:string)
output relation OUT(a:string)
NFUV("3u37ezDdBg"). NFUV("uwsUc").
NFUV("zDdBgEbgMd"). NFUV("EbgMd2DV3K").
PLEY("3u37ezDdBg"). PLEY("uwsUc").
PLEY("zDdBgEbgMd"). PLEY("EbgMd2DV3K").
OUT(t) :- NFUV(ssz), PLEY(arv), PLEY(t).
\end{lstlisting}
\end{minipage} 
&  
\begin{minipage}[c]{.13\linewidth}
\lstset{style=mystyle, frame=0}
\begin{lstlisting}[morekeywords={decl, output, symbol, number, magic}, basicstyle=\scriptsize\ttfamily\bfseries]
OUT("3u37ezDdBg"). 
OUT("uwsUc").
OUT("zDdBgEbgMd"). 
OUT("EbgMd2DV3K").
\end{lstlisting}
\end{minipage} 
&
\begin{minipage}[c]{.27\linewidth}
\lstset{style=mystyle, frame=0}
\begin{lstlisting}[morekeywords={input, printtuples}, basicstyle=\scriptsize\ttfamily\bfseries]
input relation Z(a:string)
relation PQRI(a:string)
relation PLEY(a:string)
relation NFUV(a:string)
output relation OUT(a:string) 
Z("3u37ezDdBg"). Z("uwsUc").
Z("zDdBgEbgMd"). Z("EbgMd2DV3K").
PQRI(v) :- Z(v), Z(nbj).
PLEY(o) :- PQRI(x), Z(o), Z(x), 
            PQRI(z), PQRI(x).
NFUV(q) :- Z(fym), PLEY(q).
OUT(t) :- NFUV(ssz), PLEY(arv), PLEY(t).
\end{lstlisting}
\end{minipage} 
&
\begin{minipage}[c]{.13\linewidth}
\lstset{style=mystyle, frame=0}
\begin{lstlisting}[morekeywords={decl, output, symbol, number, magic}, basicstyle=\scriptsize\ttfamily\bfseries]
-
\end{lstlisting}
\end{minipage} 
\\\bottomrule
    \end{tabular}
\end{table}

\subsection{Bugs found by \tool}

There were 5 bugs might not be detectable by queryFuzz: bug 3 (\ie shown in Section~\ref{sec:debug3}), bug 4 (\ie shown in Section~\ref{sec:debug4}), bug 5 (shown in Section~\ref{sec:debug5}), bug 8 (\ie shown in Section~\ref{sec:debug8}), and bug 13 (\ie shown in Section~\ref{sec:debug13}).
queryFuzz's method for detecting bugs in programs beyond conjunctive queries relies on the concept of monotonicity;
in short, queryFuzz applies transformations only to conjunctive queries, and the final results should not contradict their oracles, because of monotonicity. 

Bug 3  was caused by subsumption, 
which removes facts from a relation based on the conditions specified in its body. If there are no facts that meet these conditions, the subsumption will not remove any facts. This bug causes the subsumption to fail to remove any facts, even if they meet the specified conditions. This behavior is identical to the case where no facts meet the conditions, as queryFuzz is unaware of whether any facts meet the conditions, it was impossible for queryFuzz to detect it. 
Bug 4 and bug 8 were also difficult for queryFuzz to detect for the same reason. 

Bug 5 was related to equivalence relations, and it not works under magic transformation. Equivalence relation is a binary relation in Souffl\'{e} and exhibits three properties: reflexivity, symmetry, and transitivity. For example, deriving a fact \lstinline[style=lstinlinestyle]{(1, 2)} for an  equivalence relation \lstinline[style=lstinlinestyle]{c}, will result in the following results: \lstinline[style=lstinlinestyle]{c(1, 2). c(2, 1). c(1, 1). c(2, 2).}. So when given an input to the equivalence relation, that input should be a subset of the final result, but there is no way to tell if they are equal (\eg, if the fact derived for \lstinline[style=lstinlinestyle]{c} is \lstinline[style=lstinlinestyle]{(1, 1)}, the results of \lstinline[style=lstinlinestyle]{c} will be \lstinline[style=lstinlinestyle]{c(1, 1).}). So it is challenging for queryFuzz to determine whether the equivalence relation works well, resulting in its inability to detect this bug.

Bug 13 was a bug in CozoDB, which was caused by the discrepancy between the comparison functions employed in the magic sets rewrite and the binary operators utilized in the rules. 
During the magic sets rewrite, CozoDB employs range scanning on the relation \lstinline[style=lstinlinestyle]{a} instead of testing each individual value. However, the comparison function used in the range scan considers \lstinline[style=lstinlinestyle]{-0.0 >= 0} to be true, while the comparison function used in the binary operators considers it to be false.
This bug can only be identified when the program is evaluated twice, once with the magic sets rewrite and once without it.
We believe that queryFuzz would not be able to detect this bug because the transformations it applies are designed specifically for conjunctive queries, which consist of single non-recursive function-free Horn rules. 
We got feedback from the developer of CozoDB that their magic sets rewrite only applied on conjunctive queries. 
Therefore, the transformations supported by queryFuzz, such as adding an existing subgoal into the rule body, modifying a variable, or removing a subgoal from the rule body, are inadequate for preventing the magic sets rewrite in this particular scenario.